\documentclass[12pt,notitlepage,aps,preprintnumbers,superscriptaddress,nofootinbib,aps,prd,floatfix]{revtex4-2}
\usepackage{graphicx,float}
\usepackage{amsthm} 
\usepackage{amsmath,amssymb,physics}
\usepackage{url} 
\usepackage{hyperref}
\usepackage{xcolor} 
\usepackage{multirow}
\definecolor{nicered}{rgb}{0.5,0.,0.}
\definecolor{nicegreen}{rgb}{0.,0.5,0.}
\definecolor{niceblue}{rgb}{0.,0.,0.5}
\definecolor{darkpink}{rgb}{0.8,0.47,0.47}
\hypersetup{colorlinks,citecolor=nicegreen,linkcolor=nicered,urlcolor=niceblue}
\usepackage{enumitem} 
\usepackage[normalem]{ulem}
\setlist{nolistsep} 
\usepackage{setspace}
\renewenvironment{eqnarray}{\begin{equation}\begin{aligned}}
{\end{aligned}\end{equation}}

\newcommand{\GeV}{\textrm{GeV}}
\newcommand{\TeV}{\textrm{TeV}}

\newcommand{\calL}{\mathcal{L}}
\newcommand{\ipb}{\textrm{pb}$^{-1}$}
\newcommand{\ifb}{\textrm{fb}$^{-1}$}

\begin{document}

\title{Precision studies of the post-CT18 LHC Drell-Yan data\\
in the CTEQ-TEA global analysis}

\author{Ibrahim Sitiwaldi}
\email{ibrahim010@sina.com}
\affiliation{School of Physics Science and Technology, 
Xinjiang University, Urumqi, Xinjiang 830046 China\looseness=-1}
\author{Keping Xie}
\email{xiekeping@pitt.edu}
\affiliation{Pittsburgh Particle Physics, Astrophysics, and Cosmology Center, 
Department of Physics and Astronomy, University of Pittsburgh, Pittsburgh, PA 15260, USA\looseness=-1}
\author{Alim Ablat}
\email{alimablat@stu.xju.edu.cn}
\affiliation{School of Physics Science and Technology, 
Xinjiang University, Urumqi, Xinjiang 830046 China\looseness=-1}
\author{Sayipjamal Dulat}
\email{sdulat@hotmail.com}
\thanks{Corresponding Author}
\affiliation{School of Physics Science and Technology, 
Xinjiang University, Urumqi, Xinjiang 830046 China\looseness=-1}
\author{Tie-Jiun Hou}
\email{tjhou@msu.edu}
\affiliation{School of Nuclear Science and Technology, University of South China, Hengyang, Hunan 421001, China\looseness=-1}
\author{C.-P. Yuan}
\email{yuanch@msu.edu}
\affiliation{Department of Physics and Astronomy, Michigan State University, East Lansing, MI 48824, USA\looseness=-1}

\collaboration{CTEQ-TEA Collaboration}
\date{\today}

\begin{abstract}
In this study, we examine closely the impact of the post-CT18 LHC Drell-Yan data on parton distribution functions (PDFs) in the general CTEQ-TEA global analysis framework. 
We compare the two main theoretical predictions, the MCFM fixed order calculations at next-to-next-to-leading order (NNLO) and the ResBos2 NNLO matched to $q_T$ resummation up to next-to-next-to-next-to-leading logarithmic (N3LL) level. We find that the overall inclusive cross sections agree well but the fiducial distributions can differ at a percent level.
We mainly discuss the result of the ResBos2 resummation calculation which yields a smaller Monte-Carlo uncertainty, and a better description to the post-CT18 LHC Drell-Yan data.
We find that the majority of post-CT18 LHC Drell-Yan data are consistent with the ATLAS 7 TeV  $W,Z$ data, which were included in the CT18A, but not CT18, fit and increases the strange quark distribution at the small $x$ region. The noticeable exception is that the ATLAS and LHCb 8 TeV $W$ data  pull $d(\bar{d})$ quark PDFs to the opposite direction with respect to the ATLAS 7 TeV $W,Z$ data.
The inclusion of these post-CT18 LHC  Drell-Yan data sets in the CTEQ-TEA global analysis is to update the CT18 PDFs following  similar trends as CT18Z PDFs. The parton luminosities and a few phenomenological implications with the fiducial $W^\pm,Z$ and inclusive $H,t\bar{t},t\bar{t}H$ productions at the 14 TeV LHC, as examples, are presented.
\end{abstract}

\preprint{MSUHEP-23-004, PITT-PACC-2310}

\maketitle
\tableofcontents

\section{Overview}
\label{sec:intro}
The Drell-Yan mechanism, first introduced by Sidney Drell and Tung-Mow Yan~\cite{Drell:1970wh},
describes the lepton pair production through quark-antiquark annihilation in hadron collisions.
It turns out to be a cornerstone process to establish the strong interaction theory, \emph{i.e.}, quantum chromodynamics (QCD).
The charged- and neutral-current Drell-Yan productions at the Super Proton Synchrotron (SPS) led to the discoveries of the $W$~\cite{UA1:1983crd,UA2:1983tsx} and $Z$~\cite{UA1:1983mne,UA2:1983mlz} bosons,
separately observed by UA1 and UA2 collaborations. 
Meanwhile, the Drell-Yan production can also provide a precise determination of the Standard Model (SM) parameters, such as the weak mixing angle~\cite{CMS:2011utm,CMS:2018ktx,ATLAS:2018gqq}, the strong coupling~\cite{CMS:2019oeb}, and the weak boson widths and branching ratios~\cite{Camarda:2016twt}. 
The latest high-precision measurement of $W$ boson mass performed by the CDF collaboration based  on the Tevatron Run II data shows a $7\sigma$ deviation from the SM expectation~\cite{CDF:2022hxs},
which inspires numerous Beyond the Standard Model (BSM) explanations.
In modern global QCD analyzes~\cite{Accardi:2016qay,Alekhin:2017kpj,Hou:2019efy,Bailey:2020ooq,NNPDF:2017mvq,NNPDF:2021njg,ATLAS:2021vod}, the Drell-Yan production plays a crucial role in constraining the parton distribution functions (PDFs).

The accumulation of high-quality data at the Large Hadron Collider (LHC) brings the experimental uncertainties to a percent or sub-percent precision level for many SM processes, especially for the Drell-Yan production.  
It mandatorily requires theoretical precision to enter the same level to analyze these data. 
Currently, next-to-next-to-leading order (NNLO) calculations in terms of the perturbative expansion of the strong coupling become a state-of-the-art~\cite{Heinrich:2020ybq}. In some instances, even the next-to-NNLO (N3LO) accuracy for some processes (such as Drell-Yan and Higgs productions) becomes available, and for some processes is in progress~\cite{Huss:2022ful}.
Meanwhile, next-to-leading order (NLO) corrections in terms of quantum electrodynamics (QED) (or electroweak in a more general sense) are also necessary for some specific processes~\cite{Bertone:2017bme,Xie:2021equ,Cridge:2021pxm}. 

Specifically for the Drell-Yan production, the NNLO fixed order correction to the total cross section has been known for three decades~\cite{Hamberg:1990np}.
The NNLO dilepton rapidity distribution of the virtual photon produced in the Drell-Yan process came afterwards~\cite{Anastasiou:2003yy,Anastasiou:2003ds}, followed by the fully differential kinematics in the leptonic decay of the vector boson $(W, Z/\gamma^*)$,
including the $\gamma$-$Z$ interference, finite-width effects, as well as spin-correlations~\cite{Melnikov:2006di,Melnikov:2006kv,Catani:2009sm}.
Nowadays, a few public NNLO codes with different subtraction schemes are available, such as the transverse moment ($q_T$) subtraction in DYNNLO~\cite{Catani:2007vq,Catani:2009sm} and MATRIX~\cite{Grazzini:2017mhc}, the sector decomposition in FEWZ~\cite{Li:2012wna,Gavin:2012sy}, and the $N$-jettiness in MCFM~\cite{Boughezal:2016wmq,Campbell:2019dru}. 
On the other side, the $q_T$ resummation calculation for Drell-Yan vector boson production has been established even for a longer time firstly with the Collins-Soper-Sterman (CSS) formalism~\cite{Collins:1984kg}, and later reorganized by Catani, Cieri, de Florian, and Grazzini~\cite{Catani:2013tia}. Recently, the soft-collinear effective theory (SCET)~\cite{Bauer:2000ew,Bauer:2000yr,Bauer:2001yt,Bauer:2002nz,Beneke:2002ph} provides a convenient framework to perform the resummation up to higher orders~\cite{Becher:2010tm,Becher:2011xn,Becher:2012yn}.
 Based on  various techniques, the predictions of $q_T$ resummation calculation also became available in a few public codes, such as ResBos(2)~\cite{Ladinsky:1993zn,Balazs:1997xd,Isaacson:2017hgb}, DYRes~\cite{Catani:2015vma}, DYqT~\cite{Bozzi:2008bb,Bozzi:2010xn}, 
DYTurbo~\cite{Camarda:2019zyx}, and CuTe~\cite{Becher:2011xn,Becher:2020ugp}.

Recently, based on the antenna subtraction method, the Drell-Yan inclusive cross section was calculated up to N3LO~\cite{Duhr:2020seh,Duhr:2021vwj}.
Afterwards, based on the $q_T$ subtraction method, the Drell-Yan N3LO cross sections have been calculated either inclusively~\cite{Chen:2021vtu}, or fully exclusively, with the $q_T$ resummation matched to the next-to-next-to-next-to-leading logarithimic (N3LL) level~\cite{Camarda:2021ict,Chen:2022cgv} and beyond~\cite{Neumann:2022lft}.
In the meantime, the next-to-leading order electroweak (EW) corrections have been known in Refs.~\cite{Baur:2001ze,Balossini:2008cs}.
Very recently, the mixed QCD-EW corrections have also been computed for the neutral-current Drell-Yan dilepton production at hadron colliders, which are found to be within a percent level for the invariant mass and (potentially) rapidity distributions~\cite{Bonciani:2021zzf}.

In this work, we will follow the QCD global analysis presented in Refs.~\cite{Dulat:2015mca,Hou:2019efy} and 
mainly consider the NNLO QCD fixed order calculation with the MCFM~\cite{Campbell:2019dru} as well as the matched $q_T$ resummation calculation up to N3LL, provided by the ResBos2 program~\cite{Isaacson:2017hgb}.
A minor difference up to a percent level is found in these two calculations. A similar level of difference also shows up
with different NNLO subtraction methods. See Appendix F of Ref.~\cite{Hou:2019efy} or Ref.~\cite{Alekhin:2021xcu} for details.

The Drell-Yan production has been precisely measured at the LHC, mostly delivered by ATLAS, CMS, and LHCb collaborations. Many Run I and Run II data have already been closely inspected and partially included in the global analysis of parton distribution functions (PDFs) of proton performed by various groups~\cite{Alekhin:2017kpj,Hou:2019efy,Bailey:2020ooq,NNPDF:2017mvq,NNPDF:2021njg,ATLAS:2021vod}.
In comparison with the proton-antiproton colliders, such as SPS or Tevatron, the LHC as a proton-proton collider embraces an advantage to provide more insights into the proton's light sea quark decomposition, as a result of that one initial parton in the quark-antiquark annihilation must come from the sea quarks.
As mentioned, with the reduction of experimental uncertainties, especially the statistical ones, a consistent theoretical description of data becomes more and more of a challenge. 
In many scenarios, the theoretical uncertainties, especially the PDF uncertainty, become a bottleneck for precision theoretical predictions or Monte Carlo simulations~\cite{Zakharchuk:2018dme,CDF:2022hxs}.  
All these aspects motivate the present study to understand the impact of the LHC precision Drell-Yan data on the CTEQ-TEA global  QCD analysis of proton PDFs.

This paper is organized as follows. In Sec.~\ref{sec:data}, we will
summarize the measurements of Drell-Yan production made by the ATLAS, CMS and LHCb experiments at the LHC and emphasize those in the post-CT18 era. Meanwhile, we will describe the theoretical predictions for these data, with the comparison of the MCFM NNLO fixed order calculation and the ResBos2 NNLO+N3LL resummation calculation (denoted as ``N3LL" in the rest of this work). The correlation between the data and the PDFs will be presented as well.
In Sec.~\ref{sec:impact}, we will examine the individual impact of these new Drell-Yan data on the three ensembles (CT18, CT18A~\cite{Hou:2019efy}, and CT18As~\cite{Hou:2022onq}) of PDFs, both using the ePump's fast Hessian profiling technique~\cite{Schmidt:2018hvu,Hou:2019gfw}, and the CTEQ-TEA global analysis. 
The simultaneous fits of these post-CT18 DY data together with the corresponding phenomenological implications are presented in Sec.~\ref{sec:fit}.
In Sec.~\ref{sec:conclude}, we will provide our conclusion and discussion.
For completeness, some additional figures are collected in Appendix~\ref{app:supp}.

\section{Post-CT18 LHC Drell-Yan data}
\label{sec:data}

\begin{table}
\resizebox{\textwidth}{!}{ 
\begin{tabular}{c|c|c|c|c|c|c|c|c}
\hline
Data & $\sqrt{s}$ [TeV] & $\mathcal{L}_{\rm int}$~[\ifb] &  Ref. & CT18 & 
MSHT20& NNPDF3.1 & NNPDF4.0 & ATLASpdf21 \\

\hline
\multicolumn{9}{c}{\bf ATLAS}  \\
\hline

low-mass DY $m_{\ell\ell}$ & 7  & 1.6 & \cite{ATLAS:2014ape} &  &   & 0.90(6) &  0.88(6)     \\   

\hline  
high-mass DY $m_{\ell\ell}$ & 7 & 4.9 & \cite{ATLAS:2013xny}  &  & 1.45(13) & 1.54(5) &  1.68(5) \\   

\hline  
$W,Z$ & 7 & 0.035 & \cite{ATLAS:2011qdp}  &  & 1.0 (30) & 0.96(30) & 0.98(30) \\   

\hline  
$W,Z$  & 7 & 4.6 & \cite{ATLAS:2016nqi}  &  & 1.91(61) & 2.14(34)  & 1.67(61) & 1.24(55)  \\   

\hline
$Z~(p_T,~y_Z)$  & 7  & 4.7 & \cite{ATLAS:2014alx}  &  &  &   &   \\ 

\hline  
$W$ & 8  & 20.2  & \cite{ATLAS:2019fgb}  & 4.96(22) & 2.61(22) &  & [3.50](22) & 1.40(22)\\ 

\hline  
high-mass DY $(m_{\ell\ell},y_{\ell\ell})$  & 8  & 20.3 & \cite{ATLAS:2016gic} &  & 1.18(48) &  & 1.11(48) &  \\   

\hline
 $Z~(m_{\ell\ell},y_{\ell\ell},\cos\theta^*)$  & 8  & 20.2 & \cite{ATLAS:2017rue} & 1.95(188) &  1.45(59)&  &1.22 (60) & 1.13(184)  \\   

\hline
 $Z~(p_T,~m_{\ell\ell})$  & 8  & 20.3 & \cite{ATLAS:2015iiu}  &  &  & 0.93(44)  & 0.91(44)  \\ 

\hline
 $Z~(p_T,~y_Z)$  & 8  & 20.3 & \cite{ATLAS:2015iiu}  & 1.1(27) &  1.81(104)& 0.94(48)  & 0.90(48)  \\ 

\hline  
$\sigma_{W,Z}^{\rm tot}$ & 13  & 0.081 & \cite{ATLAS:2016fij}   &  &  &  &  0.80(3) \\  

\hline
$W,Z~\sigma^{\rm fid,tot}$ &	2.76	 &	0.004 & \cite{ATLAS:2019fyu} &   &	 &   &   \\

\hline
$W,Z$ 	 &	5.02 &	0.025	&\cite{ATLAS:2018pyl} 	  & 1.15(27)   & & &   \\

\hline
$Z~p_T^{\ell\ell}$ &	13	 &	36.1	& \cite{ATLAS:2019zci} &      &   &	 &        \\	

\hline
\multicolumn{9}{c}{\bf CMS}  \\ 
\hline
 $W$ asym  & 7 & 0.036 &                \cite{CMS:2011bet} &  & 0.31(24) & \\
 
 \hline 
  $Z $  & 7 & 0.036 &                \cite{CMS:2011wyd} &  & 0.51(35) & \\

\hline
$W~A_e$ & 7  & 0.84 &  \cite{CMS:2012ivw} & 0.4(11) & 0.70(11) & 0.78(11) & 0.84(11)  \\

\hline  
$W~A_\mu$ & 7  & 4.7 & \cite{CMS:2013pzl} & 0.7(11) &  & 1.75(11) & 1.70(11)  \\

\hline  
$W\to \mu\nu$  & 8 & 18.8 & \cite{CMS:2016qqr} & 1.0(11) & 0.58(22) & 1.0(22) & 1.38(22)  \\

\hline  
DY $Z~(m_{\mu\mu}, y_{\mu\mu})$  & 7  & 4.5 & \cite{CMS:2013zfg} &  & 1.09(132) & 1.27(110) & 1.36(110)  \\ 

\hline  
DY $Z~(p_T,y_{\ell\ell})$ & 8  & 19.7 & \cite{CMS:2015hyl}  &  & poor fit  & 1.32(28) &  1.41(22) \\   

\hline
$Z~\phi^*,(y_{\ell\ell},\phi^*)$  &	8 	 &	19.7	& \cite{CMS:2017lvz}  &  &  &  & \\

\hline
$Z~y_Z,p_T,\phi_\eta^*$ &	13 	 &	35.9	& \cite{CMS:2019raw} & 9.24(12) &  &  &      \\	

\hline
$W~y_W,A_W,(\eta_\ell,p_T^\ell),A_\ell$ 	 &	13 	 &	35.9	& \cite{CMS:2020cph}   &  &  & &      \\	

\hline
$Z~(m_{\ell\ell})$	 &	13 	 &	2.8	& \cite{CMS:2018mdl}    &    &   &  & \\	
 
\hline  
\multicolumn{9}{c}{\bf LHCb}  \\ 
\hline

$Z\to ee$ & 7  & 0.94 & \cite{LHCb:2012gii} &  & 2.52(9) & 1.49(9) & 1.65(9)   \\ 
\hline
$W,Z \to \mu $ & 7 &  0.037 & \cite{LHCb:2012lki} &                &  1.25(10)&                  & \\

\hline  
$W,Z \to \mu$ & 7  & 1.0 & \cite{LHCb:2015okr} & 1.6(33) &    & 1.76(29) & 1.97(29)   \\ 
\hline  
$W,Z\to \mu$ & 8  & 2.0 & \cite{LHCb:2015mad}   & 2.1(34) &    & 1.37(30) & 1.42(30)  \\ 
\hline  
$W,Z\to \mu$ & 7+8  &  & \cite{LHCb:2015mad,LHCb:2015okr}   &  & 1.48(67)  &   &    \\ 

\hline   
$Z\to ee$ & 8  & 2.0 & \cite{LHCb:2015kwa}    & 1.0(17) & 1.54(17) & 1.14(17) & 1.33(17) \\

\hline 
$W \to e$ & 8  & 2.0 & \cite{LHCb:2016zpq}   & 1.52(14) &  &  & [2.61](8)  \\   

\hline
$Z\to ee$ & 13 & 0.294 & \cite{LHCb:2016fbk}   & 1.58(17) &  &  & 1.72(15)  \\   
\hline 
$Z\to\mu\mu$	 &	13	 &	5.1 & \cite{LHCb:2016fbk} & 0.89  &  &   &   0.99(16)  \\

\hline 
$Z\to\mu\mu$	 &	13	 &	5.1 & \cite{LHCb:2021huf} & 1.27(16)  &  &   &   \\

\hline
\end{tabular}
}
\caption{Comparison of the $\chi^2/N_{\rm pt}$ for the Drell-Yan ($W/Z$) data inlcuded in the CT18~\cite{Hou:2019efy}, MSHT20~\cite{Bailey:2020ooq}, and NNPDF3.1~\cite{NNPDF:2017mvq}, 4.0~\cite{NNPDF:2021njg} global analyses. The numbers in the parentheses indicate the data points.}
\label{tab:LHCDY}
\end{table}
Since the CT14 era~\cite{Dulat:2015mca}, the CTEQ-TEA global analysis began to include the LHC Run I data, with matrix elements calculated with the CTEQ internal codes and $K$-factors extracted from the ResBos~\cite{Ladinsky:1993zn,Balazs:1997xd}, FEWZ~\cite{Li:2012wna,Gavin:2012sy}, or VRAP~\cite{Anastasiou:2003yy,Anastasiou:2003ds}.
Specifically for Drell-Yan data sets, CT14 includes the LHCb 7 TeV inclusive $W/Z$ production~\cite{LHCb:2012lki}, CMS 7 TeV muon~\cite{CMS:2013pzl} and electron~\cite{CMS:2012ivw} charge asymmetry in $W$-boson decays, and the ATLAS 7 TeV Drell-Yan ($W^\pm$ and $Z/\gamma^*$) production~\cite{ATLAS:2011qdp}.
These CMS and ATLAS 7 TeV data sets were inherited in the CT18 analysis~\cite{Hou:2019efy}, while the LHCb one is updated with a higher luminosity data~\cite{LHCb:2015okr}.
Meanwhile, more LHC Run I Drell-Yan data were included in the CT18 analysis, with the fast interpolation grid technique, APPLgrid~\cite{Carli:2010rw}, generated with MCFM-6.8~\cite{Campbell:2010ff} and aMCfast~\cite{Bertone:2014zva}, together with NNLO $K$-factors generated with MCFM-8~\cite{Boughezal:2016wmq,Campbell:2015qma} and FEWZ~\cite{Li:2012wna,Gavin:2012sy}.

After the release of the CT18 PDFs, more Drell-Yan data become available. We summarize the ones relevant to this study as follows.
\begin{itemize}
\item \textbf{ATL5WZ.} The ATLAS collaboration measured $W^\pm$ and Z boson productions in pp collisions at the $\sqrt{s}=5.02~\TeV$, which serve as an important reference for $W^\pm$ and Z boson productions in proton–lead (p+Pb) and lead–lead (Pb+Pb) collisions at the same centre-of-mass energy.
The inclusive $W^\pm$ and $Z$ boson productions were measured for the fiducial integrated and differential cross sections with an integrated luminosity of 25~\ipb~\cite{ATLAS:2018pyl}, which we will dub as ATL5WZ as follows. The fiducial phase space is defined as,
\begin{eqnarray}\label{eq:ATLWfid}
p_T^{\ell,\nu}>25~\GeV,~|\eta_{\ell}|<2.5, ~ m_T>40~\GeV,
\end{eqnarray}
for $W^\pm$ production, and 
\begin{eqnarray}\label{eq:ATLZfid}
p_T^{\ell}>20~\GeV,~|\eta_{\ell}|<2.5,~ 66<m_{\ell\ell}<116~\GeV,
\end{eqnarray}
for on-shell $Z$ production.
The transverse mass of lepton-neutrino system is defined as $m_T=\sqrt{2p_T^{\ell} p_{T}^{\nu}(1-\cos\Delta\phi)}$, where $\Delta\phi$ is the azimuthal angle between $\vec{p}_T^{\ell}$ and $\vec{p}_T^{\nu}$.\\

In the upper panels of Fig.~\ref{fig:ATL5WZ}, we present the theoretical predictions obtained with CT18 NNLO PDFs by using the
MCFM-6.8 NLO~\cite{Campbell:2010ff}, MCFM-8 NNLO~\cite{Boughezal:2016wmq,Campbell:2015qma}, and ResBos NLO fixed-order~\cite{Ladinsky:1993zn,Balazs:1997xd} as well as ResBos2 $q_T$ resummation up to the N3LL+NNLO (denoted as ``N3LL") level~\cite{Isaacson:2017hgb} calculations.
Throughout this work, the renormalization and factorization scales $(\mu_R$ and $\mu_F)$ are set to the invariant mass of the vector boson $m_Z$ or $m_W$.
We use MCFM-6.8 to generate the fast computational APPLgrid~\cite{Carli:2010rw} tables. We have checked that both the MCFM-8 and ResBos NLO fixed order calculations reproduce the MCFM-6.8 results  at the NLO, as shown in Fig.~\ref{fig:ATL5WZ}. The $K$-factors were calculated as ratios to the MCFM-6.8 prediction, as shown in the lower insets of upper panels of Fig.~\ref{fig:ATL5WZ}. We see that the  differential fiducial cross section of Drell-Yan production predicted by the MCFM NNLO fixed-order calculation can differ from the ResBos2 resummation prediction by a percent level. 
Recall that different NNLO codes, such as FEWZ, DYNNLO and MCFM (with different subtraction schemes), can yield different predictions, at the percent level, on the differential fiducial cross section of Drell-Yan production, though their predictions agree well in total inclusive cross section~\cite{Alekhin:2021xcu,Hou:2019efy}.
This is because when both leptons of the Drell-Yan pair are required to have the same minimum values of transverse momentum, the transverse momentum of the Drell-Yan pair is about zero. In that kinematic region, a resummation calculation, summing up the effect of multiple soft gluon emissions, can provide better prediction than a fixed order calculation (which yields a singular result as the transverse momentum of the Drell-Yan pair approaches to zero). Hence, in this work, we will focus on the ResBos predictions, unless specified otherwise.
In general, we found that the Monte-Carlo uncertainty of the MCFM NNLO predictions is larger than that of ResBos2, which in principle can be improved by increasing the statistics. However, with a limited resource of computation power, we get a better convergence in the Resbos2 N3LL calculation. Comparing global fits with these two types of $K$-factors, we normally found a better $\chi^2$ for the ResBos N3LL predictions. For this reason, we will mainly present the fitted results with the ResBos $K$-factors in this work.\\

A similar comparison between the experimental data and theoretical predictions, after including the post-CT18 LHC Drell-Yan data in a global fit, will be presented in Sec.~\ref{sec:fit}.\\

In the lower panels of Fig.~\ref{fig:ATL5WZ}, we show the correlation cosine~\cite{Gao:2013bia} between the ATLAS 5.02 TeV $W/Z$ boson production data and the CT18 PDFs at $Q=100~\GeV$.
We see that the ATLAS 5.02 TeV $W^+$ data gives a large correlation to the 
$\bar{d}(d)$ PDFs around $x\sim10^{-3}$. It can be understood in terms of the LO partonic process $u\bar{d} \to W^+$, which directly probes the $\bar{d}$ PDF. The large $d$ quark PDF correlation originates from the co-evolution of $d$ and $\bar{d}$ PDFs due to the $g\to d\bar{d}$ splitting. In comparison, the $\bar{u}d \to W^-$ production gives large correlation to the $\bar{u}(u)$ PDFs, with a slightly milder value than the $\cos\phi(d(\bar{d}),W^+)$. It can be understood that the valence $u$ quark PDF is better constrained than $d$ quark PDF with the deep-inelastic scattering (DIS) data. For this reason, the ATLAS 5.02 TeV $W^-$ data can also constrain the $d$ quark PDF. Finally, the ATLAS 5.02 TeV  $Z$ production data gives a large correlation to $d(\bar{d})$ PDFs and a slightly smaller correlation to $u(\bar{u})$.

\begin{figure}[!h]
    \centering
    \includegraphics[width=0.328\textwidth]{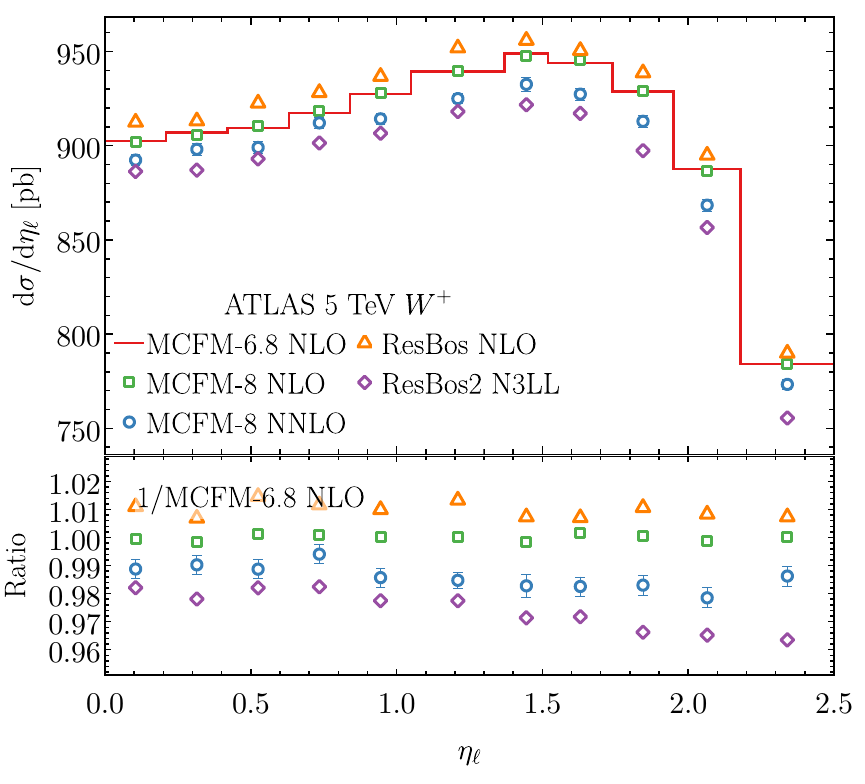}
    \includegraphics[width=0.328\textwidth]{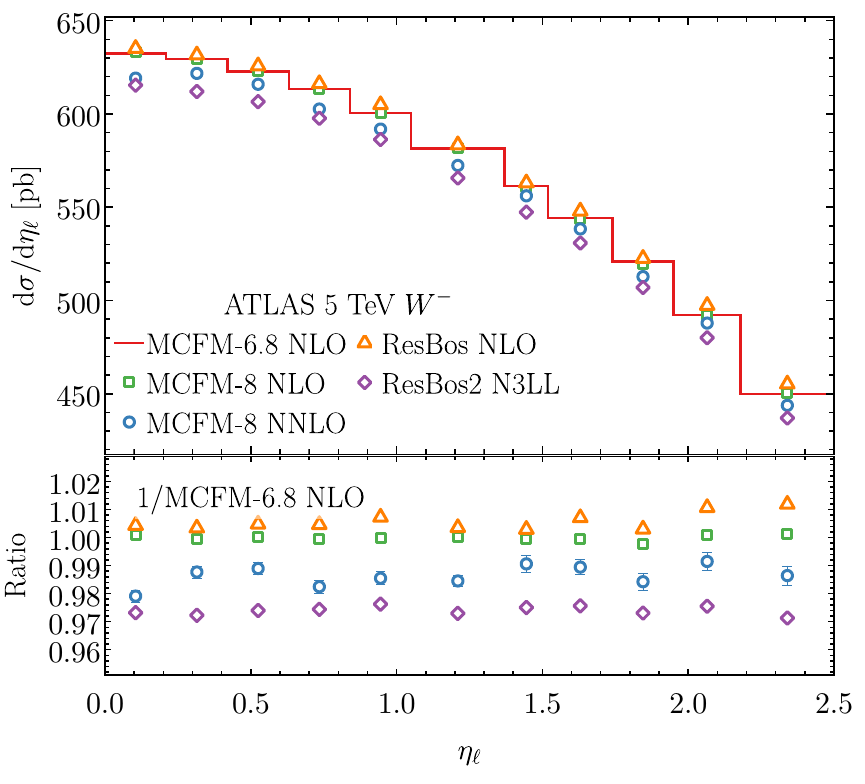}
    \includegraphics[width=0.328\textwidth]{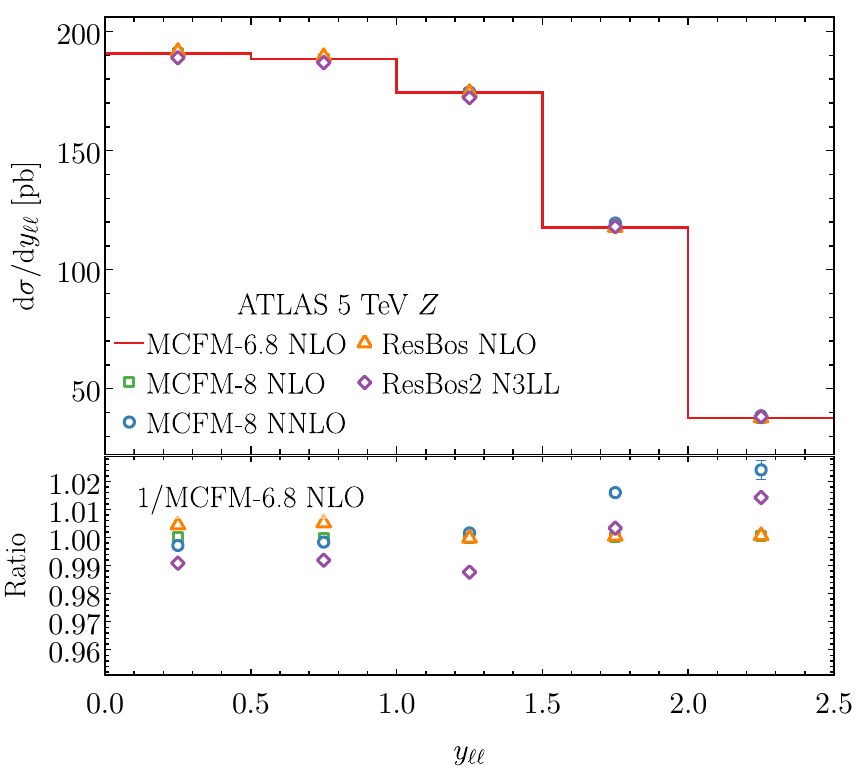}
    \includegraphics[width=0.328\textwidth]{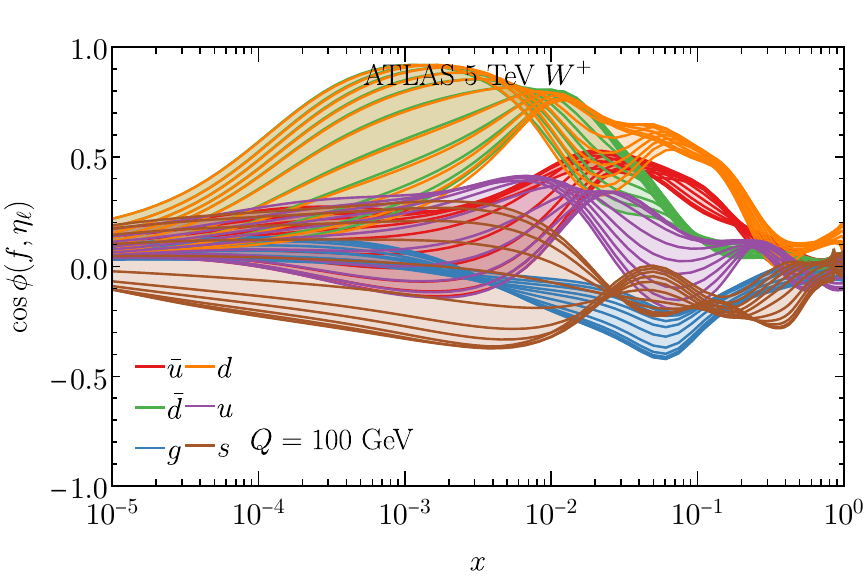}
    \includegraphics[width=0.328\textwidth]{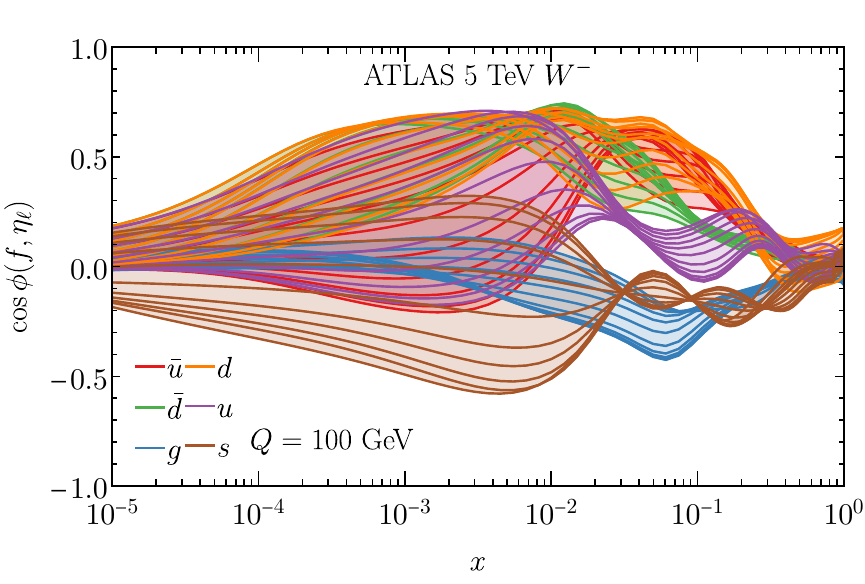}
    \includegraphics[width=0.328\textwidth]{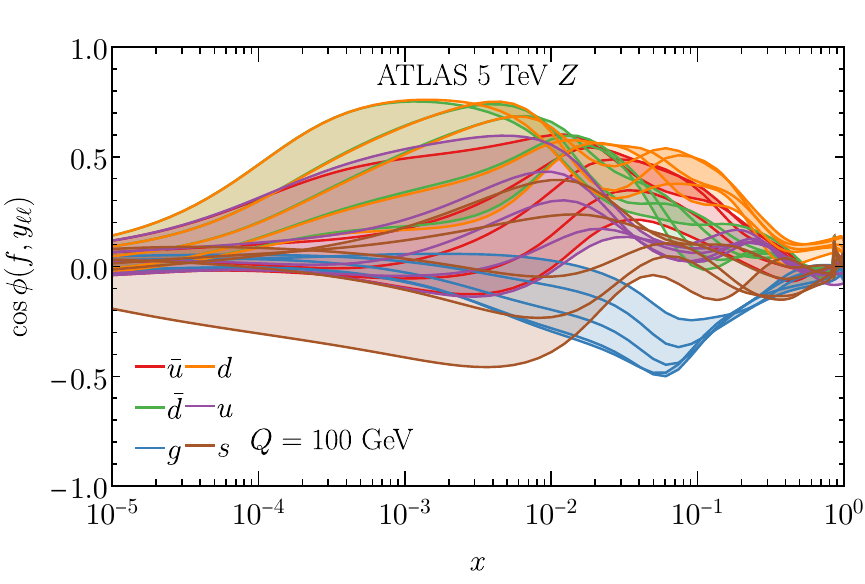}
    \caption{The comparison of CT18 predictions for the ATLAS 5.02 TeV $W/Z$ data (upper panels), and the correlation cosines between the CT18 PDFs and the  ATLAS 5.02 TeV $W/Z$ data points (lower panels).}
    \label{fig:ATL5WZ}
\end{figure}

\item \textbf{ATL8W.} The $W$ production in the muonic decay channel $(W\to \mu\nu)$ at the LHC 8 TeV was measured by the ATLAS collaboration with the 20.2~\ifb integrated luminosity~\cite{ATLAS:2019fgb}. We will call it ATL8W in this work.
The fiducial volume is defined as in Eq.~(\ref{eq:ATLWfid}), except 
the absolute muon pseudorapidity is required  as $|\eta_\ell|<2.4$.
The data were released in terms of the differential cross section as well as the charge asymmetry with respect to the muon pseudo-rapidity. Similarly, in Fig.~\ref{fig:ATL8W}, we compare the CT18 theoretical predictions in the upper panels and the corresponding correlation cosine in the lower panels. The main features of correlation already show up in the ATLAS 5.02 TeV case, while the specific strongly correlated $x$ value becomes slightly smaller due to the larger collision energy, in terms of the $x\sim(M_W/\sqrt{s})e^{\pm y}$ dependence. As we will see later, the ATL8W data can provide a strong constraint on $d(\bar{d})$-quark PDFs around $x\sim10^{-3}$.

\begin{figure}[!h]
    \centering
    \includegraphics[width=0.49\textwidth]{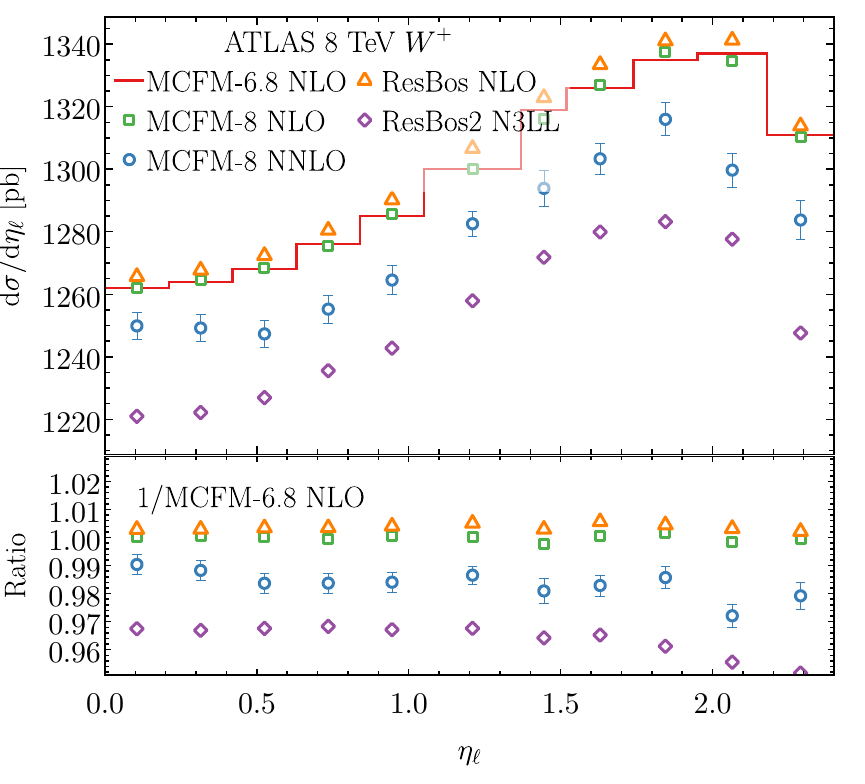}
    \includegraphics[width=0.49\textwidth]{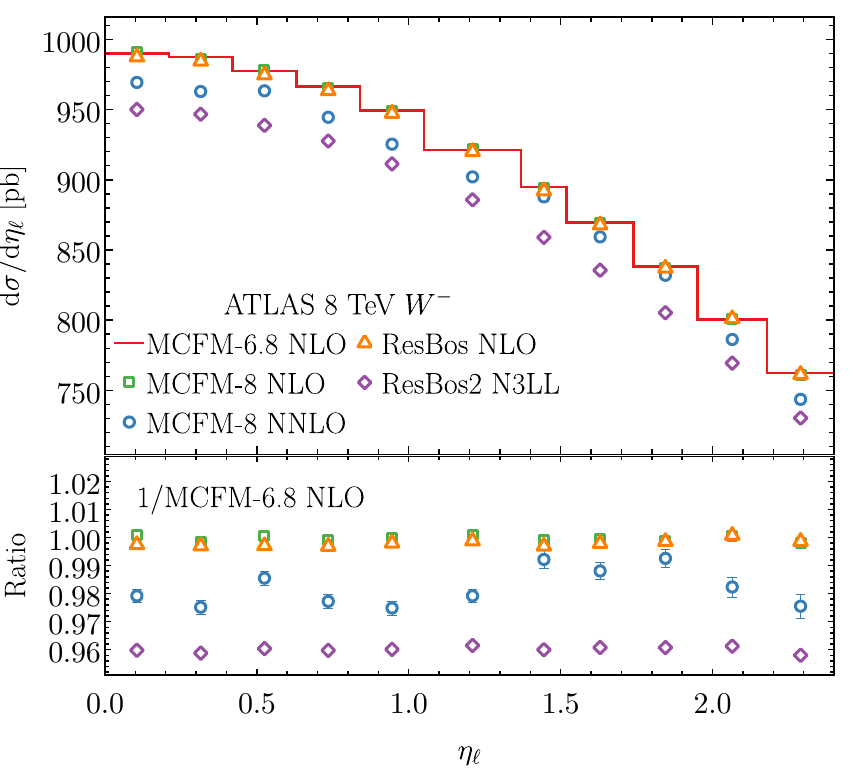} 
    \includegraphics[width=0.49\textwidth]{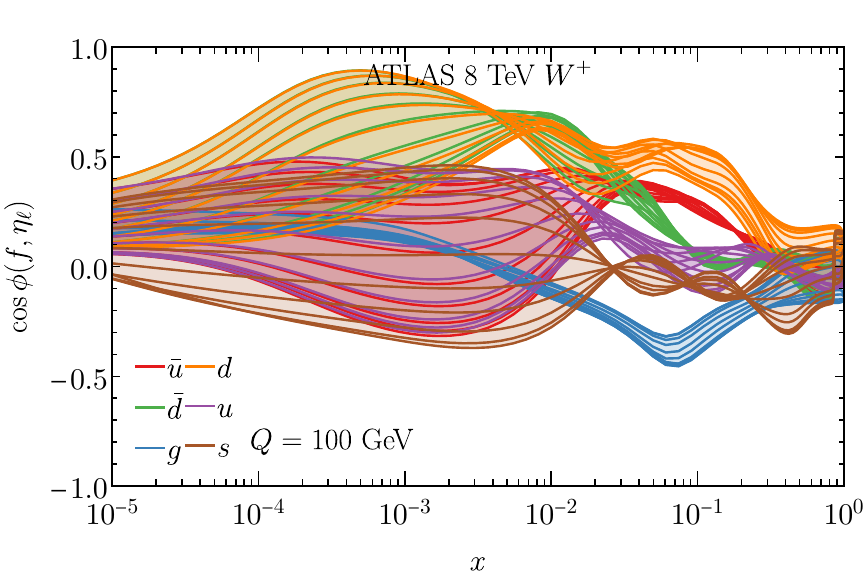}
    \includegraphics[width=0.49\textwidth]{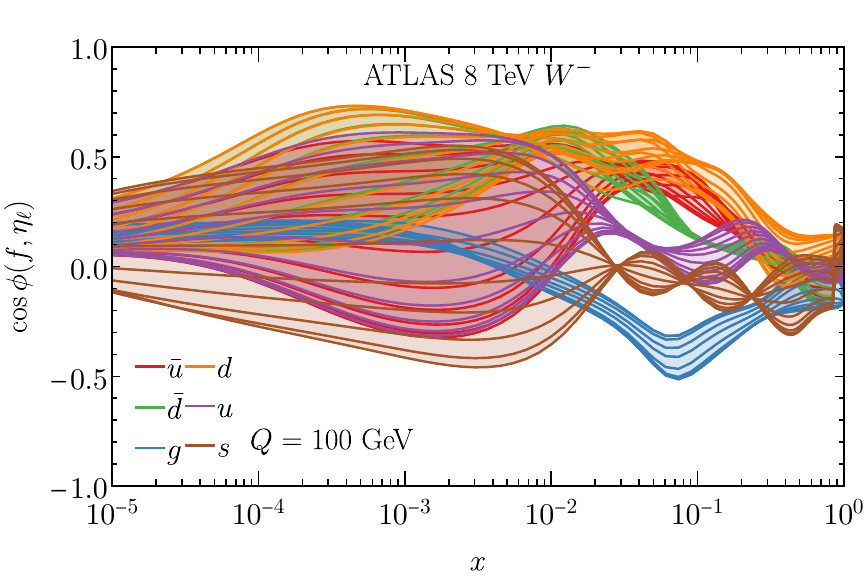}      
    \caption{Similar to Fig.~\ref{fig:ATL5WZ}, but for the ATLAS 8 TeV $W$ data.}
    \label{fig:ATL8W}
\end{figure}

\item \textbf{ATL8Z3D.} The neutral-current Drell-Yan production at the LHC 8 TeV for both electron and muon decay channels was measured by the ATLAS collaboration with an integrated luminosity of 20.2~\ifb~\cite{ATLAS:2017rue}.
For the central leptons, the fiducial region is defined as 
\begin{equation}
p_T^\ell>20~\GeV,~|\eta_\ell|<2.4,~46<m_{\ell\ell}<200~\GeV.
\end{equation}
 The data were presented as a triple differential distribution in terms of invariant mass $m_{\ell\ell}$, rapidity of dilepton $y_{\ell\ell}$, and cosine of the Collin-Soper angle $\cos\theta^*$\footnote{The cosine of the Collins-Soper angle is defined as
$\cos\theta^*=\frac{p_{z,\ell\ell}}{m_{\ell\ell}|p_{z,\ell\ell}|}\frac{p_1^+p_2^--p_1^-p_2^+}{\sqrt{m_{\ell\ell}^2+p_{T,\ell\ell}^2}}$, where $p_{i}^{\pm}=E_i\pm p_{z,i}$ and $i=1,2$ denote the negatively and positive charged leptons~\cite{Collins:1977iv}.}, which we name as ATL8Z3D. Similar to the ATLASpdf21~\cite{ATLAS:2021vod} analysis, we don't consider the high rapidity electron channel in this work.\\

For this data set, we obtain the theoretical predictions directly from the ATLAS collaboration, which was employed in the ATLASpdf21 PDF determination~\cite{ATLAS:2021vod} with the xFitter framework~\cite{Alekhin:2014irh}. In this analysis, the NLO predictions are calculated by using APPLgrid~\cite{Carli:2010rw}, generated with the MCFM-6.8~\cite{Campbell:2010ff}, while NNLO QCD and NLO EW corrections are folded into the $K$-factors with \textsc{NNLOjet}~\cite{Currie:2016bfm}. Meanwhile, the $ K$-factors provided by the ATLAS collaboration also incorporate the acceptance of the lepton fiducial cuts for every $(m_{\ell\ell},y_{\ell\ell},\cos\theta^*)$ bin. The theoretical predictions in comparison with data before and after including this data set in global fits will be presented later.
We remind readers that the MSHT20 and NNPDF4.0 treated this data differently from ATLASpdf21 and this study. Namely,  in their PDF analyses, they have summed over the $\cos\theta^*$ bins to obtain the double differential distributions in terms of $(m_{\ell\ell},y_{\ell\ell})$, which correspond to a smaller number of data points.\\

In Fig.~\ref{fig:ATL8Z3D}, we show the correlation cosine of the ATLAS 8 TeV $Z$ 3D data and the CT18 PDFs. Similar to ATLAS 5.02 TeV $Z$ case, we obtain somewhat larger correlation to the $u(\bar{u})$, $d(\bar{d})$ and $s(\bar{s})$\footnote{The CT18  fit assumes $s=\bar{s}$ at the starting scale $Q_0(=1.3~\GeV)$.} PDFs. 
This can be understood that this triple differential distribution can provide more exclusive information to constrain the PDFs in terms of the $(x,Q)$ kinematics.

\begin{figure}[!h]
    \centering
    \includegraphics[width=0.32\textwidth]{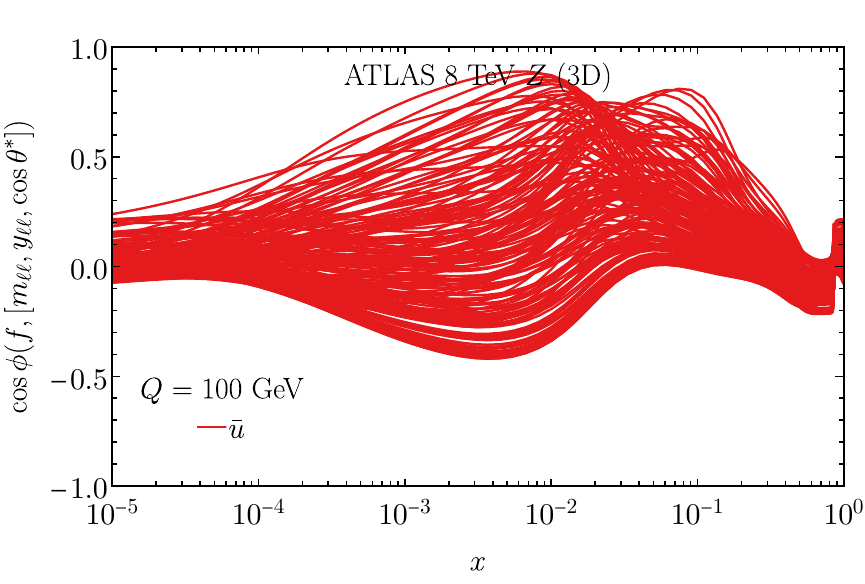}
    \includegraphics[width=0.32\textwidth]{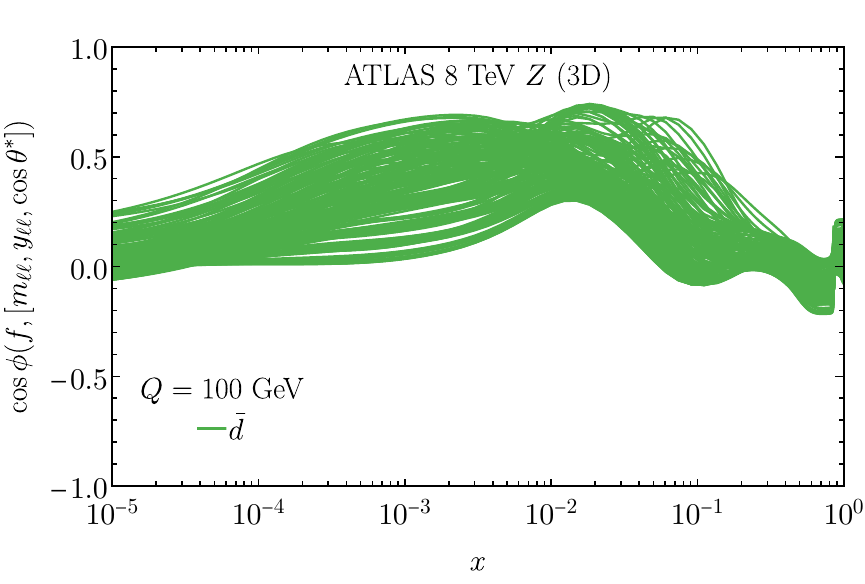}
    \includegraphics[width=0.32\textwidth]{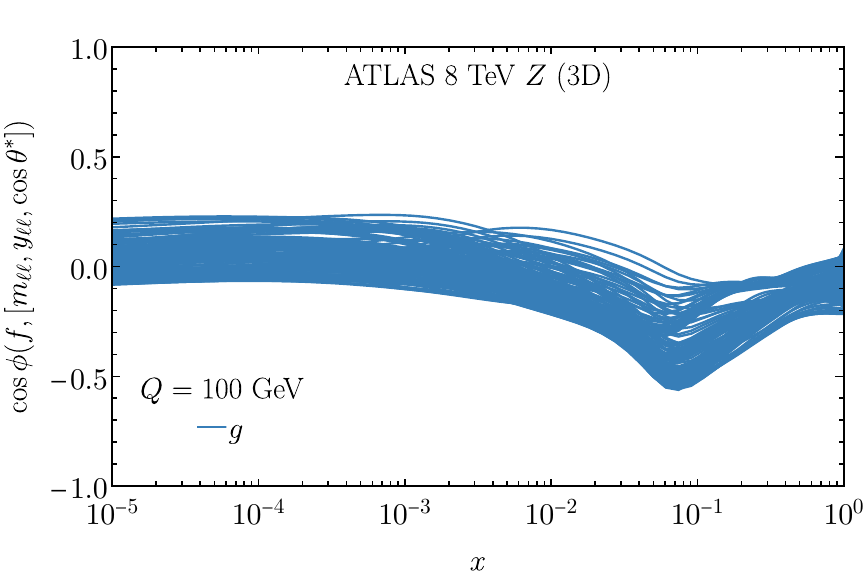}
    \includegraphics[width=0.32\textwidth]{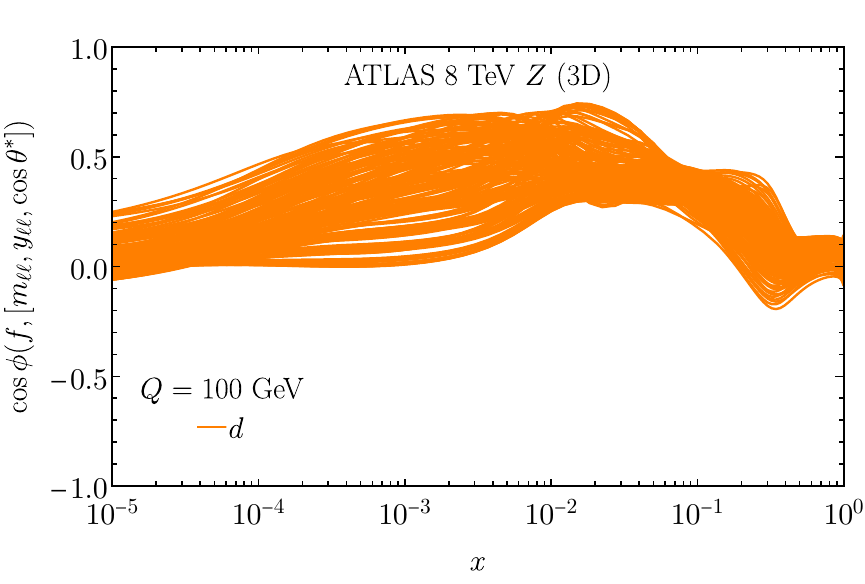}
    \includegraphics[width=0.32\textwidth]{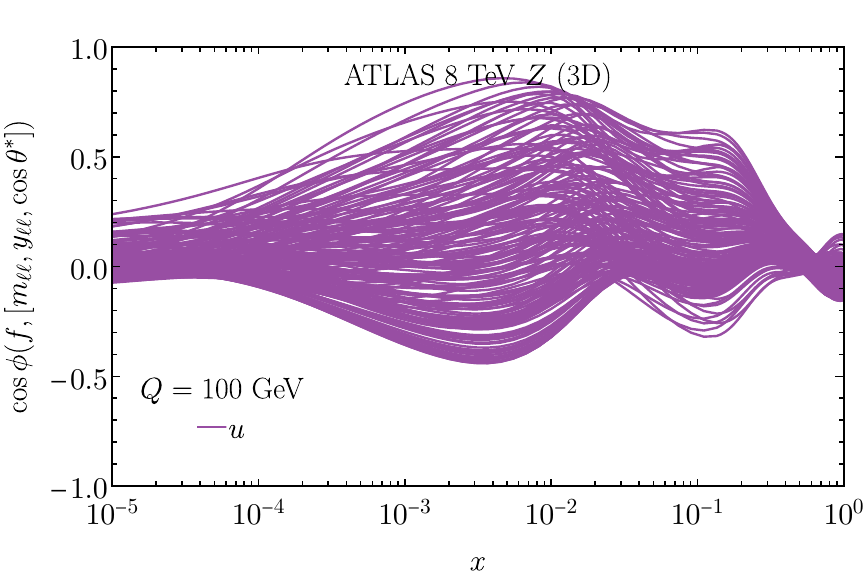}
    \includegraphics[width=0.32\textwidth]{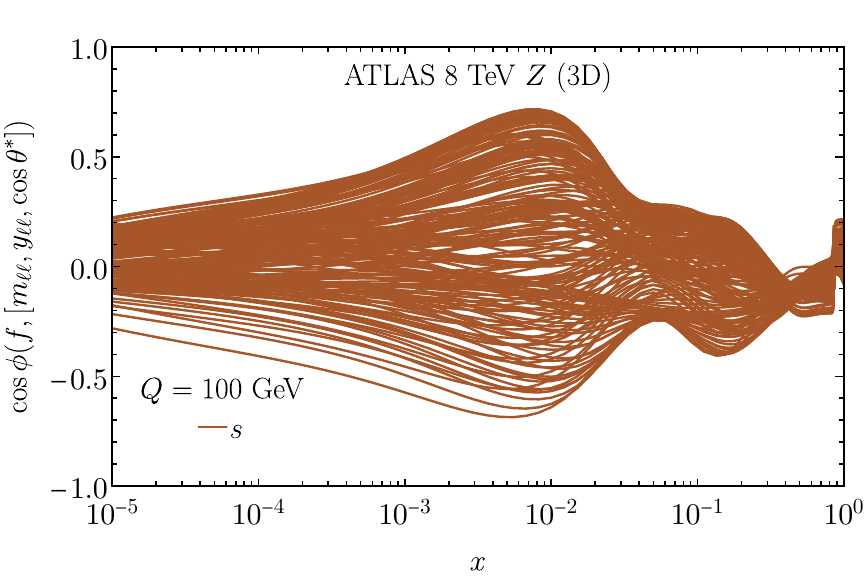}   
    \caption{The correlation cosines between the CT18 PDFs and the ATLAS 8 TeV $Z$ 3D data.}
    \label{fig:ATL8Z3D}
\end{figure}

\item \textbf{CMS13Z.} The differential cross sections for $Z$ boson production at the LHC 13 TeV were measured by the CMS collaboration with an integrated luminosity of 35.9~\ifb~\cite{CMS:2019raw}, which is denoted as CMS13Z in this work.
The fiducial phase space requires leptons to have $p_T^\ell>25~\GeV$, $|\eta_{\ell}|<2.4$ and $|m_{\ell\ell}-M_Z|<15~\GeV$. The data were presented in terms of one-dimensional distributions of rapidity $y_Z$, transverse momentum $p_T^Z$, and the optimized angular variable $\phi_{\eta}^*$\footnote{The optimized angular variable is defined as
$\phi_{\eta}^*=\tan(\frac{\pi-\Delta\phi}{2})/\cosh(\Delta\eta)$,
where $\Delta\eta$ and $\Delta\phi$ are the difference in pseudo-rapidity and azimuthal angle between the two leptons~\cite{Banfi:2010cf}. It probes similar physics as the boson transverse momentum, but with a better experimental resolution~\cite{Banfi:2012du}.} of the lepton pairs, respectively. 
The double differential distribution in terms of $(y_Z,p_T^Z)$ was also presented. In this study, we will mainly focus on the rapidity distribution, while other distributions are left for a future study that focuses on the vector boson transverse momentum $(p_T)$ distributions. 
In Fig.~\ref{fig:CMS13Z}, we display our theoretical predictions and the correlation cosine, which shows a large correlation to the strangeness PDF, while anti-correlation to the $u(\bar{u})$ PDFs.

\begin{figure}[!h]
    \centering
    \includegraphics[width=0.49\textwidth]{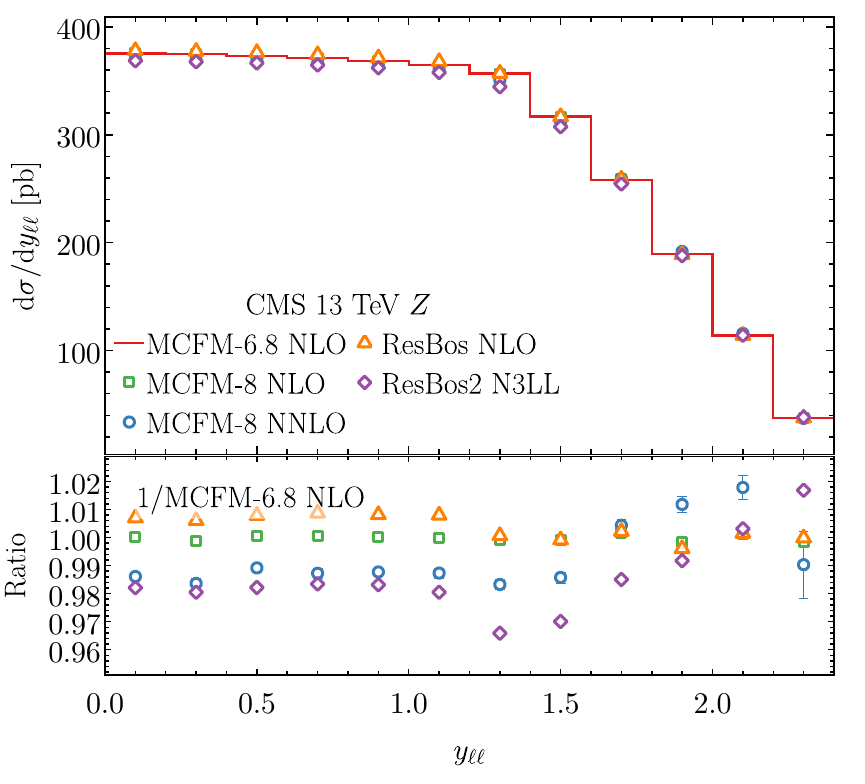}
    \includegraphics[width=0.49\textwidth]{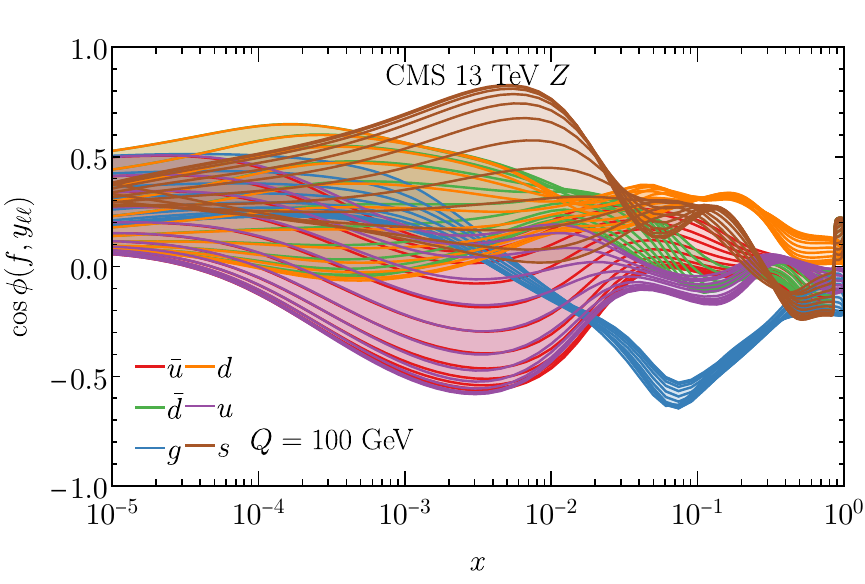}    
    \caption{Similar to Fig.~\ref{fig:ATL5WZ}, but for the CMS 13 TeV $Z$ data.}
    \label{fig:CMS13Z}
\end{figure}

\item \textbf{LHCb8W.} The $W\to e\nu$ production at the 8 TeV LHC 8 in the forward region was measured by the LHCb with an integrated luminosity of 2~\ifb~\cite{LHCb:2016zpq}. We name this data set as LHCb8W in this work. The fiducial region is defined as $2.0<\eta<4.25$ and $p_T^e>20~\GeV$, without additional requirements on the missing energy or transverse mass. The data were presented as differential cross section  and charge ratio  as a function of the electron pseudorapidity.
In this study, we focus on the absolute differential cross sections.
The CT18 predictions on the LHCb 8 TeV $W$ production cross section and the correlation cosine are shown in Fig.~\ref{fig:LHCb8W}.
We note a large variation of various theory predictions for the forward boundary bins, such as the last ones in the upper panels of Fig.~\ref{fig:LHCb8W}.
This arises mainly because the boundary phase space constrains the Monte Carlo statistics and results in a large theoretical uncertainty. At a higher order, the opening up of new phase space will yield a large QCD correction. As explored later, the inclusion of the boundary bins will lead to very poor fits. Therefore, we will exclude them from our canonical fits.\\

The correlation cosines of the LHCb 8 TeV $W$ data show similar behavior as the ATLAS 8 TeV $W$ data. However, the sensitive $x$ value becomes smaller as a result of the exponential suppression in $x\sim (M_W/\sqrt{s})e^{-y}$, for being produced at the forward (pseudo)rapidity region.

\begin{figure}[!h]
    \centering
    \includegraphics[width=0.49\textwidth]{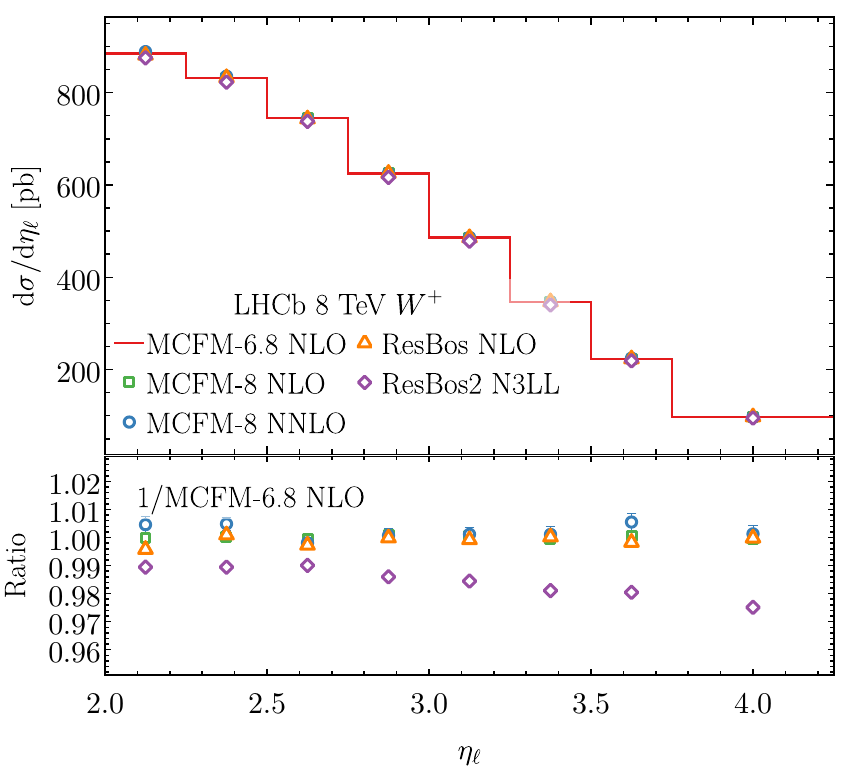}
    \includegraphics[width=0.49\textwidth]{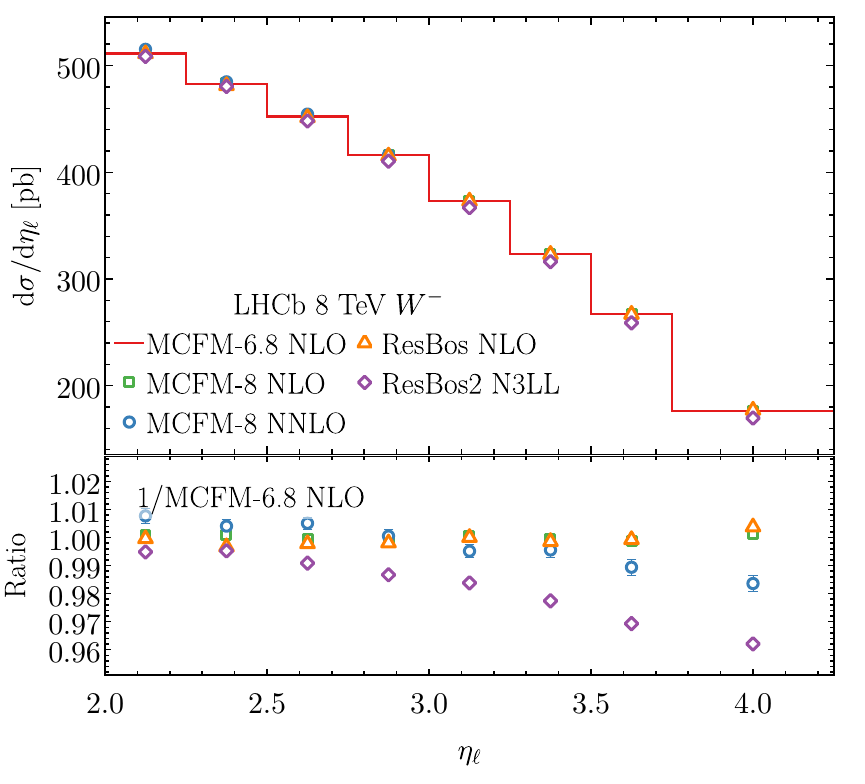} 
    \includegraphics[width=0.49\textwidth]{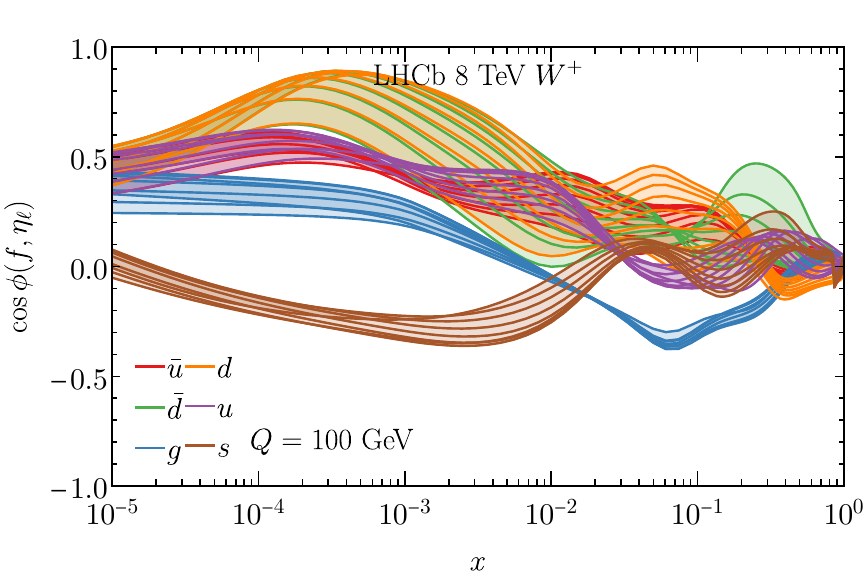}
    \includegraphics[width=0.49\textwidth]{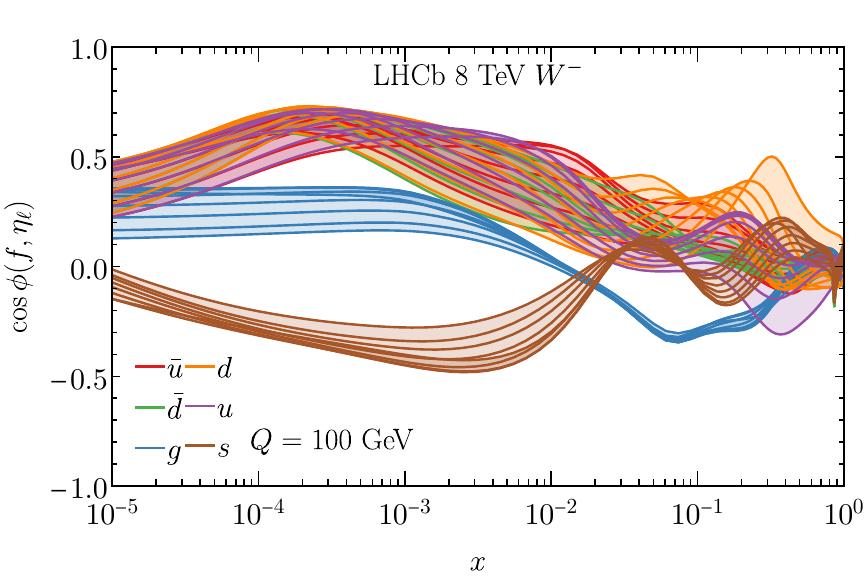}    
    \caption{Similar to Fig.~\ref{fig:ATL5WZ}, but for the LHCb 8 TeV $W$ data.    
}
    \label{fig:LHCb8W}
\end{figure}

\item \textbf{LHCb13Z.} The forward $Z$ boson production in the 13 TeV $pp$ collision in both electron and muon decay channels were measured by the LHCb collaboration, with an integrated luminosity 0.29~\ifb~\cite{LHCb:2016fbk}, which we name as LHCb13Z.
The fiducial volume is defined as $2.0<\eta_{\ell}<4.5,~p_T^\ell>20~\GeV$ and $60<m_{\ell\ell}<120~\GeV$. The differential cross sections were measured in terms of the rapidity, transverse momentum, and optimized angular variable $\phi_{\eta}^*$. Afterward, the muon decay channel data were updated with a higher integrated luminosity, 5.1~\ifb~\cite{LHCb:2021huf}. Our theoretical predictions as well as the correlation cosines are shown in Fig.~\ref{fig:LHCb13Z}. Similarly, we obtain large uncertainty for the boundary bins (the first and last ones), which will be excluded in our canonical  fits. In comparison with the CMS 13 TeV data, the LHCb 13 TeV $Z$ data show stronger correlations to the $u(\bar{u})$ and $d(\bar{d})$ PDFs  in the small-$x$ region.\\

As to be discussed below, the CT18 PDFs can describe the lower luminosity data~\cite{LHCb:2016fbk} very well 
as a result of the relatively large statistical uncertainty, with $\chi^2/N_{\rm pt}$ shown in Tab.~\ref{tab:chi2ePump}.
In another word, this data set has a very limited impact on our global fit. In contrast, the updated one with higher luminosity~\cite{LHCb:2021huf} can provide reasonable constraints on PDFs. For this reason, we will only include the updated data in  our canonical analysis.

\begin{figure}[!h]
    \centering
\includegraphics[width=0.49\textwidth]{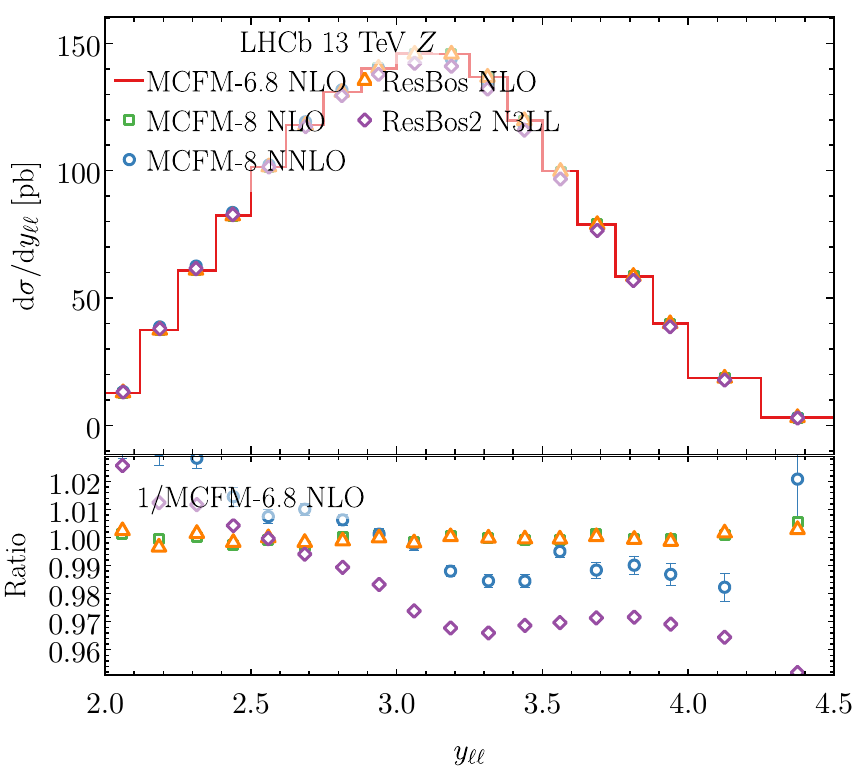}  
\includegraphics[width=0.49\textwidth]{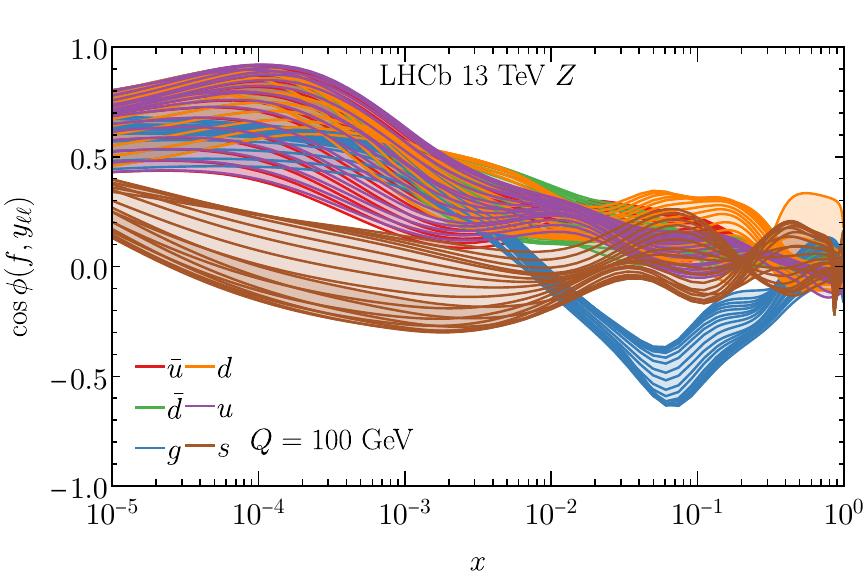}     
\caption{Similar to Fig.~\ref{fig:ATL5WZ}, but for the LHCb 13 TeV $Z$ data.      
}
    \label{fig:LHCb13Z}
\end{figure}

\end{itemize}

We summarize all these new LHC Drell-Yan data sets as well as those included in the previous CT18 PDF fit~\cite{Hou:2019efy} in Tab.~\ref{tab:LHCDY}. Meanwhile, we also compare the fitting quality $\chi^2/N_{\rm pt}$ and the corresponding number of data points included in the global analyses of the MSHT20~\cite{Bailey:2020ooq}, NNPDF3.1~\cite{NNPDF:2017mvq}, and NNPDF4.0~\cite{NNPDF:2021njg}.

In this study, we will not consider the high-mass Drell-Yan data, such as ATLAS 7 TeV~\cite{ATLAS:2013xny} and CMS 13 TeV~\cite{CMS:2018mdl} invariant mass of the dilepton pair $(m_{\ell\ell})$ distribution, the ATLAS 8 TeV~\cite{ATLAS:2016gic} and CMS 7 TeV~\cite{CMS:2013zfg} double differential $(m_{\ell\ell},y_{\ell\ell})$ distribution. A good description of these data sets requires the inclusion of EW corrections, which incorporates photon as a parton content~\cite{Xie:2021equ}. This will be left for a future study on updating the photon PDF of the proton. Furthermore, we don't consider the transverse momentum distribution of $Z$ boson either. As studied in Ref.~\cite{Isaacson:2017hgb} and references therein, the low $p_T^Z$ data requires a $q_T$ resummation calculation, which involves a nonperturbative transverse-momentum-dependent (TMD) parameter determination. The high $p_T^Z$ data suffer a large EW correction as studied in the CT18 analysis~\cite{Hou:2019efy}. We leave these aspects to future dedicated studies.

\section{Impacts on the CT18 PDFs} 
\label{sec:impact}

\begin{table}[!h]
\centering
\begin{tabular}{c|c|c|c|c|c|c|c}
\hline
\multirow{2}{*}{ID} & \multirow{2}{*}{Experiment} 
& \multirow{2}{*}{$N_{\rm pt}$} 
& \multicolumn{5}{c}{$\chi^2/N_{\rm pt}$} \\
\cline{4-8}
&  &  & Pre-fit$^\dagger$ & ePump$^\dagger$ & CT18 &  CT18A & CT18As \\
\hline
215 & ATLAS 5.02 TeV $W,Z$ & 27 & 1.15 & 0.96 & 1.07 & 0.74 & 0.71 \\
211 & ATLAS 8 TeV $W$ & 22 & 4.96 & 2.98 & 2.46 & 2.72  & 2.49 \\
214 & ATLAS 8 TeV $Z$ 3D & 188 &1.95& 1.18 & 1.16 & 1.13 & 1.14\\
212 & CMS 13 TeV $Z$ & 12 &9.24& 2.93 & 2.75 & 1.89 & 2.02\\
216 & LHCb 8 TeV $W$ & (16)14 &(3.48)1.52& (3.24)1.45 & (2.81)1.33 & (1.89)1.45 & (3.00)1.52\\
    & LHCb 13 TeV $Z$  & 18 & 0.89 & 0.88 & 0.99 & 0.92 & 0.90 \\
213 & LHCb 13 TeV $Z\to\mu^+\mu^-$ & (18)16 &(2.39)1.27& (2.33)1.17 & (2.55)1.12 & (2.49)1.12 & (2.28)0.87\\
\hline
\end{tabular}
\caption{Comparisons of $\chi^2/N_{\rm pt}$ for the post-CT18 LHC Drell-Yan data which are included one by one in the fit.
The numbers in the parentheses correspond to the case when including all the LHCb  data points, while the outside ones are for the case  when excluding two boundary bins of the LHCb data.
$^\dagger$The $\chi^2/N_{\rm pt}$ values for the pre-fit and ePump  are based on the CT18 PDFs.
}
\label{tab:chi2ePump}
\end{table}

Before performing global fits with the inclusion of all these post-CT18 LHC Drell-Yan data sets simultaneously, we would like to examine the impact of the individual data set on the CT18 PDFs. This can be done by performing the ePump updating~\cite{Schmidt:2018hvu,Hou:2019gfw} or a global fitting by including one  new data set at a time on the top of CT18. Note that the CT18+ATL7WZ fit is the same as the CT18A which includes the ATLAS 7 TeV $W,Z$ precision data set (named as ATL7WZ)~\cite{ATLAS:2016nqi}.
The comparisons of $\chi^2/N_{\rm pt}$ from CT18's pre-fit, ePump updating and global fitting are summarized in Tab.~\ref{tab:chi2ePump}.

In comparison with the ePump updating, in genral, the $\chi^2/N_{\rm pt}$ decreases slightly in the global fitting, reflecting the non-negligible impact of these new data sets. 
Two exceptions happen to the ATL5WZ and LHCb13Z data sets, due to the pull of other data sets in the  global fitting. 
Nevertheless, both data sets can be well described by CT18 with $\chi^2/N_{\rm pt}$ about 1.
Considering the pre-fit values of $\chi^2/N_{\rm pt}\sim1$, the impact of these two data sets is expected to be minimal, which will be  illustrated later.

For comparison, we also present the global fitted $\chi^2/N_{\rm pt}$ with the setup of CT18A and CT18As in Tab.~\ref{tab:chi2ePump}.
Recall that 
CT18A is CT18 with the inclusion of ATL7WZ precision data which shows sizable tension with other data sets, such as the HERA I+II combined data and neutrino DIS di-muon data~\cite{Hou:2019efy}.
On top of CT18A, the CT18As fit includes more degrees of freedom in the strangeness parameterization to allow $s$ not equal to $\bar s$ at the initial scale $Q_0 =1.3~\GeV$ which relaxes the above-mentioned tension~\cite{Hou:2022onq}.
In comparison with the fitted CT18 $\chi^2/N_{\rm pt}$, we see the fitted CT18A ones generally have smaller values,
suggesting the consistency between the ATL7WZ with these new DY data sets. 
In contrast, the $\chi^2/N_{\rm pt}$ for the ATL8W data increases from 2.46 to 2.72, as an indication of some
tension. With a more flexible strangeness parameterization in CT18As, this tension can be relaxed down to $\chi^2/N_{\rm pt}=2.49$. 
A minor increase of $\chi^2/N_{\rm pt}$  value is observed when using
the LHCb8W data to update  CT18As as compared to CT18 and CT18A.
But, considering the small variation with respect to the ATL8W one, the impact of LHCb8W is expected to be small, which will be examined more closely later.

\begin{figure}[!h]
\centering
\includegraphics[width=0.49\textwidth]{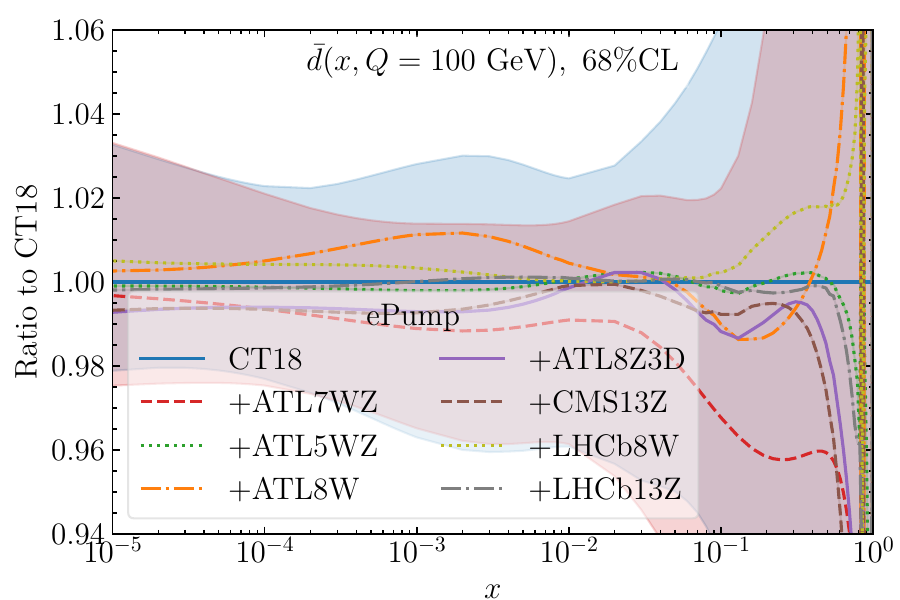}   
\includegraphics[width=0.49\textwidth]{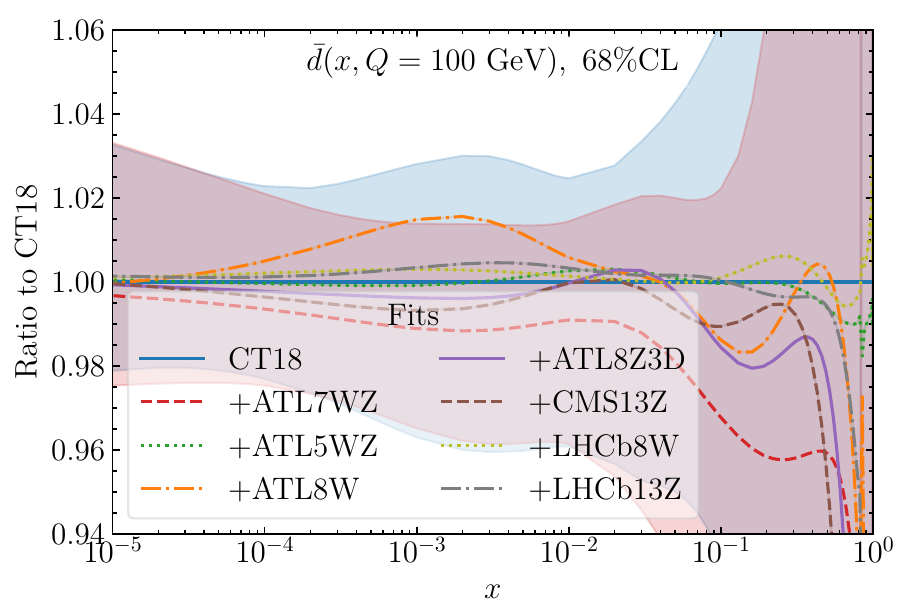}
\includegraphics[width=0.49\textwidth]{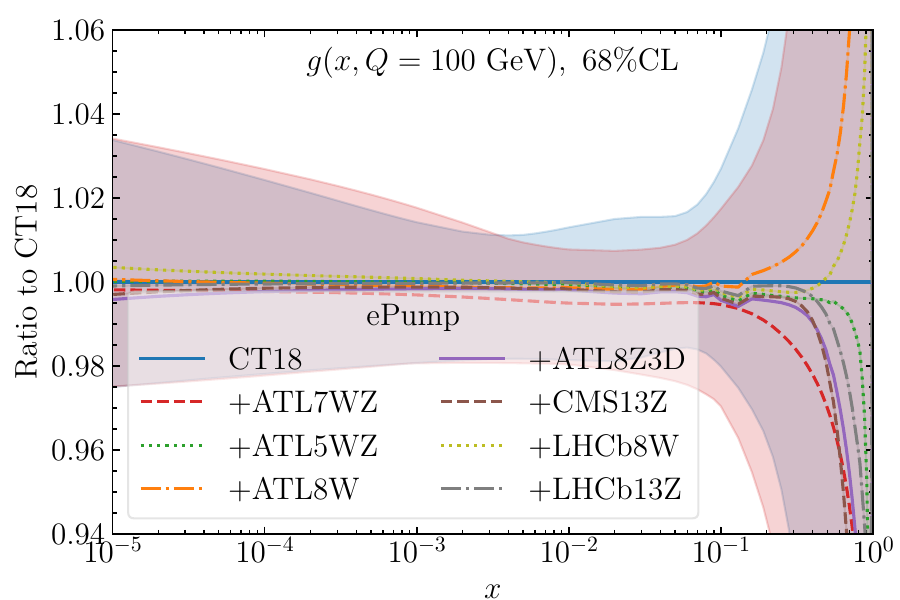} 
\includegraphics[width=0.49\textwidth]{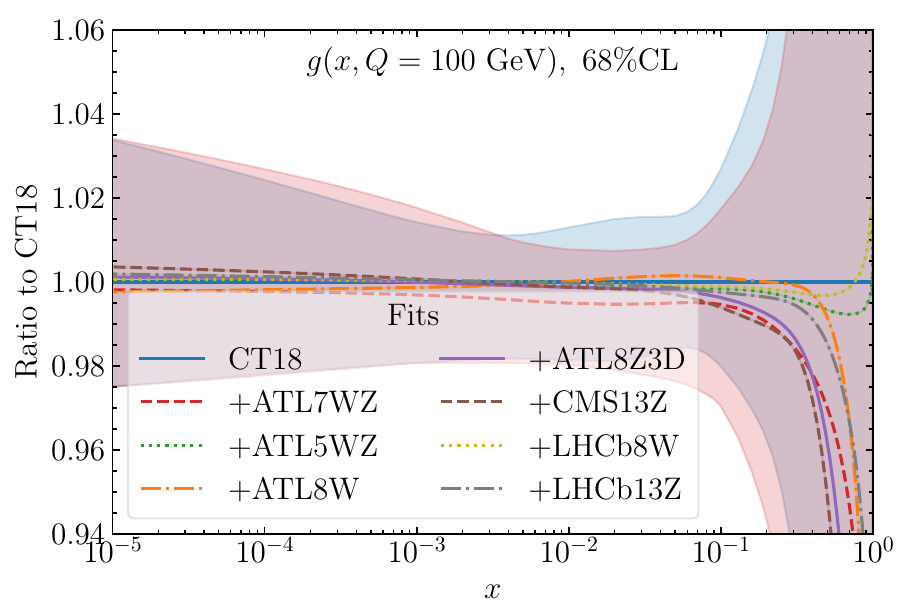}
\includegraphics[width=0.49\textwidth]{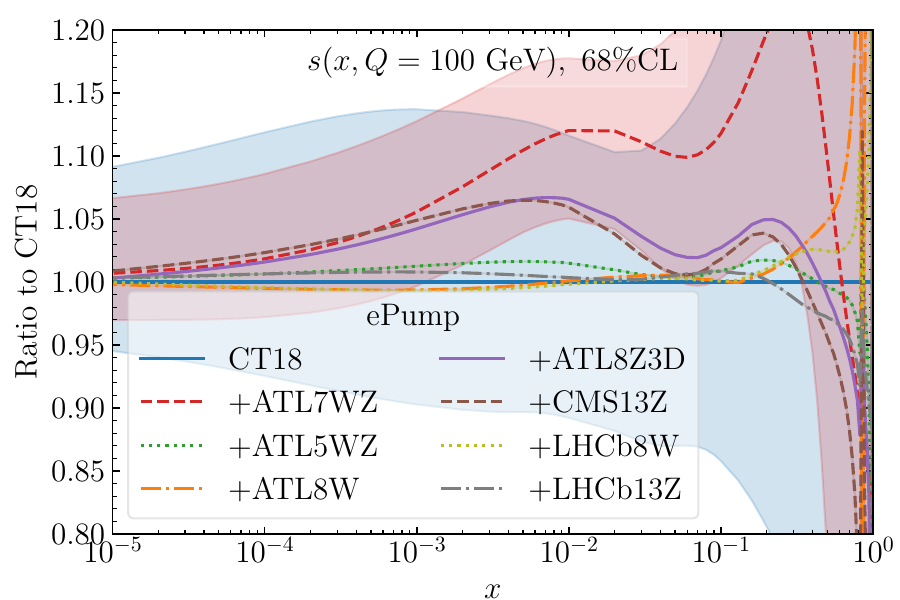}
\includegraphics[width=0.49\textwidth]{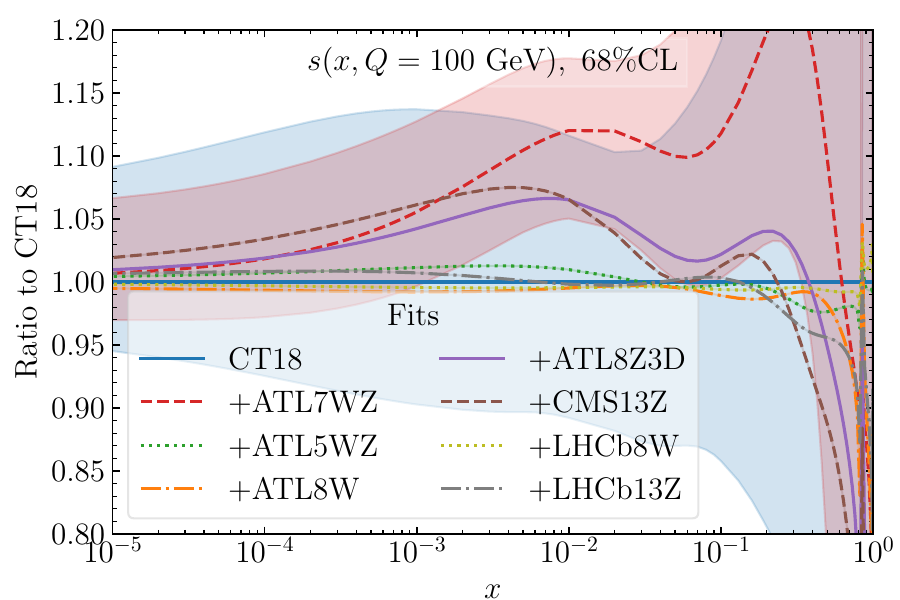}
\caption{The impact of post-CT18 LHC Drell-Yan data sets on updating the CT18 $\bar{d}$ and $s$ PDFs at $Q=100~\GeV$, by including the individual data set one by one with the ePump (left) and global fitting (right). 
The two error bands correspond to the CT18 and CT18+ATL7WZ=CT18A PDFs, respectively.
}
\label{fig:FitOne}
\end{figure}

In Fig.~\ref{fig:FitOne}, we show the ePump updated and the global fitted $\bar{d},g,s$ PDFs at $Q=100~\GeV$, as an example to demonstrate the impact of post-CT18 LHC Drell-Yan data sets on the CT18 PDFs.
A similar comparison for other PDF flavors is collected in Fig.~\ref{fig:FitOne2} of Appendix~\ref{app:supp}.
Here, we also include the reference PDF error bands for the CT18+ATL7WZ fit, \emph{i.e.} CT18A, while only central PDFs for other fits.

As shown in Fig.~\ref{fig:FitOne}, the ATL7WZ data set in the CT18A fit softens the $\bar{d}$ PDF, but enhances the  $s(\bar s)$ PDF at $x$ around $10^{-3}-10^{-2}$.
Most of the post-CT18 LHC Drell-Yan data sets also prefer larger  $s(\bar s)$  PDF than what CT18 predicts,
implying consistency with the ATL7WZ data.
The strongest impacts come from the ATL8Z3D and CMS13Z data, even though the strength are milder than ATL7WZ data.
In comparison with the fitted PDFs, the ePump updated ones give similar results, with only slightly smaller pulls for CT18+CMS13Z $s(x)$ PDF, as a result of the limitation of the Hessian linear approximation~\cite{Schmidt:2018hvu,Hou:2019efy}. 

However, by looking at the $\bar{d}$ PDF, we realize that both the ATL8W and LHCb8W data pull the CT18 $\bar{d}$ PDF to the opposite direction at $x$ around $\sim10^{-3}$, with respect to ATL7WZ and other post-CT18 LHC Drell-Yan data. 
It suggests a tension of the ATL8W (and LHCb8W) data with other data sets, such as ATL8Z3D.
We also note that when the LHCb8W data is included to update the CT18 PDFs, cf. Fig.~\ref{fig:FitOne}, an opposite pull on $\bar{d}$ PDF occurs at $x\sim 0.3$ where the PDF uncertainty is large.

\section{Global fits}
\label{sec:fit}

\subsection{$\chi^2$ and PDFs for individual flavors}

\begin{table}[!h]
\hspace*{-10pt}
\begin{tabular}{c|c|c|c|c|c|c|c|c}
\hline
\multirow{2}{*}{ID} & \multirow{2}{*}{Experiment} 
& \multirow{2}{*}{$N_{\rm pt}$} 
& \multicolumn{6}{c}{$\chi^2/N_{\rm pt}$} \\
\cline{4-9}
&  &  & CT18 &  CT18A & CT18As &  ATLASpdf21 &   MSHT20 & NNPDF4.0  \\
\hline
215 & ATLAS 5.02 TeV $W,Z$ & 27 & 0.81 & 0.71 & 0.71 & -- & -- & -- \\
211 & ATLAS 8 TeV $W$ & 22 & 2.45 & 2.63 & 2.51 & 1.41 &  2.61 & [3.50]  \\
214 & ATLAS 8 TeV $Z$ 3D$^\dagger$ & 188 & 1.12 & 1.14 & 1.18  & 1.13(184) & 1.45(59) & 1.22(60)  \\
212 & CMS 13 TeV $Z$ & 12 & 2.38 & 2.03 & 2.71 &-- & -- & --\\
216 & LHCb 8 TeV $W$ & 14 & 1.34 & 1.36 & 1.43 & --&-- & --\\
213 & LHCb 13 TeV $Z$  & 16 & 1.10 & 0.98 & 0.83 &-- & -- & -- \\
\hline
248 & ATLAS 7 TeV $W,Z$ & 34 & 2.52 & 2.50 & 2.30 & 1.24(55) & 1.91(61) & 1.67(61) \\
\hline
\multicolumn{3}{c|}{Total 3994/3953/3959 points}  & 1.20  & 1.20 & 1.19 &  -- & -- & -- \\
\hline
\end{tabular}
\caption{Similarly to Tab.~\ref{tab:chi2ePump}, but with global fits by  including the post-CT18 LHC Drell-Yan data sets simultaneously. We include the MSHT20~\cite{Bailey:2020ooq},  NNPDF4.0~\cite{NNPDF:2021njg} and ATLASpdf21~\cite{ATLAS:2021vod} results for comparison.
$^\dagger$Different from ATLASpdf21~\cite{ATLAS:2021vod} and our treatment of the ATLAS 8 TeV $Z$ data, \emph{i.e.}, directly fitting the triple differential distributions of $(m_{\ell\ell},y_{\ell\ell},\cos\theta^*)$, MSHT20 and NNPDF4.0 have summed over the $\cos\theta^*$ bins and resulted in double differential distributions of $(m_{\ell\ell},y_{\ell\ell})$ with the number of data points indicated in parentheses, respectively. 
}
\label{tab:chi2Fit}
\end{table}

Now, we can simultaneously fit these new post-CT18 LHC Drell-Yan data sets (denoted as ``CT18+nDY", etc) to study
their impact on the CTEQ-TEQ PDFs. Within the CT18, CT18A, and CT18As framework, we present the corresponding fitted $\chi^2/N_{\rm pt}$ for each of the post-CT18 LHC Drell-Yan data set and the combined data sets in Tab.~\ref{tab:chi2Fit}. We included the ATL7WZ data set as well in this comparison. As discovered already in the individual fits, the $\chi^2$ for each set follows the same trend as Tab.~\ref{tab:chi2ePump}.
That is to say, except for the ATLAS and the LHCb 8 TeV $W$ data sets, the $\chi^2$ decreases from CT18 to CT18A, and to CT18As, reflecting the general consistency among these data sets. 
The $\chi^2/N_{\rm pt}$ values of ATL8W and LHCb8W increase from updating CT18 to CT18A, as a result of the tension with the ATL7WZ data.
The updated CT18As can reduce the $\chi^2$ for these two sets, due to the additional degree of freedom in the $s$ and $\bar s $ PDFs. 

\begin{figure}[!h]
    \centering
    \includegraphics[width=0.49\textwidth]{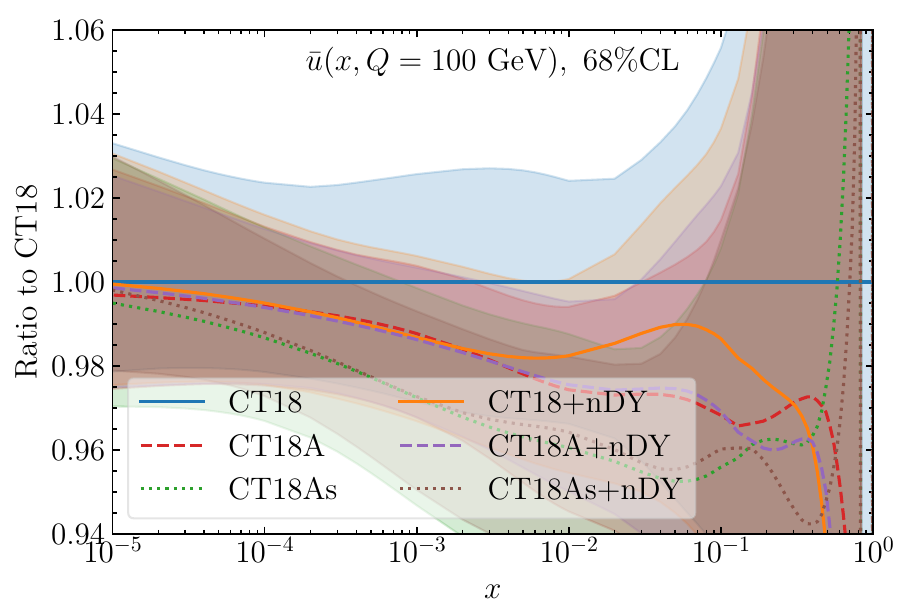}
    \includegraphics[width=0.49\textwidth]{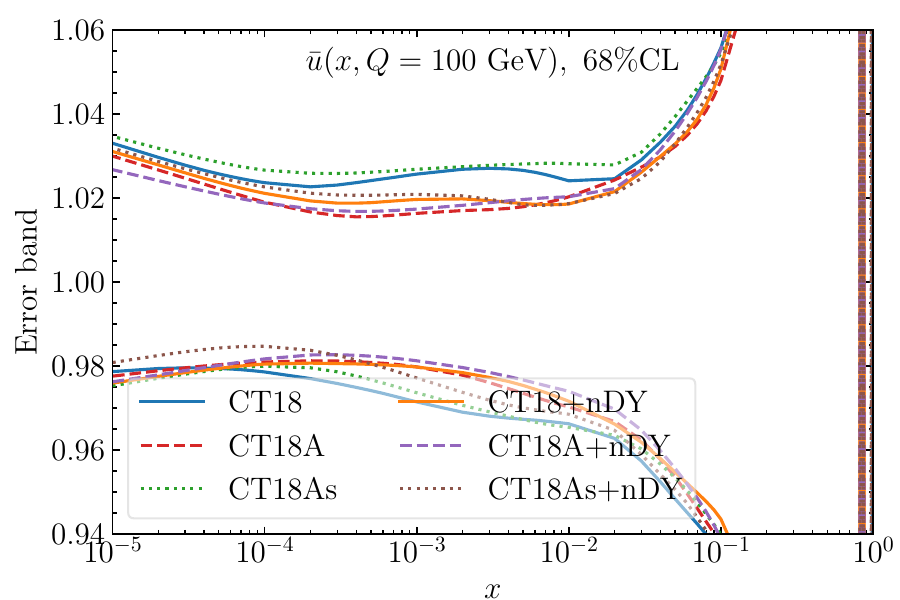}     
    \includegraphics[width=0.49\textwidth]{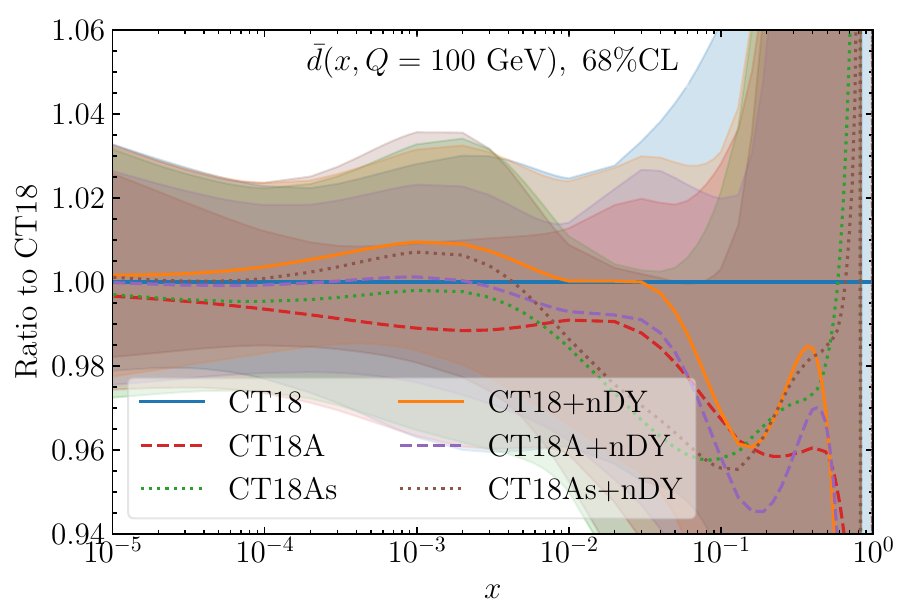}
    \includegraphics[width=0.49\textwidth]{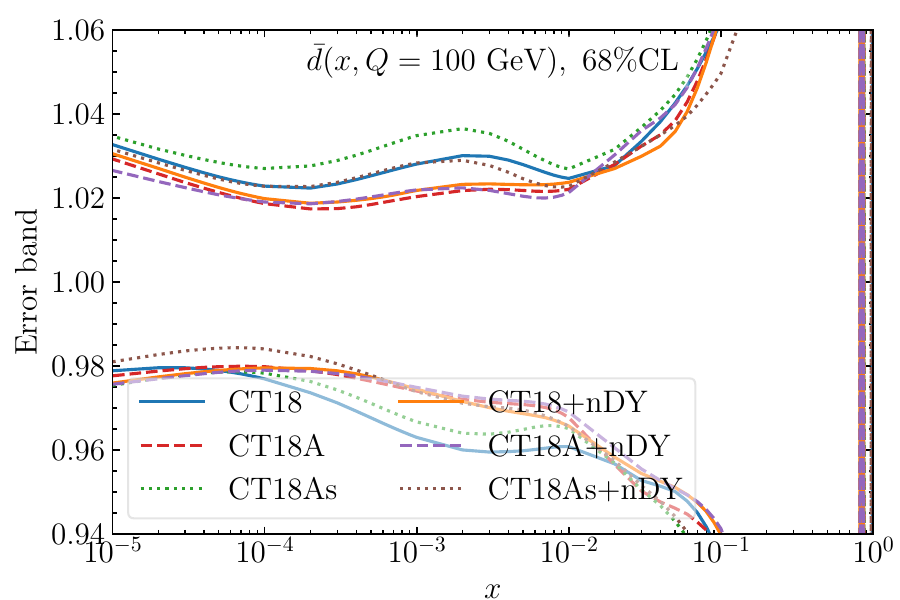}    
    \includegraphics[width=0.49\textwidth]{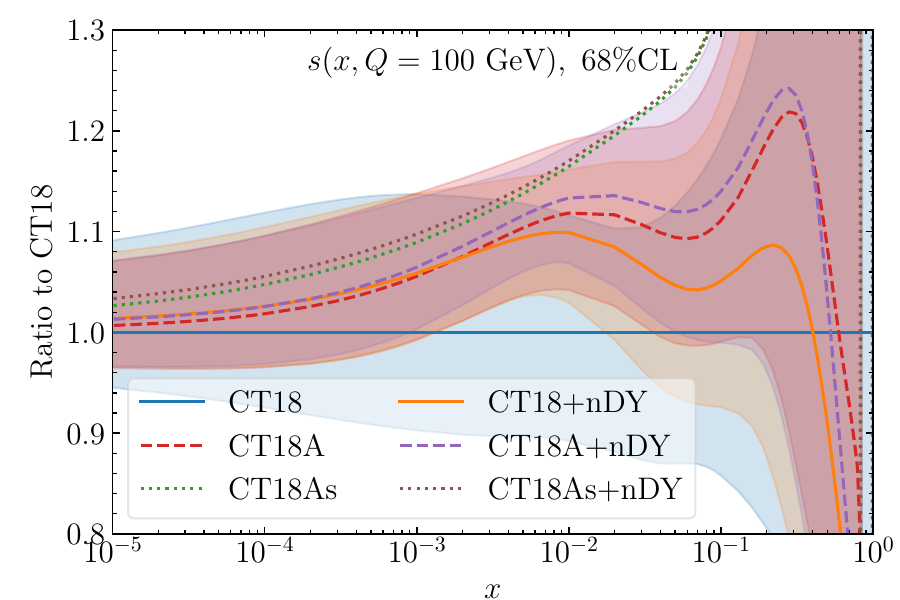}
    \includegraphics[width=0.49\textwidth]{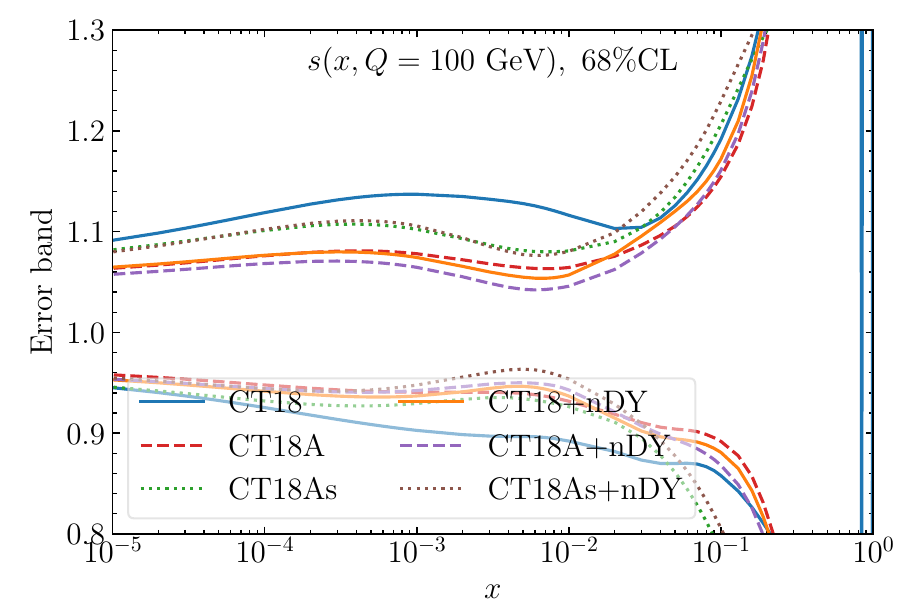}    
    \caption{Comparison of $\bar{u}$, $\bar{d}$, and $s$  PDFs at $Q=100~\GeV$ for various fits.}
    \label{fig:FitAll}
\end{figure}

\begin{figure}[!h]
    \centering
    \includegraphics[width=0.49\textwidth]{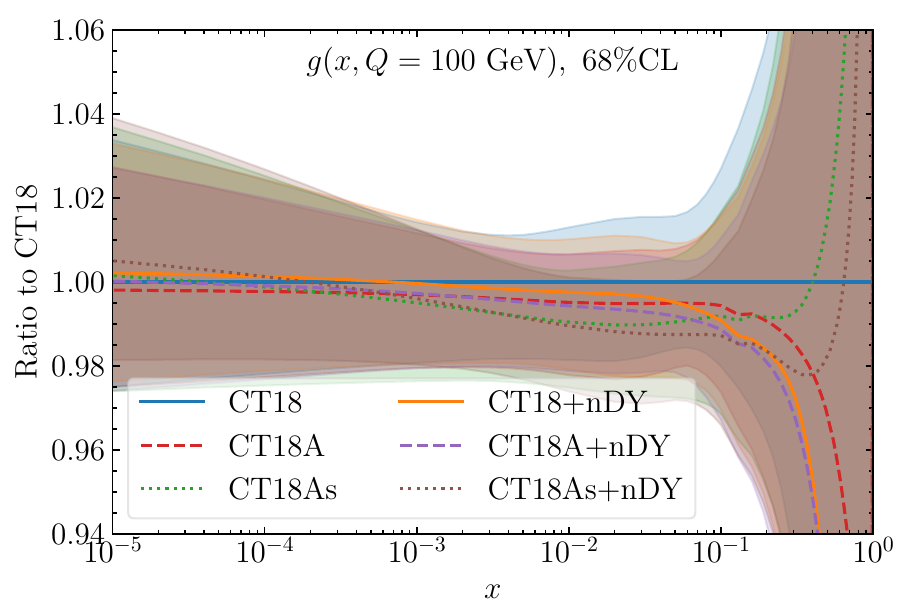}   
    \includegraphics[width=0.49\textwidth]{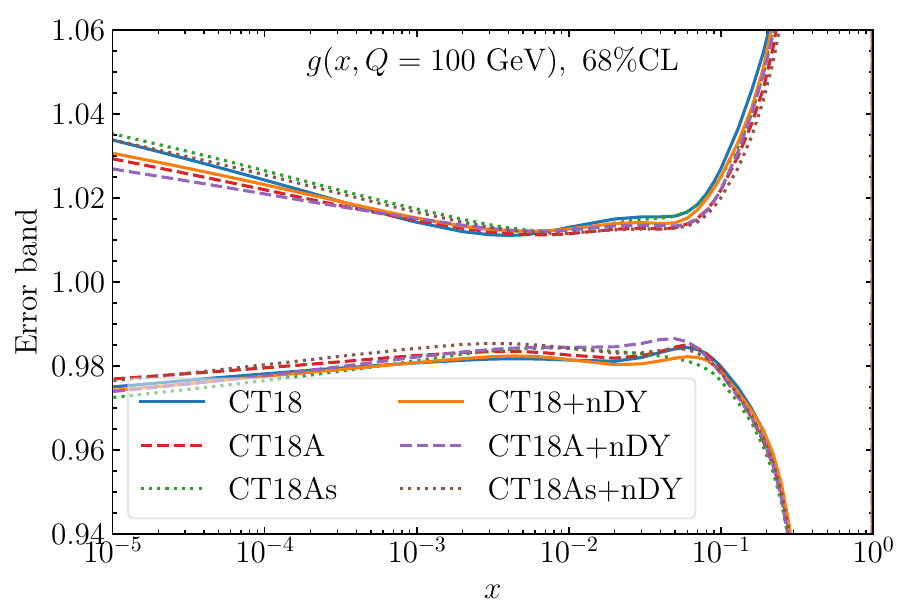}          
    \includegraphics[width=0.49\textwidth]{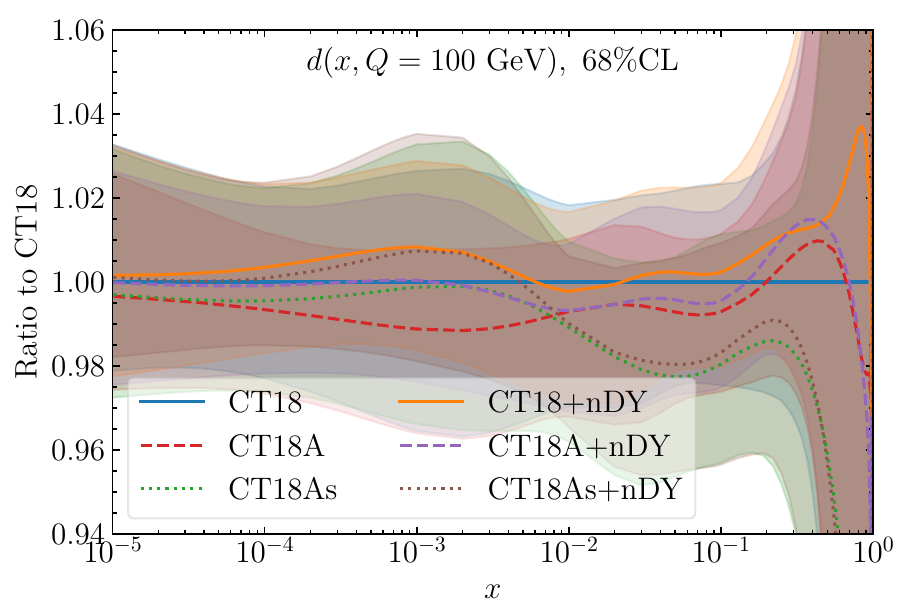}
    \includegraphics[width=0.49\textwidth]{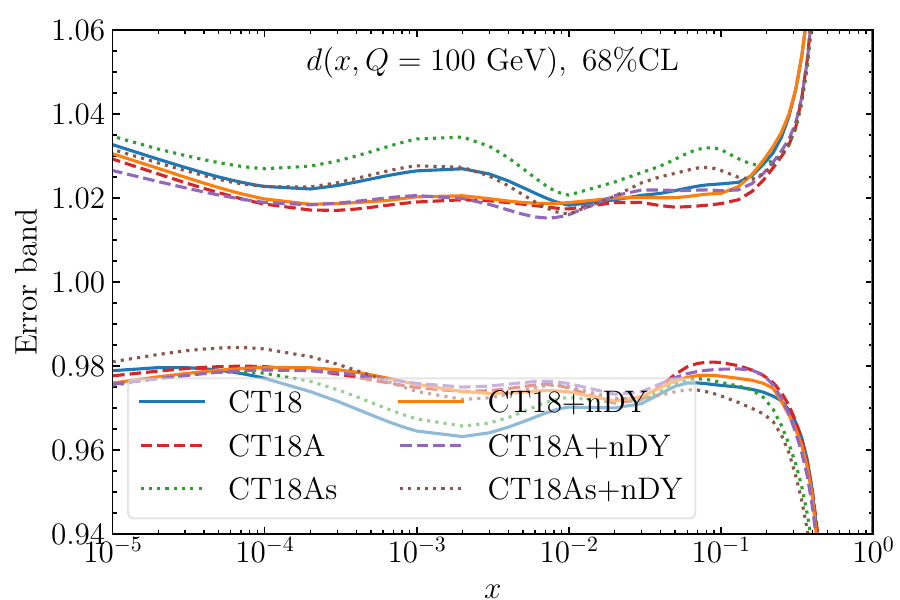}    
    \includegraphics[width=0.49\textwidth]{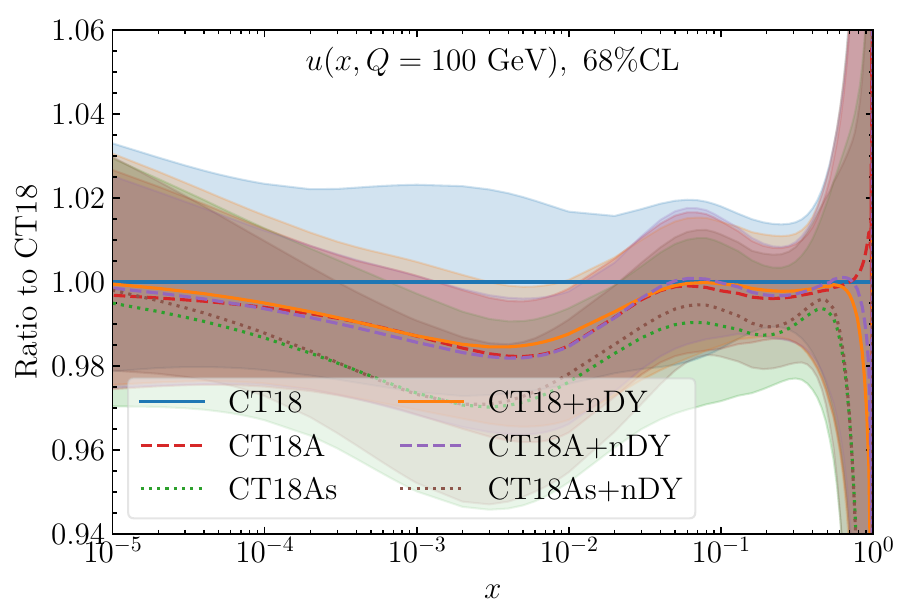}    
    \includegraphics[width=0.49\textwidth]{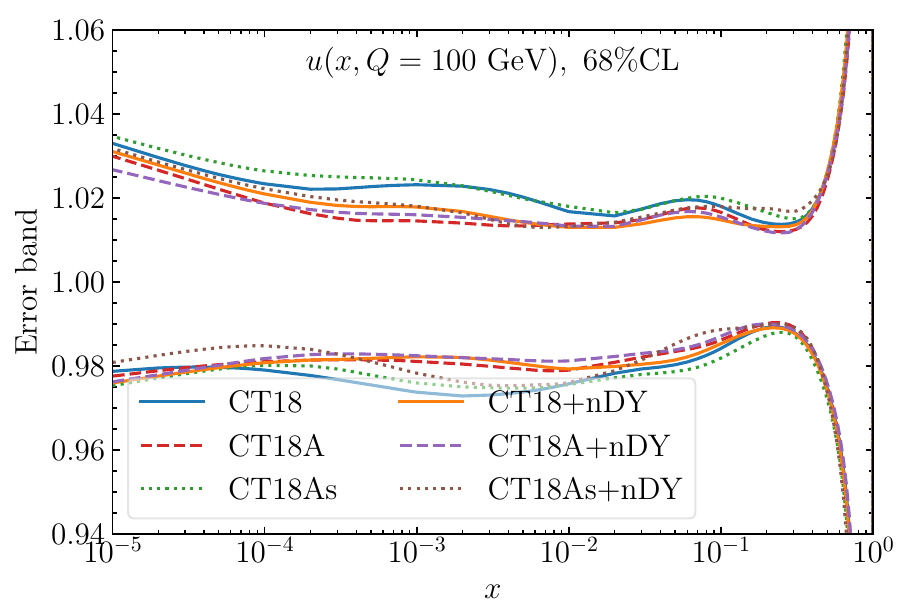}        
    \caption{Similar to Fig.~\ref{fig:FitAll}, but for $g$, $d$ and $u$ PDFs at $Q=100~\GeV$.}
    \label{fig:FitAll2}
\end{figure}

We compare the fitted PDFs for sea quarks $\bar{u},\bar{d}$ and $s$ at $Q=100~\GeV$, before and after including the post-CT18 LHC Drell-Yan data sets in Fig.~\ref{fig:FitAll}, and the comparisons for other light flavors $g$,$u$ and $d$ are shown in Fig.~\ref{fig:FitAll2}.
We display the flavor singlet $\Sigma=\sum_{i}(q_i+\bar{q}_i)$ in Fig.~\ref{fig:sglt} of Appendix~\ref{app:supp}.
Recall that the ATL7WZ data in the CT18A fit enhances the $s(\bar s)$ PDFs but reduces the $d(\bar{d})$, which is shown again in Fig.~\ref{fig:FitAll}. In comparison, the post-CT18 Drell-Yan data sets in the ``CT18+nDY" fit pull the strangeness in the same direction but with a milder distance. 
The ``CT18A+nDY" accumulates both the impacts from ATL7WZ and post-CT18 LHC Drell-Yan data sets.
However, when examining the $d(\bar{d})$ flavor more carefully, we see that the post-CT18 LHC Drell-Yan data pull CT18 PDFs in the opposite direction with respect to the CT18A PDFs. This behavior shows up already in the individual fits as shown in Fig.~\ref{fig:FitOne}. It is driven by the tension between the ATL8W data and ATL7WZ data. The inclusion of ATL7WZ and post-CT18 LHC Drell-Yan data sets in the CT18A+nDY fit 
yields $d(\bar{d})$ PDFs lying in between CT18+nDY and CT18A, as a consequence of the tension between ATL7WZ and ``nDY" data sets.

In comparison with the CT18A, the additional freedom of the $s(\bar{s})$ PDFs in CT18As allows
the $s(\bar{s})$ to be pulled further~\cite{Hou:2022onq}.
As a result, the ATL8Z3D $\chi^2/N_{\rm pt}$ can increase slightly, as shown in Tab.~\ref{tab:chi2Fit}, indicating that the relaxed tension mentioned above is not completely resolved.
Under such an assumption, the impacts of the post-CT18 LHC Drell-Yan data sets on the 
$s(\bar{s})$ PDFs, as well as other flavors, becomes minor in most scenarios, with only a mild enhancement of the $d(\bar{d})$ PDFs around $x\sim10^{-3}$, as shown in Figs.~\ref{fig:FitAll}-\ref{fig:FitAll2}.
The impacts of all these Drell-Yan data sets  on the gluon PDF is limited, which contributes through higher-order corrections to the Drell-Yan process. Details of the comparison can be found in Fig.~\ref{fig:FitAll2}.

We also include the comparison of the $\bar{d}$ and $s$ PDF error bands in Fig.~\ref{fig:FitAll}. 
Starting with strangeness, we see the ATL7WZ data set shrinks the PDF error band in the  $x\in[10^{-4},10^{-1}]$ range. A similar impact can be obtained with the post-CT18 LHC Drell-Yan data sets, shown as the error band of ``CT18+nDY", which imposes a slightly stronger constraint around $x\sim10^{-2}$ as a result of  including more data. The impacts from both ATL7WZ and post-CT18 LHC Drell-Yan data sets are combined in the ``CT18A+nDY". With the additional freedom of strangeness asymmetry in CT18As, the $s(\bar{s})$ PDF error band becomes larger in comparison with the CT18A. Under such a scenario, the change of the strangeness error band due to the post-CT18 LHC Drell-Yan data sets is quite limited.

When examining the $d(\bar{d})$ flavor, the CT18A shrinks the error bands in the range of $x\in[10^{-4},10^{-2}]$. The post-CT18 LHC Drell-Yan data sets play the same role as ATL7WZ data in CT18A, which yields a similar error band to ``CT18+nDY" as the CT18A. Under such a condition, the accumulation of post-CT18 LHC Drell-Yan data sets only contributes a minor reduction of the error band in the range $x\in[10^{-3},10^{-1}]$ in the negative direction of Hessian asymmetric errors.
Similar to the strangeness PDF, the additional freedom in CT18As enlarges the $d(\bar{d})$ error bands. On top of CT18As, the inclusion of the post-CT18 LHC Drell-Yan data sets can reduce the PDF error bands.

\subsection{Flavor ratios and strangeness asymmetry}

\begin{figure}[!h]
    \centering
    \includegraphics[width=0.49\textwidth]{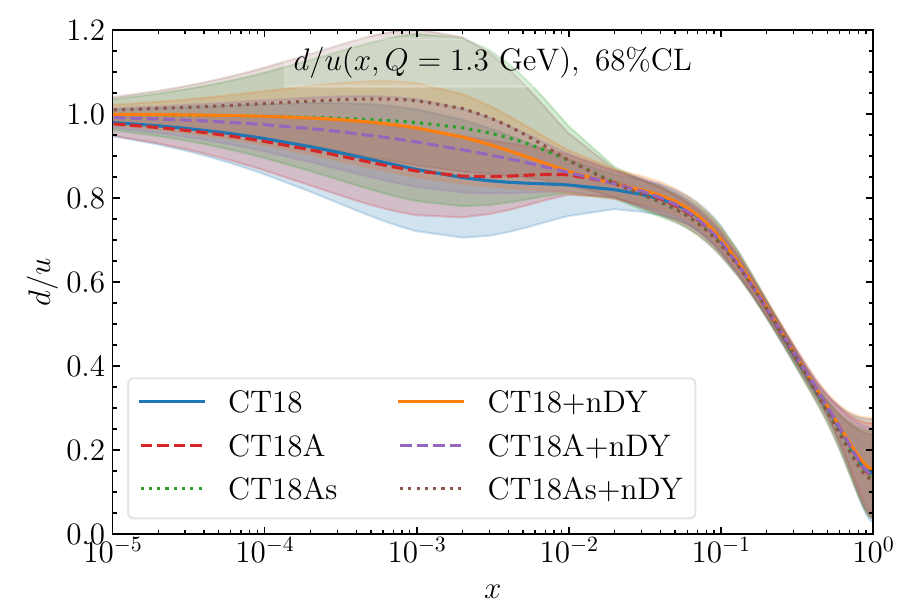}
    \includegraphics[width=0.49\textwidth]{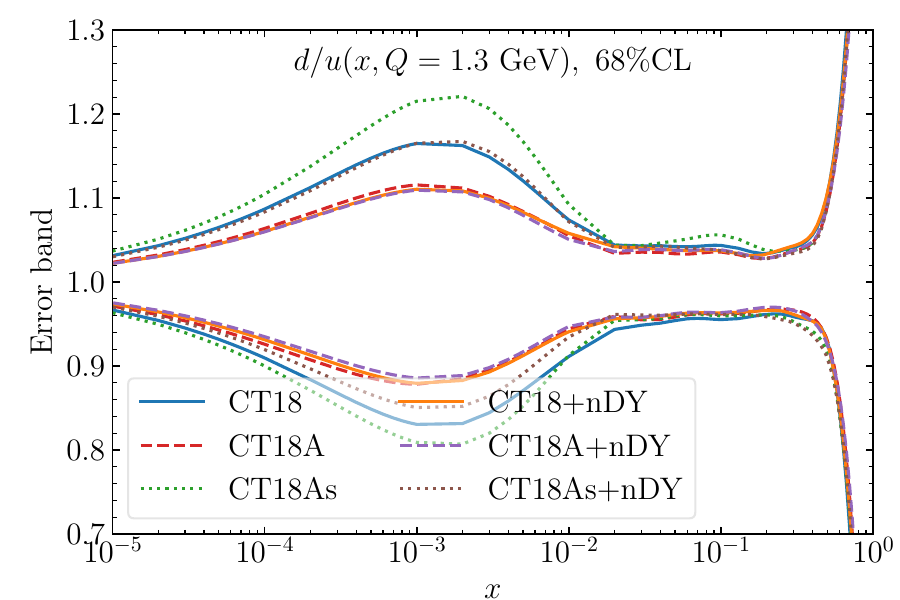}    
    \includegraphics[width=0.49\textwidth]{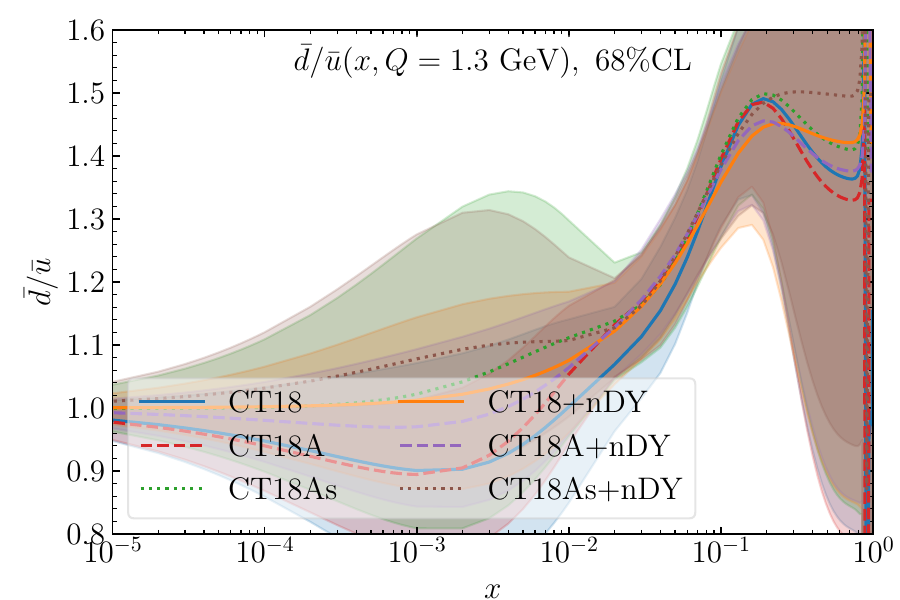}
    \includegraphics[width=0.49\textwidth]{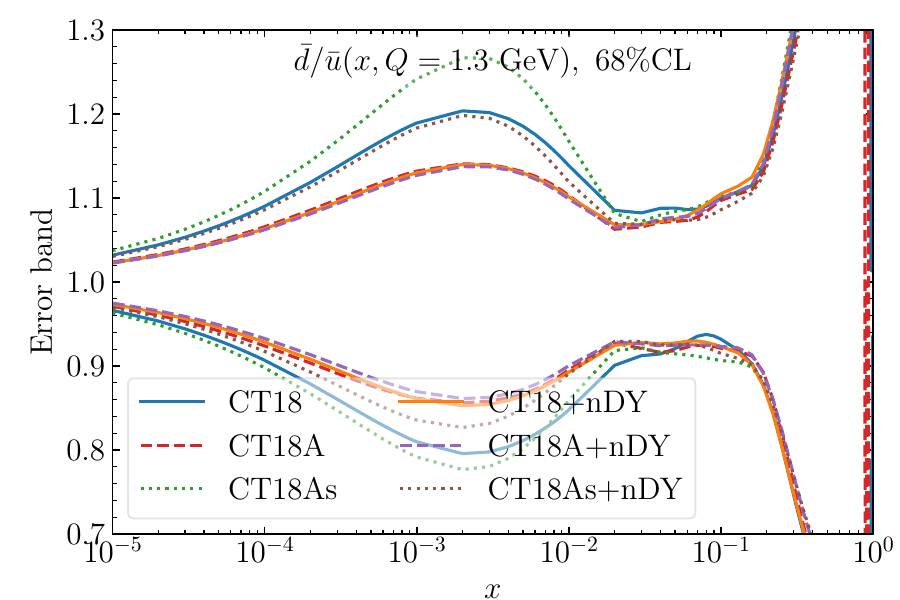}
    \includegraphics[width=0.49\textwidth]{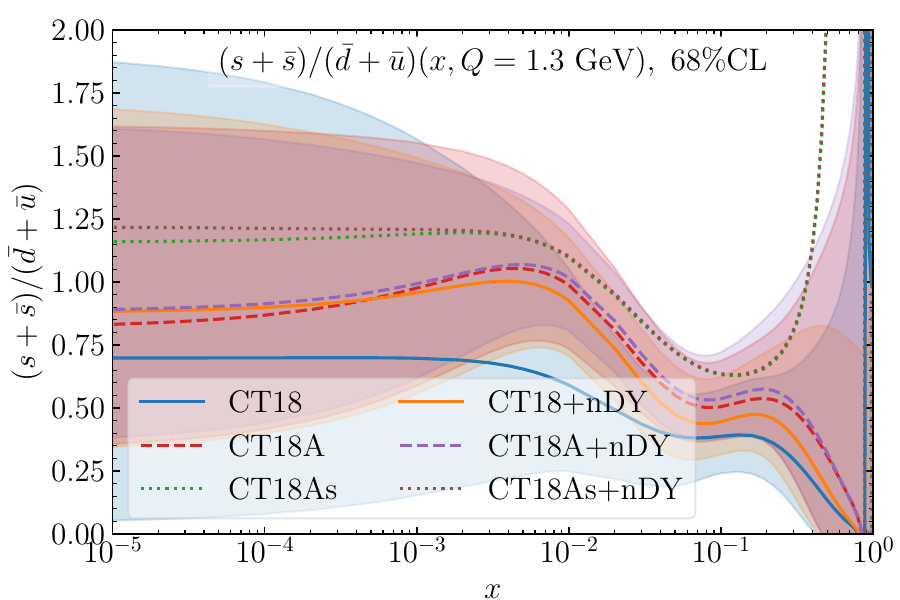}
    \includegraphics[width=0.49\textwidth]{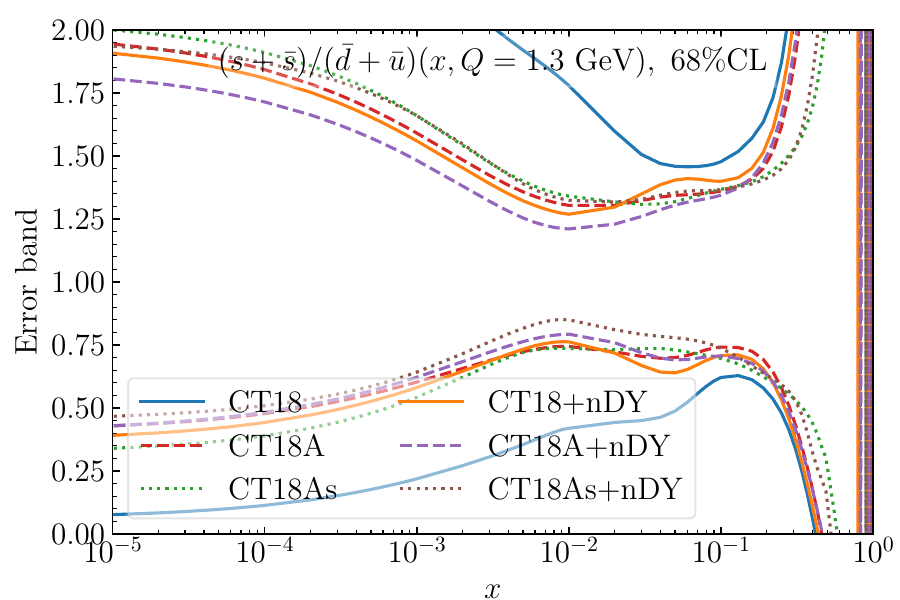}        
    \caption{Similar to Fig.~\ref{fig:FitAll}, but for the flavor ratios $d/u,\bar{d}/\bar{u}$ and $R_s=(s+\bar{s})/(\bar{d}+\bar{u})$ at $Q=1.3~\GeV$.}
    \label{fig:fitCT18r}
\end{figure}

Besides the individual flavors, we also examine the impact of the post-CT18 LHC Drell-Yan data sets on the flavor ratios $d/u$, $\bar{d}/\bar{u}$ and $(s+\bar{s})/(\bar{d}+\bar{u})$ in Fig.~\ref{fig:fitCT18r}.
Instead of a high scale such as $Q=100~\GeV$, here we show them at the starting scale $Q_0=1.3~\GeV$, which shows a more pronounced effect from the post-CT18 LHC Drell-Yan data sets. The main features can be deduced from the individual flavors already shown in Fig.~\ref{fig:FitAll} and Fig.~\ref{fig:FitAll2}. 
For example, the post-CT18 LHC Drell-Yan data enhance the $R_s=(s+\bar{s})/(\bar{d}+\bar{u})$, with the extent slightly smaller than the impact of
the ATL7WZ in  CT18A. Both impacts are accumulated in the CT18A+nDY sets. With a more flexible parameterization of strangeness PDFs in CT18As, the impact of post-CT18 LHC Drell-Yan data is quite minimal. 
The inclusion of the post-CT18 LHC Drell-Yan data enhances the $d/u$ and $\bar{d}/\bar{u}$ ratios at $x\sim10^{-3}$, which mainly originates from the ATL8W's modification to the $d(\bar{d})$ PDF.
This feature is different from the CT18A, of which the $d/u$ and $\bar{d}/\bar{u}$ ratios at $x\sim10^{-3}$ are more or less the same as CT18, after the inclusion of ATL7WZ data. 
The impact on the ratio uncertainties can be understood similarly.

\begin{figure}[!h]
    \centering
    \includegraphics[width=0.49\textwidth]{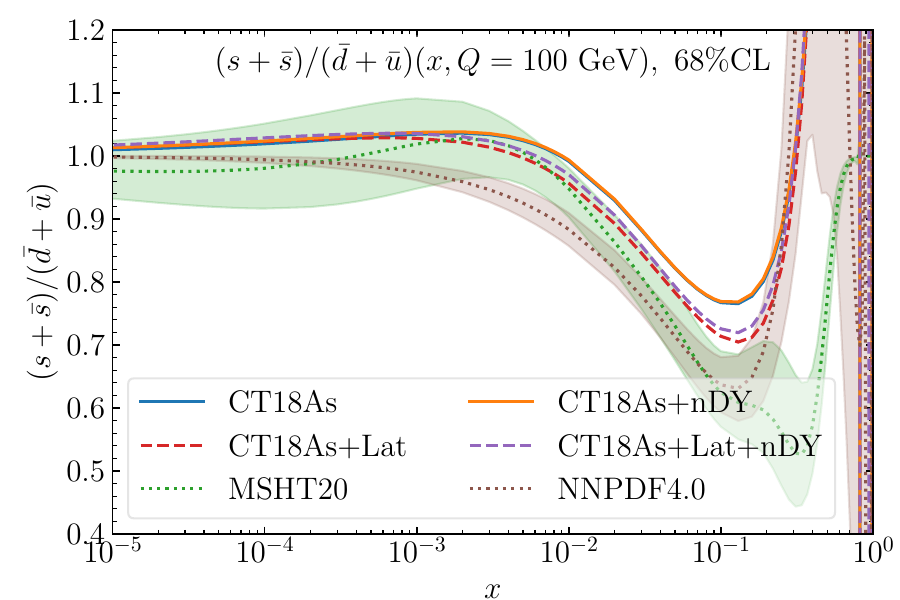}    
    \includegraphics[width=0.49\textwidth]{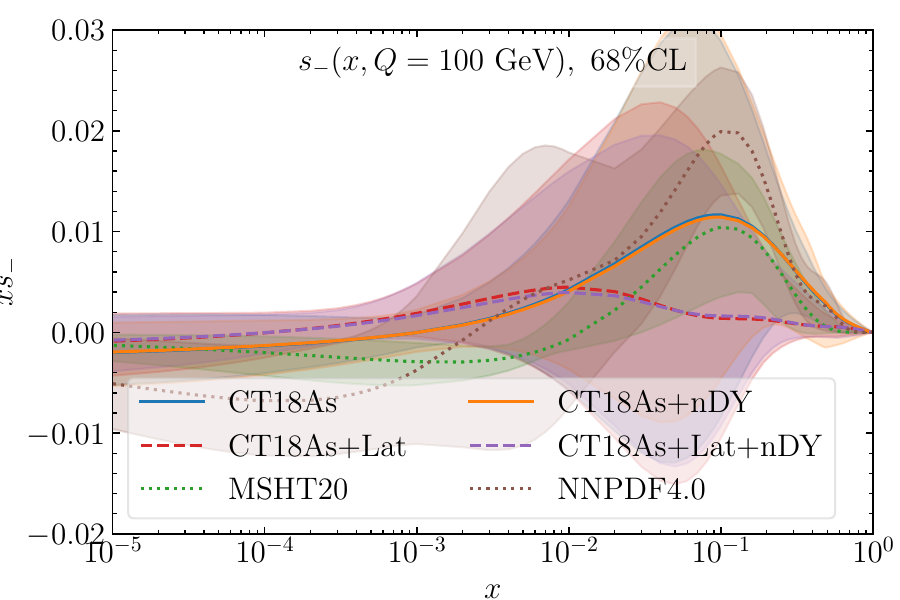}
    \caption{Comparison of the ratio $R_s\equiv(s+\bar{s})/(\bar{d}+\bar{u})$ (left) and the strangeness asymmetry $s_-=s-\bar{s}$ (right)  of CT18As(+Lat/nDY), MSHT20 and NNPDF4.0  PDFs at $Q=100~\GeV$.}
    \label{fig:str}
\end{figure}

Finally, we compare the $R_s$ as well as the strangeness asymmetry $s_{-}(x)=s(x)-\bar{s}(x)$ among the CT18As(+Lat\footnote{Throughout this work, we use the notation CT18As+Lat to represent the CT18As$\_$Lat fit presented in Ref.~\cite{Hou:2022onq}, for convenience.}~\cite{Hou:2022onq}/nDY), MSHT20, and NNPDF4.0 in Fig.~\ref{fig:str}.
We remind that the CT18As+Lat fit~\cite{Hou:2022onq} includes additional lattice data~\cite{Zhang:2020dkn}, which constrain the strangeness asymmetry in the large $x$ region.
Similar to the $s$ PDF in Fig.~\ref{fig:FitAll} or $R_s$ ratio in Fig.~\ref{fig:fitCT18r}, the post-CT18 LHC Drell-Yan data sets give a very minor impact on the strangeness asymmetry, both with and without the lattice data. 
Both MSHT20 and NNPDF4.0 give softer $R_s$ ratios than CT18(+nDY), with an overall agreement within the corresponding error band. 
As emphasized in Ref.~\cite{Hou:2022onq}, the inclusion of the lattice $s_{-}$ data in the CT18As+Lat fit gets modifications on both the central fit and the error bands of $R_s$ and $s_{-}(x)$, as shown in Fig.~\ref{fig:str}.
The $R_s$ of CT18As+Lat gets closer to that of MSHT20 and NNPDF4.0, while the strangeness asymmetry $s_{-}$ of CT18As is closer to zero at large $x$ than that of MSHT20 and NNPDF4.0, due to the constraint from the lattice data. 
Furthermore, this conclusion remains when the post-CT18 LHC Drell-Yan (nDY) data is included in various CT18As(+Lat) global fits.

\subsection{Data descriptions}

\begin{figure}[!h]
    \centering
    \includegraphics[width=0.49\textwidth]{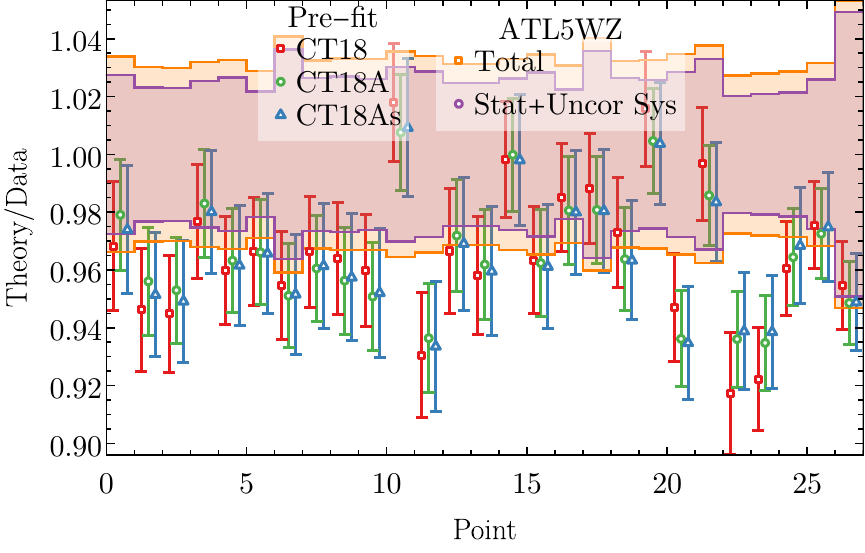}
    \includegraphics[width=0.49\textwidth]{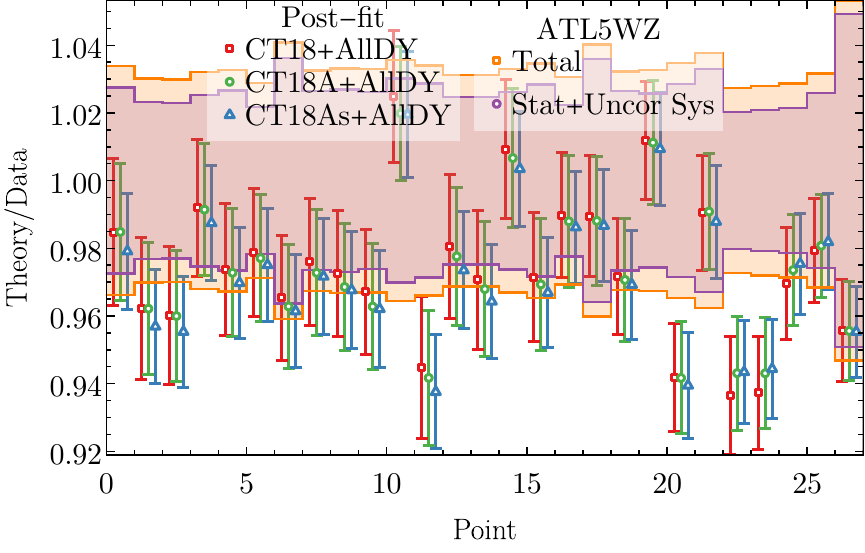}
    \includegraphics[width=0.49\textwidth]{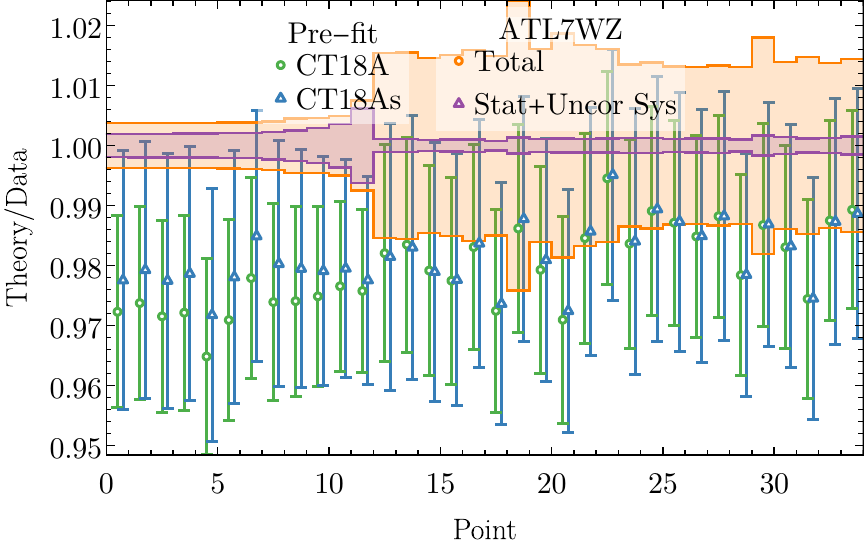}
    \includegraphics[width=0.49\textwidth]{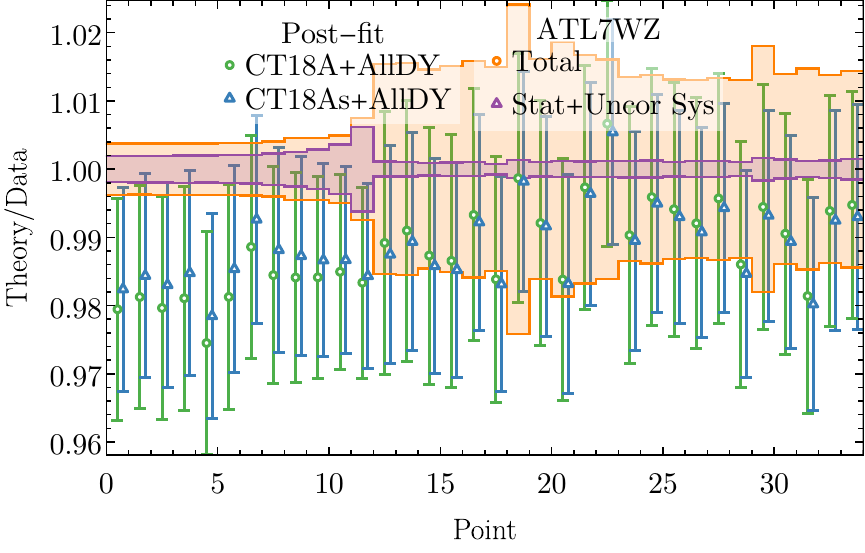}    
    \includegraphics[width=0.49\textwidth]{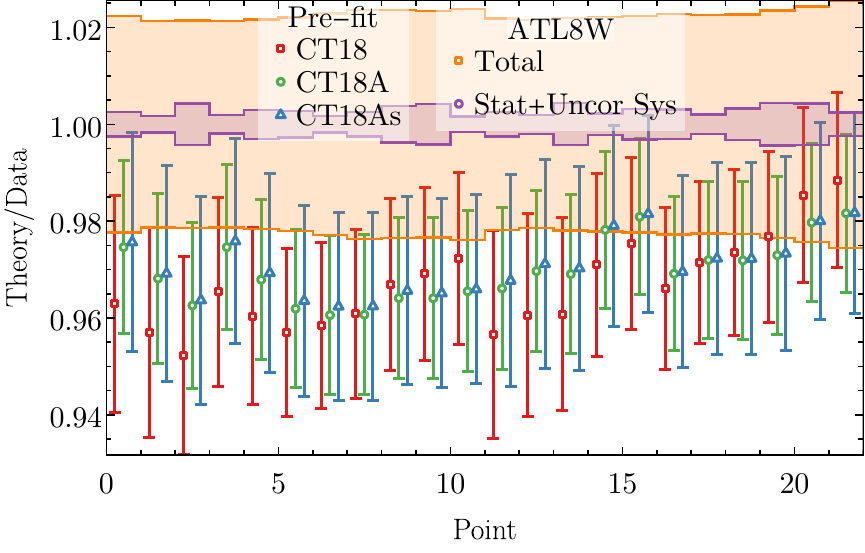}
    \includegraphics[width=0.49\textwidth]{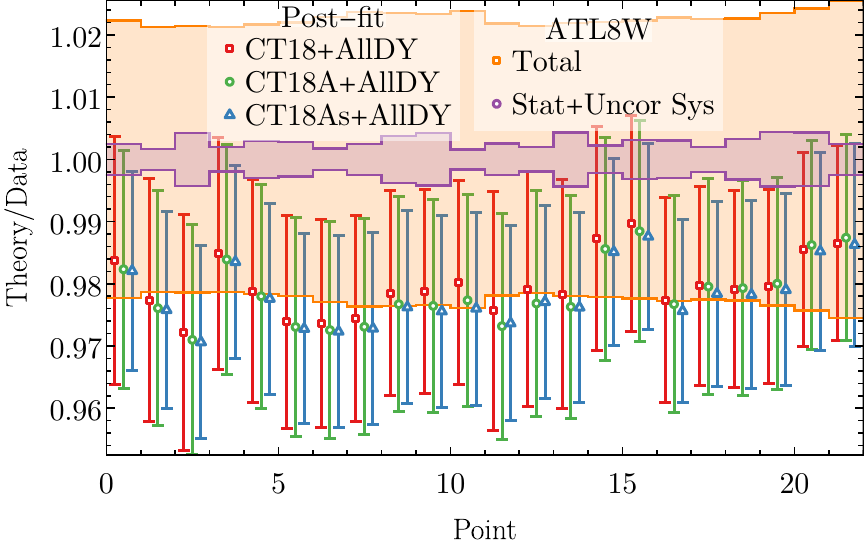}
    \caption{Comparisons of various theoretical predictions of CT18, CT18A, CT18As PDFs (left) and  CT18+nDY, CT18A+nDY, CT18As+nDY PDFs (right)
 and the ATLAS 5.02, 7, and 8 TeV $W,Z$ data. The total error of each experimental data point is presented as the shaded yellow band, which is much larger than the sum of statistical and uncorrelated systematic errors in quadrature.
}
\label{fig:thATL}
\end{figure}

In Figs.~\ref{fig:thATL}, we compare the theoretical predictions for the ATLAS 5.02, 7, and 8 TeV $W,Z$ productions from the CT18, CT18A, and CT18As fits before and after including the post-CT18 LHC Drell-Yan data. We provide similar comparisons for the ATL8Z3D, CMS13Z and LHCb data in
Figs.~\ref{fig:thATL8Z3D}, \ref{fig:thCMS}, and \ref{fig:thLHCb} of Appendix~\ref{app:supp}, respectively.
Here, we show the raw data, instead of the shifted ones, as the systematical shifts vary in terms of the corresponding fitted PDFs. As we expected, the general agreement of the fits with experimental data becomes better after the inclusion of these post-CT18 LHC Drell-Yan data sets. Meanwhile, the degree of impact of each data set usually depends on both the degree of agreement before the fit as well as the  size of
uncorrelated uncertainty (quadrature sum of the statistical and uncorrelated systematical uncertainties) of the data. 
For example, we obtain a significantly larger pull on the $d(\bar{d})$ PDFs from the ATL8W data, with respect to the ATL5WZ, mainly driven by the smaller uncorrelated uncertainties, as shown in Fig.~\ref{fig:thATL}.
Of course, the number of data points makes a difference as well, such as in the ATL8Z3D case as shown in Fig.~\ref{fig:thATL8Z3D}. 

Also, we observe that the theoretical predictions of these Drell-Yan data are generally smaller than the experimental data, It suggests an overall consistency among these data sets, which results in the same directional pulls, \emph{e.g.}, enhancing $s$-PDF at $x$ around $10^{-3}$,
as shown in Fig.~\ref{fig:FitOne}. As we know in terms of the correlation cosine, the pull of $W$ production is mainly reflected on the $d(\bar{d})$ and $u(\bar{u})$ PDFs. As a consequence, the ATL8W data will lift the $d$ and $\bar{d}$ PDFs upwardly, as shown in Fig.~\ref{fig:FitOne} and Fig.~\ref{fig:FitOne2}, respectively. However, this does not show up in the ATL7WZ data, due to the different pull of the $Z$ data. 

In comparison with the ATL7WZ pull in the CT18A PDFs, we obtain a similar pull on the $s$ PDF from the CMS13Z data, hinted by the similar trend of comparison between the theoretical prediction and experimental data, as shown
in Fig.~\ref{fig:thCMS}. In comparison, the impact of the LHCb 13 TeV $Z$ data is quite minimal. We compare the corresponding theory predictions and data in Fig.~\ref{fig:thLHCb} for completeness.

\subsection{Phenomenological implications}

\begin{figure}[!h]
    \centering
    \includegraphics[width=0.49\textwidth]{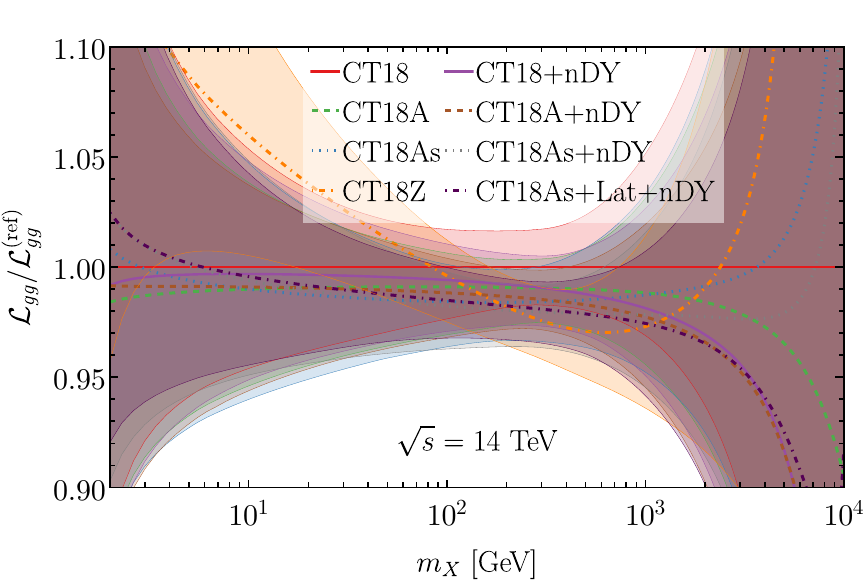}
    \includegraphics[width=0.49\textwidth]{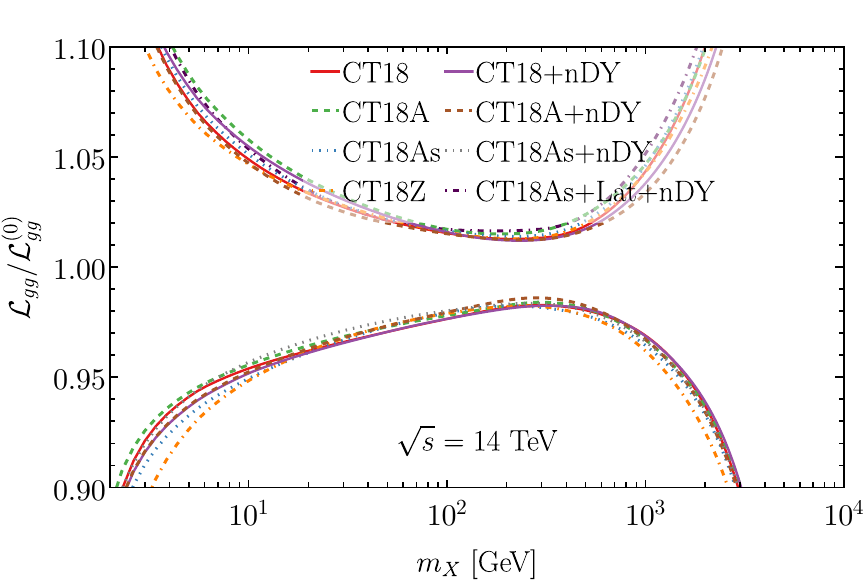}
    \includegraphics[width=0.49\textwidth]{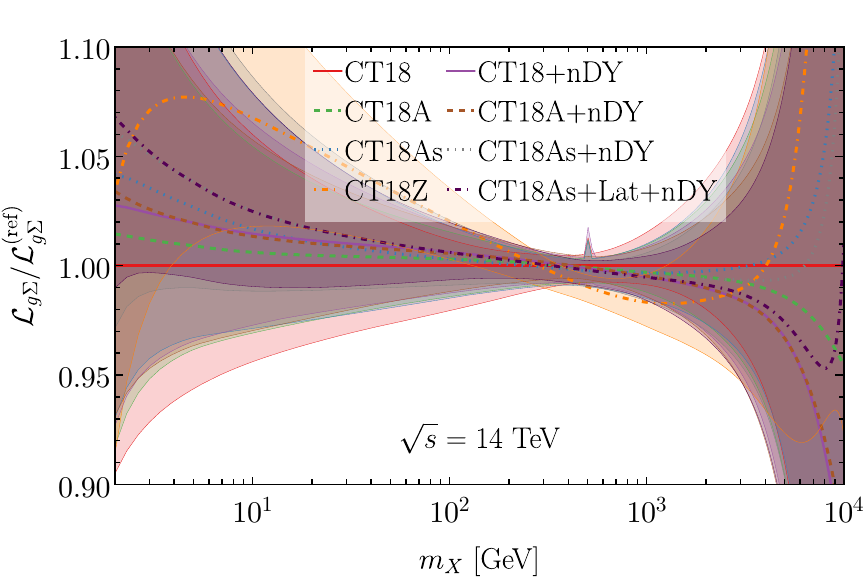}
    \includegraphics[width=0.49\textwidth]{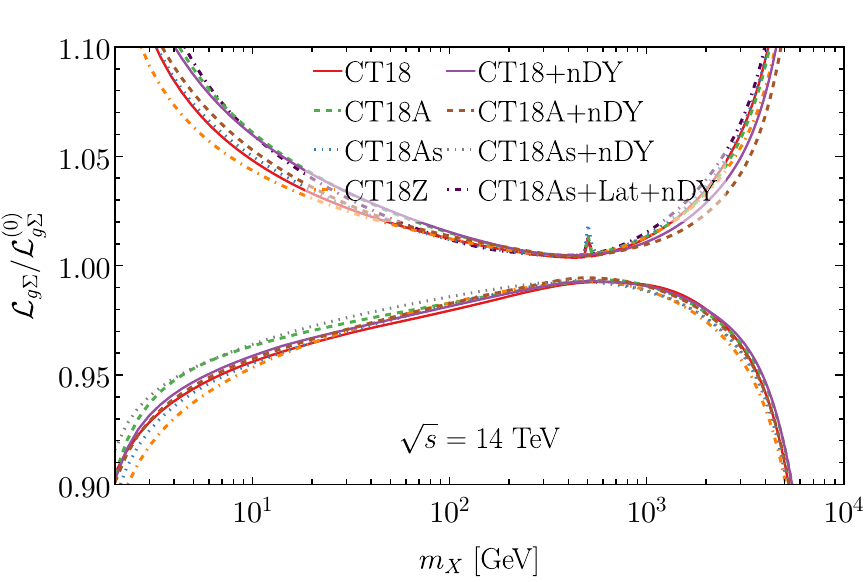}    
    \includegraphics[width=0.49\textwidth]{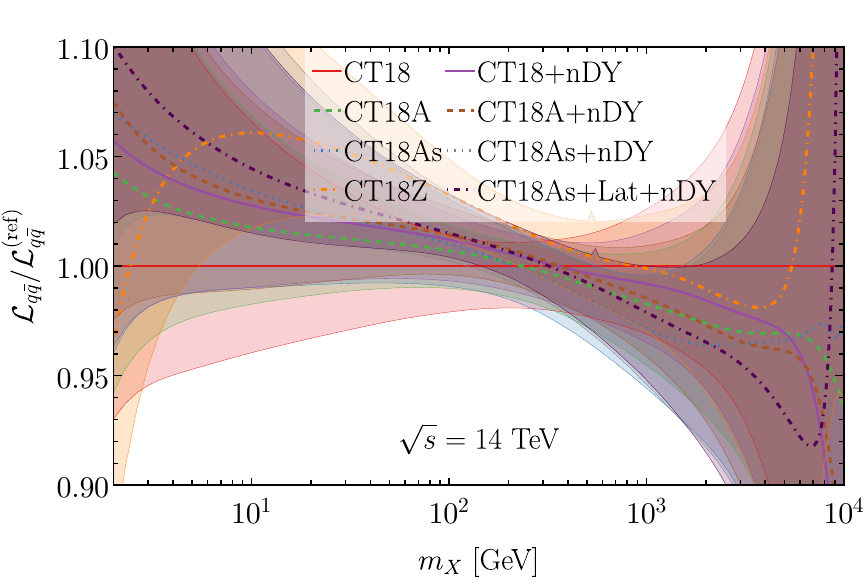}
    \includegraphics[width=0.49\textwidth]{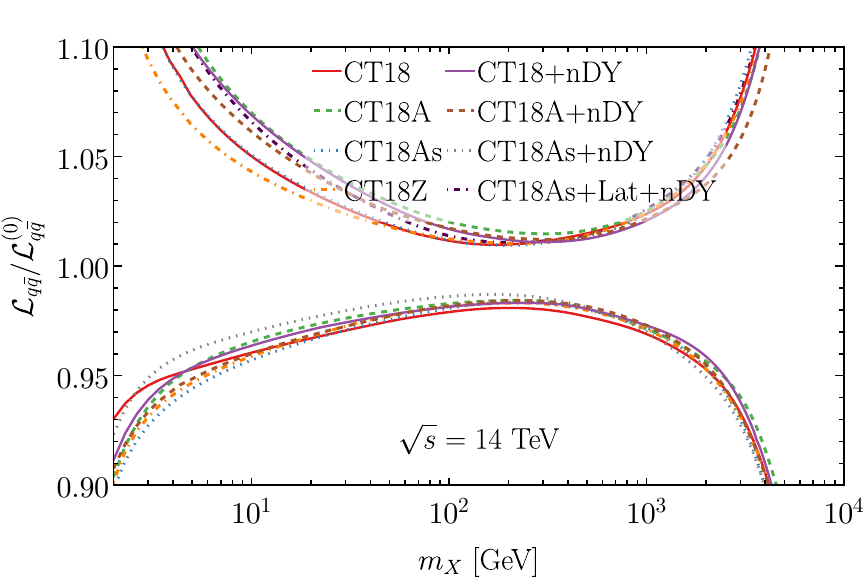}
    \includegraphics[width=0.49\textwidth]{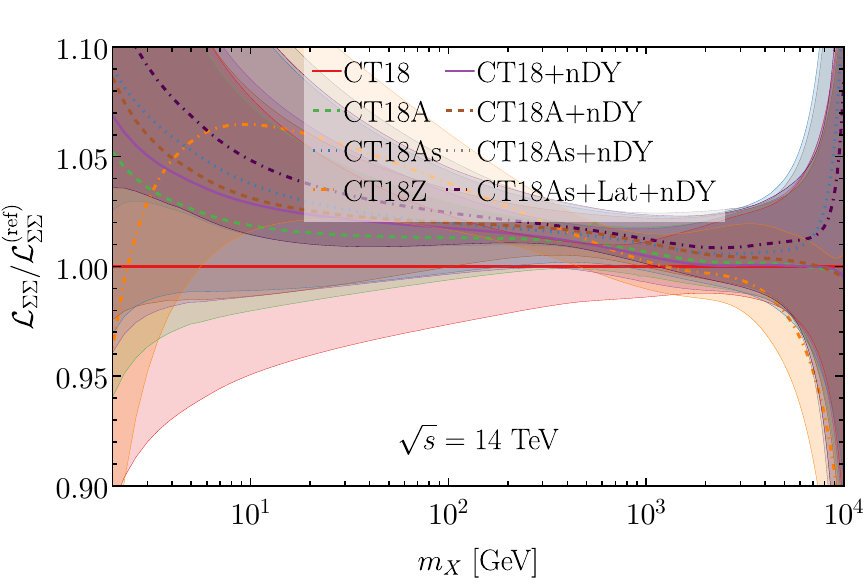}
    \includegraphics[width=0.49\textwidth]{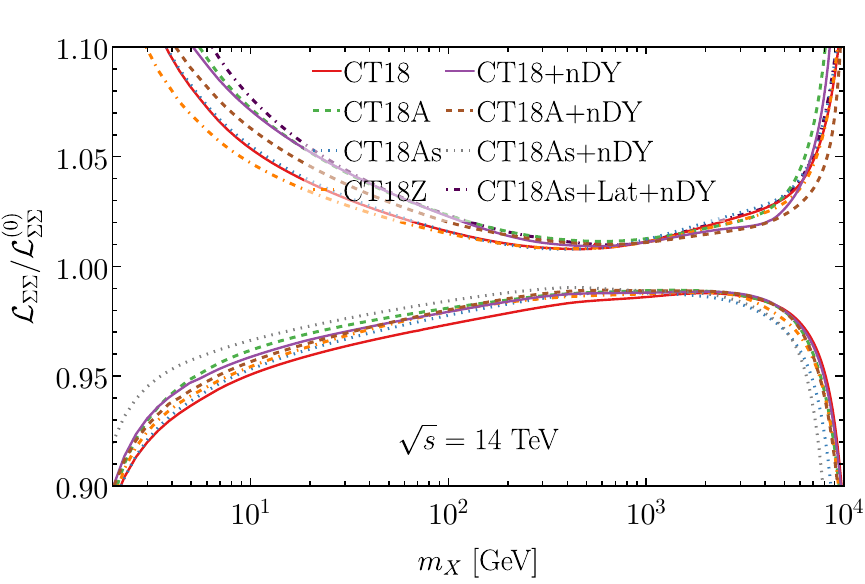}       
    \caption{Comparison of the parton luminosities (left) and and their uncertainties (right) at the $14$ TeV LHC, computed with the CT18, CT18A and CT18As PDFs and the PDFs after including the post-CT18 LHC Drell-Yan (nDY) data sets.
}
    \label{fig:lumi}
\end{figure}

We present the comparison of the PDF parton luminosities before and after including the post-CT18 LHC Drell-Yan data in the CT18, CT18A, and CT18As, as well as CT18Z and CT18As+Lat+nDY (accumulating Lattice~\cite{Hou:2022onq} and nDY data) fits in Fig.~\ref{fig:lumi}.
The parton luminosity is defined as~\cite{Campbell:2006wx}
\begin{equation}
\mathcal{L}_{ij}(s,m_X^2)=\frac{1}{s}\frac{1}{1+\delta_{ij}}\int_{\tau}^1\frac{\dd x}{x}[f_{i}(x,Q^2)f_{j}(\tau/x,Q^2)+(i\leftrightarrow j)],
\end{equation}
where $\tau=m_X^2/s$, with $\sqrt s$ being the center of mass energy of the collider, and factorization scale $Q^2=m_X^2$.
Let us recall the parton luminosities of CT18A~\cite{Hou:2019efy} at low invariant mass region:
 (i) the gluon-gluon luminosity $(\mathcal{L}_{gg})$ is reduced; 
 (ii) both the quark-antiquark luminosity $(\mathcal{L}_{q\bar{q}}=\sum_{i}\mathcal{L}_{q_i\bar{q}_i})$ and the singlet-singlet luminosity $(\mathcal{L}_{\Sigma\Sigma})$ are enhanced; 
 (iii) the gluon-singlet luminosity $(\mathcal{L}_{g\Sigma})$ is almost unchanged. The parton luminosity uncertainties of CT18A are slightly smaller than the CT18. 
 
In comparison with the  $gg$ luminosity of CT18 PDFs, 
 the post-CT18 LHC Drell-Yan data sets  pull the CT18+nDY $gg$ luminosity in the same direction as CT18A  $gg$ luminosity at low and high invariant mass  $m_X$ region. 
However, the the amount of change in ${q\bar{q}}$ luminosity is opposite, \emph{i.e.}, larger (smaller) change at low (high) invariant mass $m_X$ region. In comparison, the change in the ${g\Sigma}$ luminosity and ${\Sigma\Sigma}$ luminosity become larger than the CT18A's difference from CT18.
In many cases, the parton luminosities of CT18A+nDY  resemble CT18+nDY, except in the TeV region. Furthermore, the parton luminosities of CT18As+nDY are quite similar to CT18A+nDY. 
The parton luminosity uncertainties of CT18As+nDY PDFs are  slightly larger than CT18A+nDY as a result of additional degrees of freedom in the strangeness PDFs.
We also note that the changes from both CT18 PDFs to CT18As+nDY PDFs (e.g., $s$, $g$, $\bar d$) and the corresponding parton luminosities are quite similar to the changes from CT18 to CT18Z PDFs ~\cite{Hou:2019efy}.

\begin{figure}[!h]
    \centering
    \includegraphics[width=0.49\textwidth]{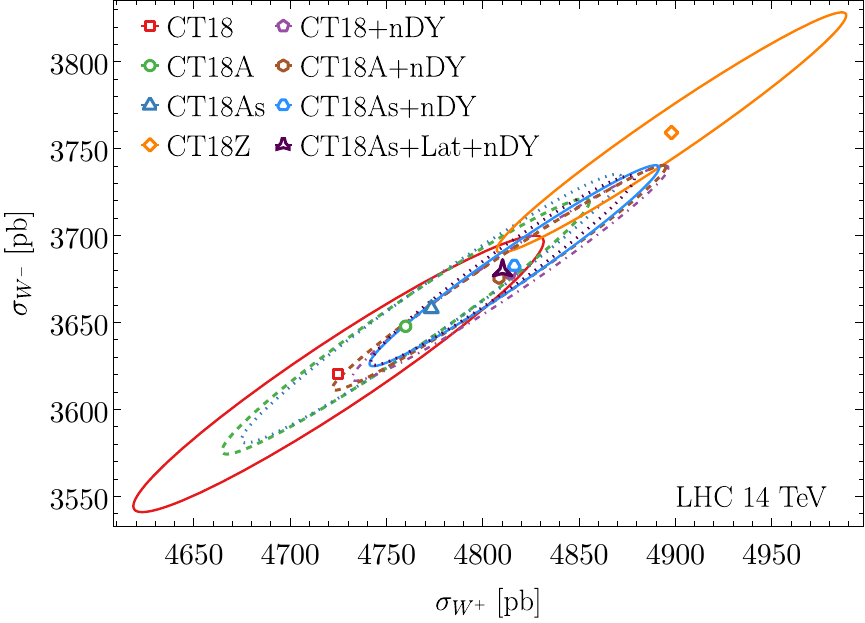}
    \includegraphics[width=0.49\textwidth]{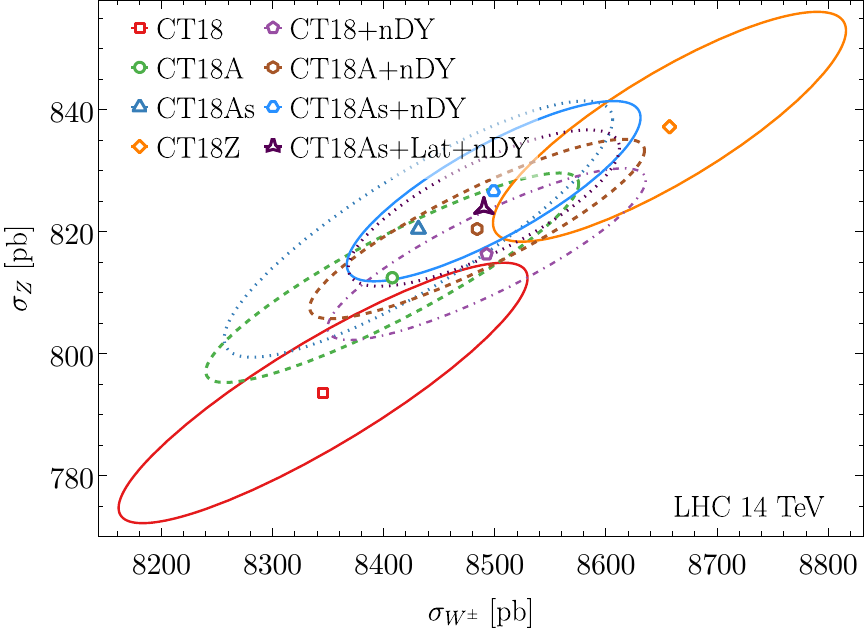}
    \includegraphics[width=0.49\textwidth]{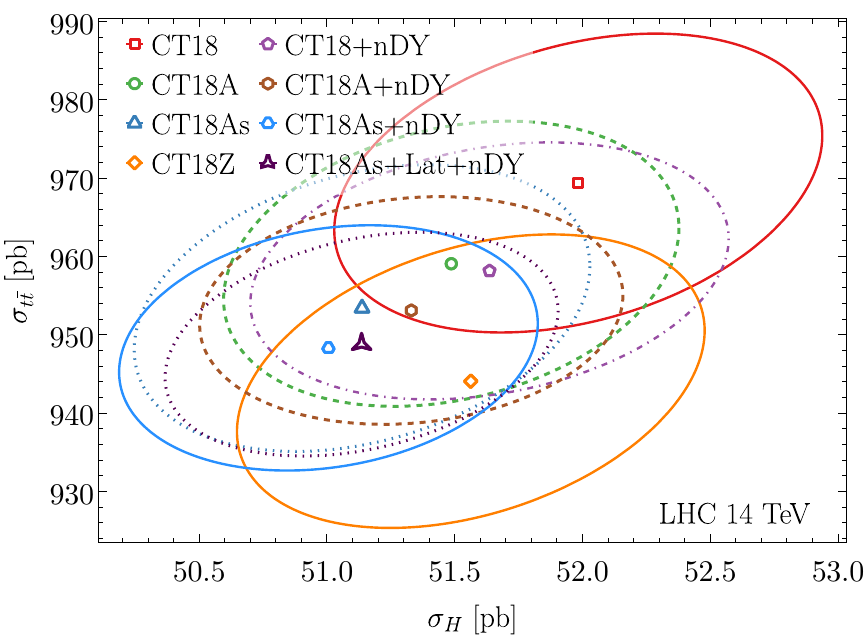}
    \includegraphics[width=0.49\textwidth]{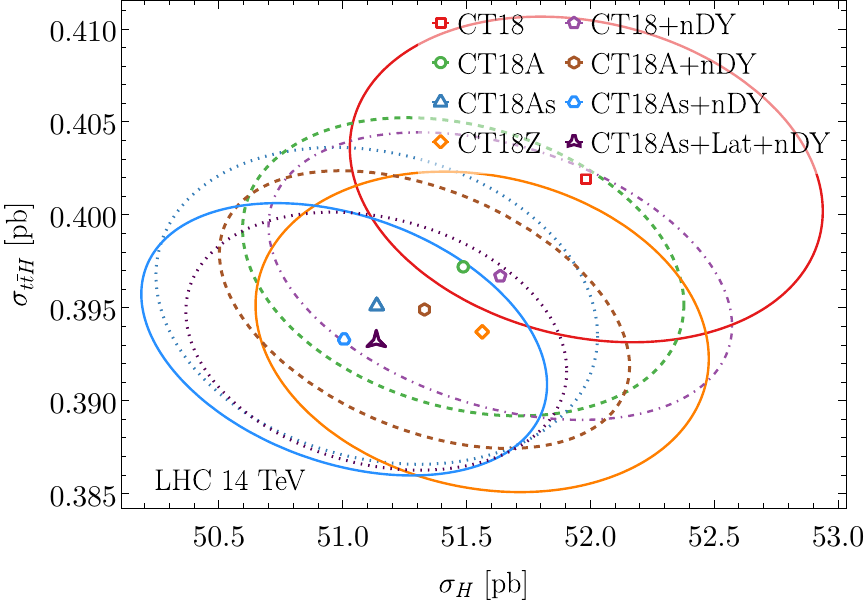}    
    \includegraphics[width=0.49\textwidth]{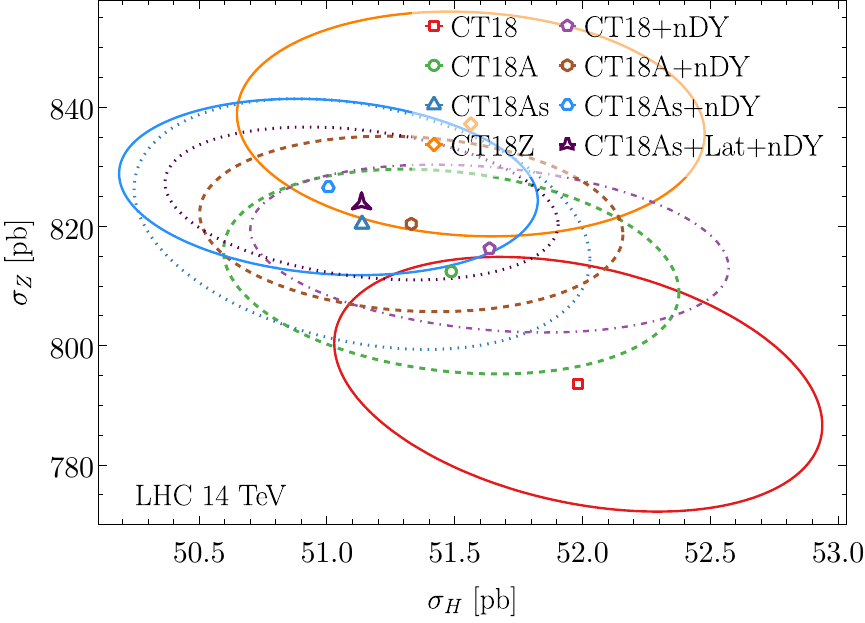}
    \includegraphics[width=0.49\textwidth]{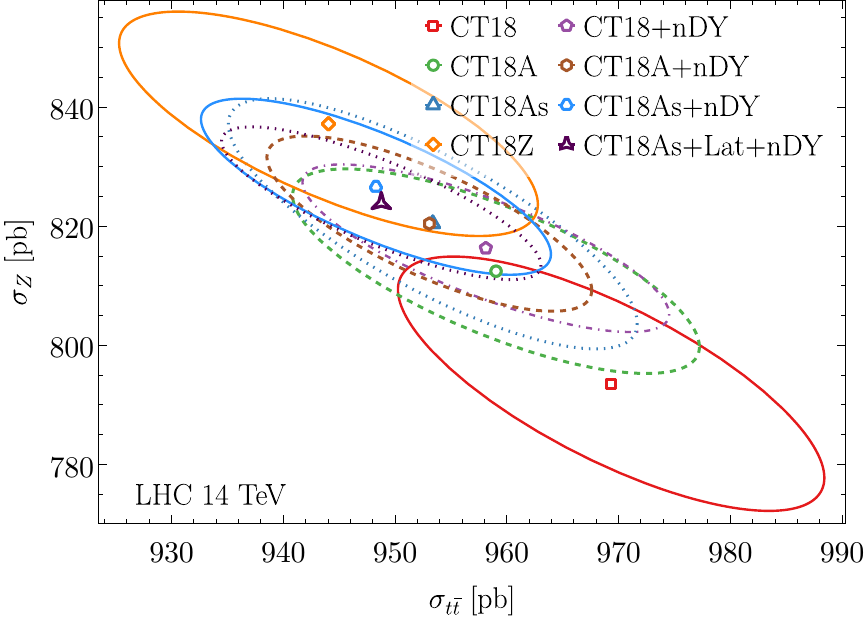}     
    \caption{The 68\% CL correlation ellipses among the fiducial $W^\pm,Z$ and the inclusive $H$, $t\bar{t}$ and $t\bar{t}H$ productions at the  14 TeV LHC. See the text for its details.}
    \label{fig:ellipse}
\end{figure}

As examples of phenomenological implications, we present the correlation ellipses of the LHC 14 TeV fiducial $W^\pm,Z$ and inclusive Higgs boson ($H$), top-quark pair $(t\bar t)$, and associated $t\bar{t}H$ productions in Fig.~\ref{fig:ellipse}, as well as the corresponding correlation cosine to various PDF flavors at $Q=100~\GeV$ in Fig.~\ref{fig:CosPhiInc} and  Fig.~\ref{fig:CosPhiRS} of Appendix~\ref{app:supp}.
The fiducial $W^\pm,Z$ cross sections correspond to the same kinematic cuts adopted in the ATLAS 13 TeV measurement~\cite{ATLAS:2016fij},
\begin{eqnarray}
W^{\pm}: & \quad p_T^{\ell,\nu}>25~\GeV,~ |\eta_\ell|<2.5, ~m_T>50~\GeV,\\
Z: &\quad  p_T^\ell>25~\GeV, ~|\eta_l|<2.5, ~ 66<m_{\ell\ell}<116~\GeV.
\end{eqnarray}
The calculation is performed with the NLO APPLgrid~\cite{Carli:2010rw} with the NNLO $K$-factors from the MCFM~\cite{Campbell:2019dru}.
The Higgs, $t\bar{t}$, and $t\bar{t}H$ productions refer to the full phase space, without decays. 
The $t\bar{t}$ cross section is calculated with Top++~\cite{Czakon:2011xx} at NNLO with soft gluons resummed up to the NNLL level and factorization and renormalization scales are set to the top-quark mass $m_t$. The Higgs boson production is calculated with ggHiggs~\cite{Bonvini:2016frm} at N3LO with the threshold resummation up to N3LL, and scale as the Higgs mass $m_H$.
The $t\bar{t}H$ associated production is calculated with MadGraph\_aMC@NLO~\cite{Frederix:2018nkq} interfacing with PineAPPL fast interpolation grid~\cite{Carrazza:2020gss} at the NLO, and the scale as the partonic collision energy $\sqrt{\hat{s}}$.

Taking the CT18 as a reference, we see that both the ATLAS 7 TeV $W,Z$ data and the post-CT18 LHC Drell-Yan data sets pull the $W^\pm$ and $Z$ boson cross sections to a larger value. The correlations in $(W^+,W^-)$ and $(W^\pm,Z)$ are unchanged, with the reduction of corresponding uncertainties, reflecting the change in the quark-antiquark luminosity. 
In comparison, Higgs, top-quark pair, and associated $t\bar{t}H$ cross sections are pulled to the smaller values, as a result of the reduction of $gg$ luminosity.
We also observe the reduction of the error band  with respect to the CT18, due to the inclusion of the post-CT18 LHC Drell-Yan data sets in the fits. 

Generally speaking, the theory prediction of CT18+nDY deviates from CT18 
in a similar way as how CT18Z deviates from CT18, as shown in Fig.~\ref{fig:ellipse}. In most cases, the predicted cross sections of CT18As+nDY lie between those of CT18 and CT18Z,  except that 
$\sigma_H({\rm CT18}) > \sigma_H({\rm CT18Z}) > \sigma_H({\rm CT18As+nDY})$.
This can be understood by noting the very similar pattern of difference in various flavor PDFs between CT18As+nDY and CT18Z, when comparing to CT18, \emph{e.g.}, $s$ and $g$ PDFs, as shown in Fig.~\ref{fig:CT18Z} of Appendix~\ref{app:supp}.
Excluding CT18Z, the CT18As+nDY PDFs generate the largest deviations from the CT18 predictions.
As first pointed out in the CTEQ6.6 analysis~\cite{Nadolsky:2008zw}, the mutual dispositions of the $(W^\pm,Z)$ error ellipse, \emph{i.e.}, the top right panel of Fig.~\ref{fig:ellipse}, can be associated to the differences among the strangeness and gluon PDFs.
The direction parallel to the semi-minor axis, associated with the relative cross-section ratio $R_{Z/W^\pm}=\sigma_{Z}/\sigma_{W^\pm}$, is most closely identified with the strange PDF, as shown in terms of the correlation cosine in Fig.~\ref{fig:CosPhiRS} (left).
Meanwhile, the semi-major axis, \emph{i.e.}, the ``$\sigma_{Z} +\sigma_{W^\pm}$" direction, is most related to the gluon PDF, as shown with respect to the correlation cosine in Fig.~\ref{fig:CosPhiRS} (right). 

When looking closely at the correlation cosines between various cross sections and flavor PDFs in Fig.~\ref{fig:CosPhiInc}, we realize that the large anti-correlation between the $W^\pm,Z$ cross sections and gluon-PDF happens at $x\sim10^{-1}$, where a large positive correlation is observed for the $t\bar{t}$ and $t\bar{t}H$ cross sections. In comparison, the biggest correlation between the Higgs cross section and gluon-PDF occurs at a smaller $x$ value, around $x\sim10^{-2}$. We also present the $(Z,H/t\bar{t})$ correlation ellipses in Fig.~\ref{fig:ellipse}, which shows a large anti-correlation between the $Z$ and $t\bar{t}$ cross sections. In comparison, the $(Z,H)$ anti-correlation is relatively weaker.
Again, after excluding CT18Z, the largest deviation from the CT18 predictions occurs when using the CT18As+nDY PDFs. The inclusion of additional lattice constraint on strangeness asymmetry in  CT18As+Lat+nDY yields a reduced difference, to a relatively minor extent.

\section{Discussion and Conclusion}
\label{sec:conclude}

Since the release of the CT18 family of PDFs~\cite{Hou:2019efy},  a large number of
LHC precision data  become available. 
In this work, we perform a detailed study on  the impact of
the post-CT18 LHC Drell-Yan precision data on the PDFs in the framework of CT18 global analysis. 

From the ATLAS measurements, we have considered the 5.02 TeV TeV $W,Z$ production (ATL5WZ)~\cite{ATLAS:2018pyl}, the 8 TeV $W$ production (ATL8W)~\cite{ATLAS:2019fgb}, and the 8 TeV triple differential distribution neutral-current Drell-Yan $Z$ production data $\dd^3\sigma/(\dd m_{\ell\ell}\dd y_{\ell\ell}\dd\cos\theta^*)$ (ATL8Z3D)~\cite{ATLAS:2017rue}.
Similarly, we have included the CMS 13 TeV $Z$ boson production (CMS13Z)~\cite{CMS:2019raw}, and the LHCb 8 TeV $W$ boson production (LHCb8W)~\cite{LHCb:2016zpq} as well as the 13 TeV $Z$ boson production (LHCb13Z)~\cite{LHCb:2016fbk,LHCb:2021huf}. 
In this work, we mainly focus on the (pseudo)rapidity distributions around the peak region. The high-mass invariant mass distributions and the transverse momentum measurements require dedicated treatments of the inclusion of photon PDF in calculations of the electroweak corrections and the nonperturbative transverse momentum dependent (TMD) parton density (to account for the effect of multiple soft gluon emissions), respectively, which are left for future studies. 

We have compared the MCFM NNLO fixed order predictions and the ResBos2 matched $q_T$ N3LL resummation calculations for the distributions we considered in this work. The difference is generally at a percent level, as a reflection of the impact of fiducial cuts on different high-order treatments.
With a limited computational resource, we generally obtain larger Monte-Carlo uncertainties for MCFM NNLO calculation than the ResBos2 resummation calculation, which will propagate to the global fitting and lead to a worse description (larger $\chi^2$) of data. For this reason, the final presentations of this work are mainly based on the ResBos2 resummation calculation.

We have identified the correlation between the experimental data points and the PDF flavors in a large momentum faction range $x\in[10^{-5},1]$, with the scale choice $Q=100~\GeV$, as an example. 
We find that the $W^+$ boson data can provide significant constraint on the $d(\bar{d})$ PDFs around $x\sim10^{-3}$, while the $W^-$ on the $u(\bar{u})$ PDFs. In comparison, the $Z$ boson production can potentially constrain the $s(\bar s)$ PDFs at $x\sim 10^{-2}$.
With respect to the single differential distribution, the triple one in the ATLAS 8 TeV $Z$ production can provide additional information.
In addition, the LHCb measurements in the forward region can extend the sensitivity to the lower-$x$ region. 

With the ePump's PDF updating and the CT global fitting, we have examined the impact of the post-CT18 LHC Drell-Yan data sets, one at a time,
on the CT18, CT18A and CT18As PDFs. 
We remind the reader that the CT18A analysis includes the ATLAS 7 TeV $W,Z$ precision data (ATL7WZ)~\cite{ATLAS:2016nqi} as an alternative
fit due to its apparent tension with other pre-existing data in the CT18, and the CT18As fit includes additional degrees of freedom in the strangeness PDF parameterization, which relaxes this tension.
Overall, the impact of the post-CT18 LHC Drell-Yan data on the CT18 strangeness PDF is consistent with ATL7WZ  in the CT18A fit. However, when examining more closely the $\bar{d}(d)$ PDFs, we obtain an opposite pull, driven by the ATL8W data, suggesting the tension with the ATL7WZ data. A similar tension from the LHCb8W data is observed at large $x$ around $x\sim0.3$, but with a minor impact due to the large PDF uncertainty in the large $x$ region. With the additional shape parameter in the $s(\bar s)$ PDFs of the CT18As fit,
the tension between the ATL8W and ATL7WZ is relaxed.  

We have also performed global fits to simultaneously include all
these post-CT18 LHC Drell-Yan data sets, based on the CT18, CT18A and CT18As setups. The general feature agrees with the individual fits, \emph{i.e.}, 
the post-CT18 LHC Drell-Yan data sets pull the $ s(\bar s)$ PDFs  in the same direction as CT18A fit, but with a weaker strength. In contrast, the post-CT18 LHC Drell-Yan data sets pull the $\bar{d}(d)$ PDFs to the opposite direction of the  ATL7WZ data. The simultaneous inclusion of the ATL7WZ and post-CT18 LHC Drell-Yan data sets in the CT18A+nDY fit accumulates the pull on $ s(\bar s)$ PDFs, but balances the impact on $\bar{d}(d)$ PDFs.
With the additional degrees of freedom in the CT18As fit, the impact of the post-CT18 LHC Drell-Yan data on the $ s(\bar s)$ PDFs is quite minimal, while the impact on $\bar{d}(d)$ PDFs remains.
Regarding the PDF uncertainties, the post-CT18 LHC Drell-Yan data can significantly reduce the $ s(\bar s)$ PDF error bands, similar to the ATL7WZ data. The more flexible parameterization of the $s(\bar s)$ PDFs  in CT18As can enlarge the $\bar{d}(d)$ and $s(\bar{s})$ PDF error bands. Under such a condition, the post-CT18 LHC Drell-Yan data can help shrink the $d(\bar{d})$ PDF uncertainties but leave the $ s(\bar s)$ almost unchanged.
With the inclusion of the lattice $s_{-}$  data~\cite{Zhang:2020dkn} in CT18As+Lat, the result of global fit CT18As+Lat+nDY is similar to CT18As+nDY, except the strangeness asymmetry $s_{-}$ in the large-$x$ region.

Finally, we have examined the impact of post-CT18 LHC Drell-Yan data on the parton luminosities as well as the corresponding phenomenological implications, by considering the correlation among $W^+$, $W^-$, $Z$, $H$, $t\bar t$ and $t {\bar t}H$ at the 14 TeV LHC, as examples. 
We find that the post-CT18 LHC Drell-Yan data normally enhance (reduce) the quark-related parton luminosities (such as $\mathcal{L}_{g\Sigma,q\bar{q},\Sigma\Sigma}$) at low (high) invariant mass region. In comparison, the $gg$ luminosity $\calL_{gg}$ is reduced.
Normally, the luminosity uncertainties shrink due to the constraining power of the post-CT18 LHC Drell-Yan data sets. The specific size of reduction can depend on the nonpertubative parameterization forms of PDFs.
As a consequence of the quark-antiquark luminosity, the predicted cross sections for $W^\pm$ and $Z$ productions at the 14 TeV LHC are enhanced, while their corresponding error bands are reduced.
On the other hand, the total cross sections of Higgs, top-quark pair, and associated $t\bar{t}H$ processes are pulled to the smaller values, as a result of the reduction of $gg$ luminosity.

As emphasized in the CT18 global analysis~\cite{Hou:2019efy} as well as some follow-up studies~\cite{Courtoy:2022ocu}, the inclusion of more high-precision data from the LHC does not necessarily yield the more precise PDFs, due to the pulls in different direction originated from tensions among the data sets.
For this reason, we have released another fit such as CT18A PDFs to incorporate the additional ATLAS 7 TeV $W,Z$ precision data (ATL7WZ), which is found in tension with some existing data sets, such as the HERA I+II combined DIS data and the neutrino DIS dimuon production.
In this study, we found that some of the post-CT18 LHC Drell-Yan data are consistent with the ATL7WZ data, such as ATL8Z3D and CMS13Z. But some tension with the ATL7WZ data is still found among other data sets, such as the ATL8W data and, to a lesser extent, LHCb8W. 
With a more flexible non-perturbative parameterization of PDFs,
this tension can be relaxed to some extent but not completely resolved. 
A full understanding of this tension is beyond the scope of this work, which is left for future work, especially the next round of CTEQ-TEA global analysis, to which this study will provide essential inputs.

\begin{acknowledgements}
We would like to thank Yao Fu as well as other CTEQ-TEA colleagues for many helpful discussions.
The work of S. Dulat and I. Sitiwaldi are supported by the National Natural Science Foundation of China under Grant No.11965020 and Grant No. 11847160, respectively.
The work of KX is supported by the U.S. Department of Energy under grant No. DE-SC0007914, the U.S. National Science Foundation under Grants No. PHY-1820760, and also in part by the PITT PACC. 
The work of C.-P. Yuan is supported by the U.S. National Science Foundation under Grant No. PHY-2013791. C.-P. Yuan is also grateful for the support from the Wu-Ki Tung endowed chair in particle physics.
\end{acknowledgements}

\appendix

\section{Supplementary figures}
\label{app:supp}

We collect many supplementary figures in this appendix, with the corresponding detailed explanations to be found in the main text.

\begin{figure}[!h]
\centering
\includegraphics[width=0.49\textwidth]{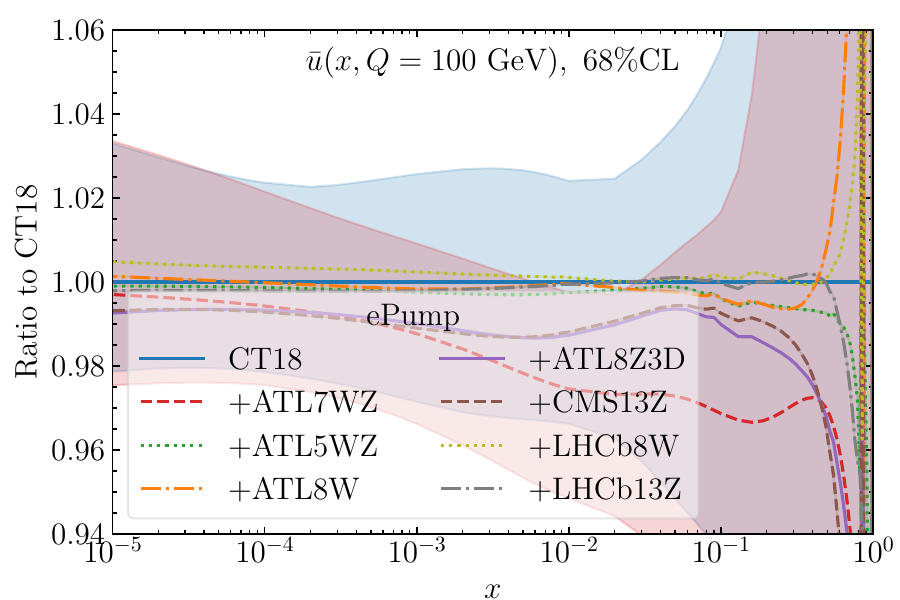} 
\includegraphics[width=0.49\textwidth]{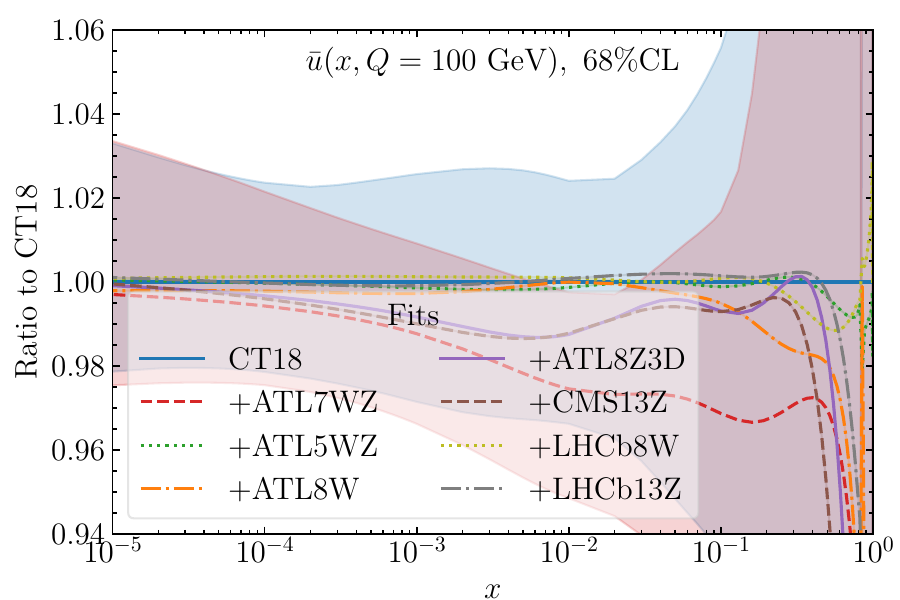}
\includegraphics[width=0.49\textwidth]{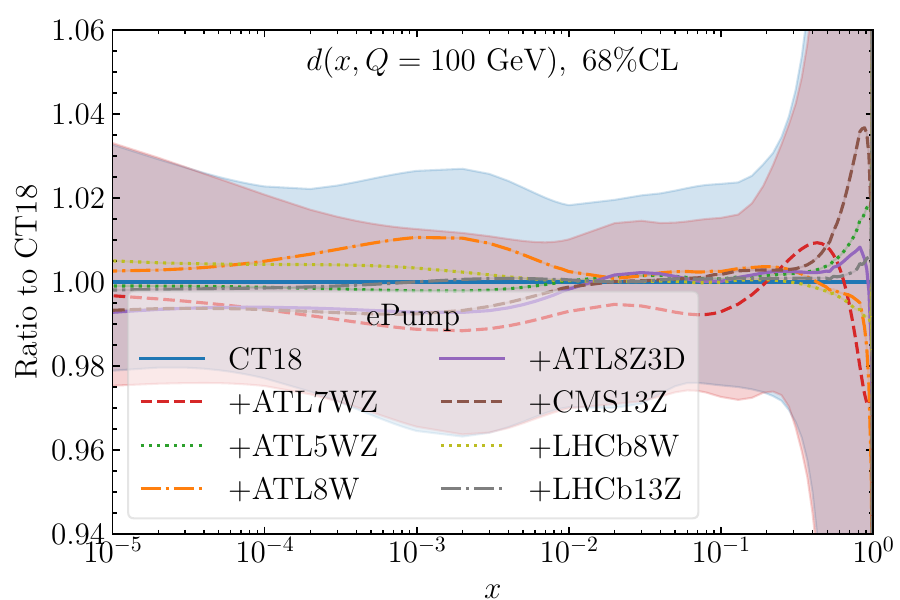}
\includegraphics[width=0.49\textwidth]{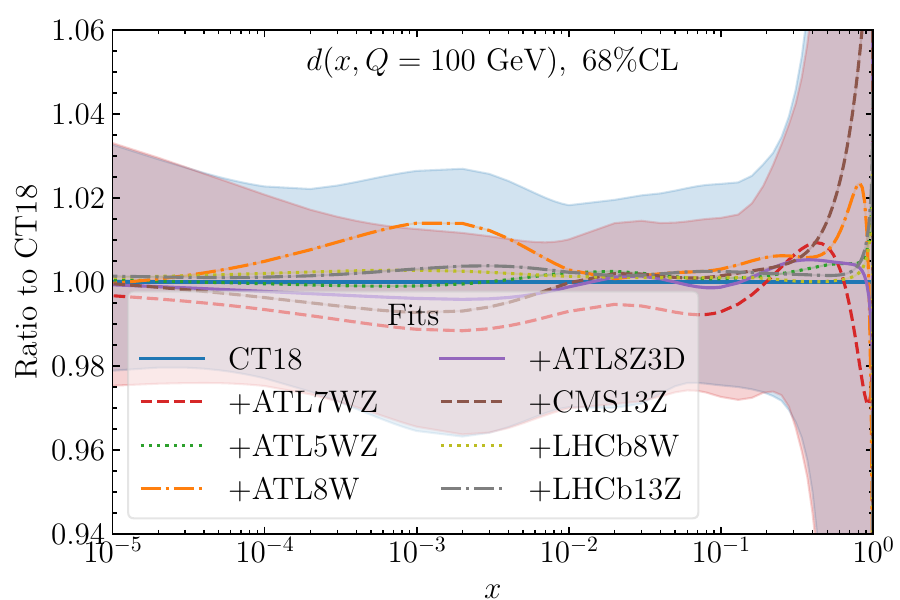}
\includegraphics[width=0.49\textwidth]{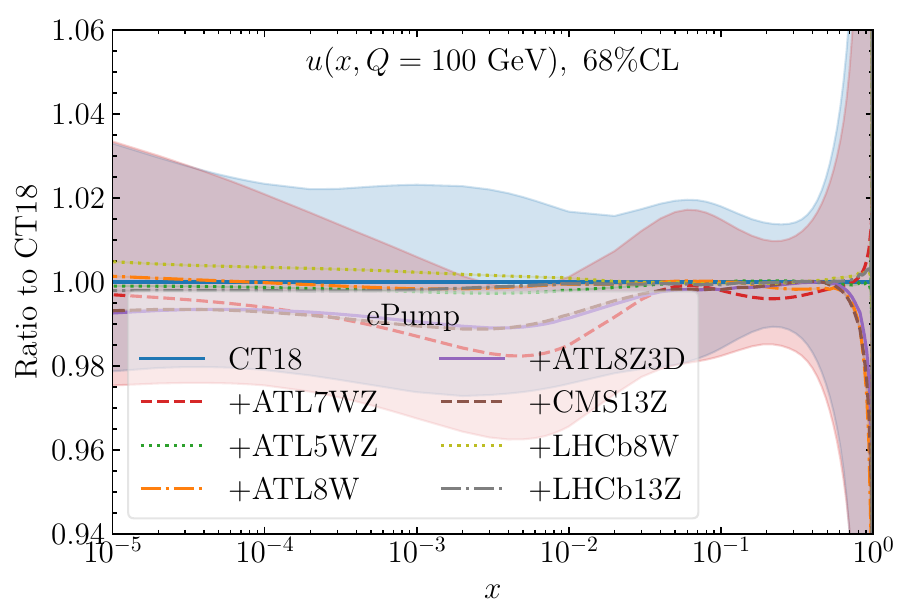}
\includegraphics[width=0.49\textwidth]{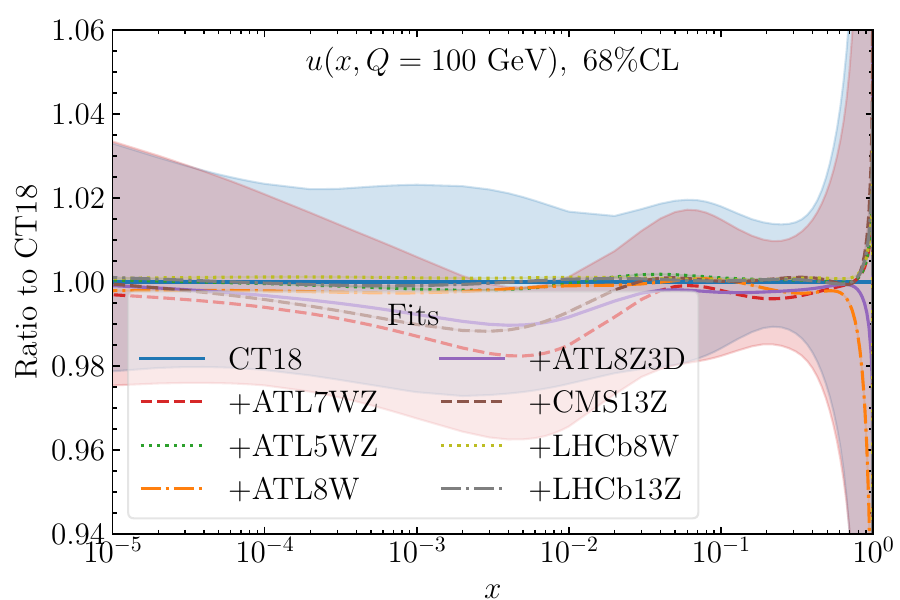}
\caption{Similar to Fig.~\ref{fig:FitOne}, but for $\bar{u},g,d,u$ PDFs at $Q=100~\GeV$.}
\label{fig:FitOne2}
\end{figure}

\begin{figure}[!h]
    \centering
    \includegraphics[width=0.49\textwidth]{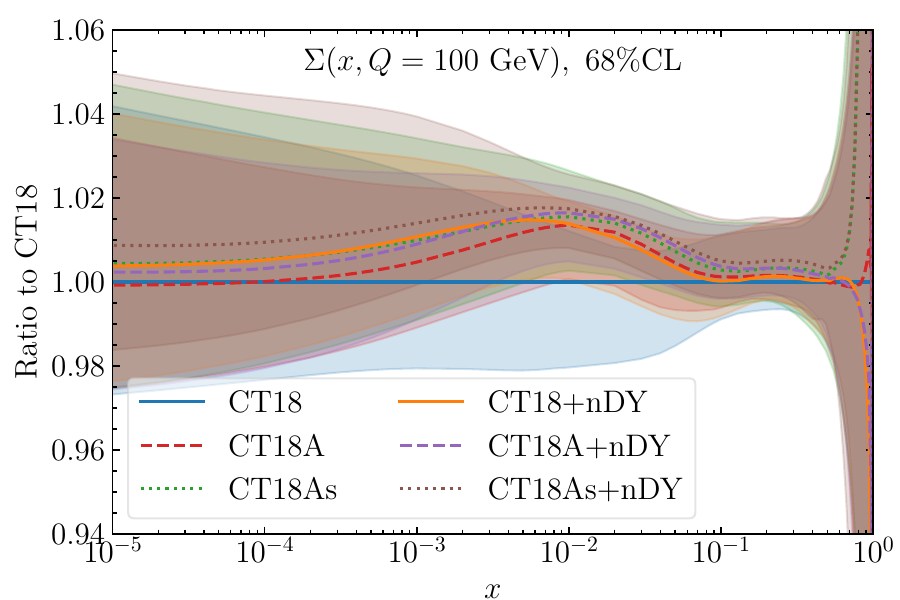}
    \includegraphics[width=0.49\textwidth]{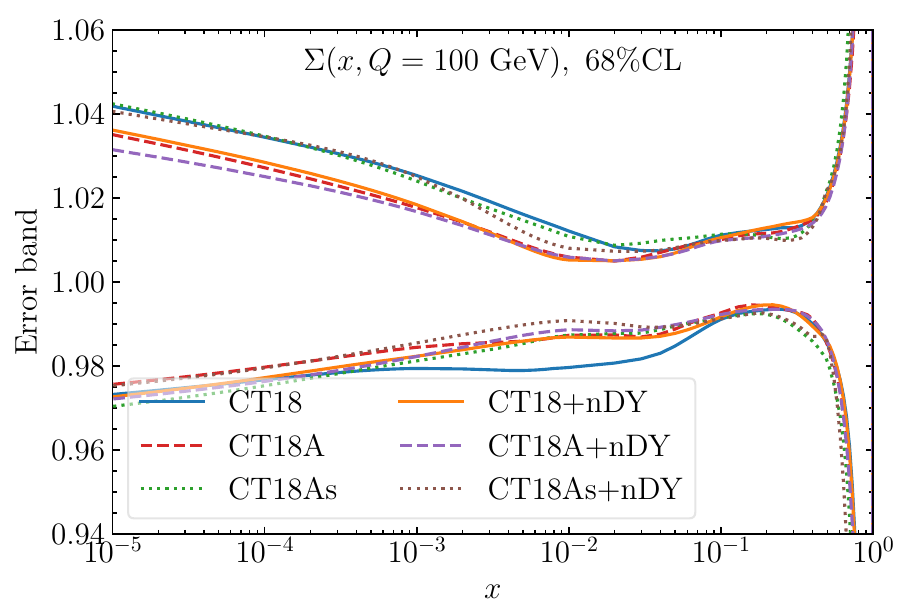}     
    \caption{Similar to Fig.~\ref{fig:FitAll}, but for the flavor singlet $\Sigma$.}
    \label{fig:sglt}
\end{figure}

\begin{figure}[!h]
    \centering
    \includegraphics[width=0.49\textwidth]{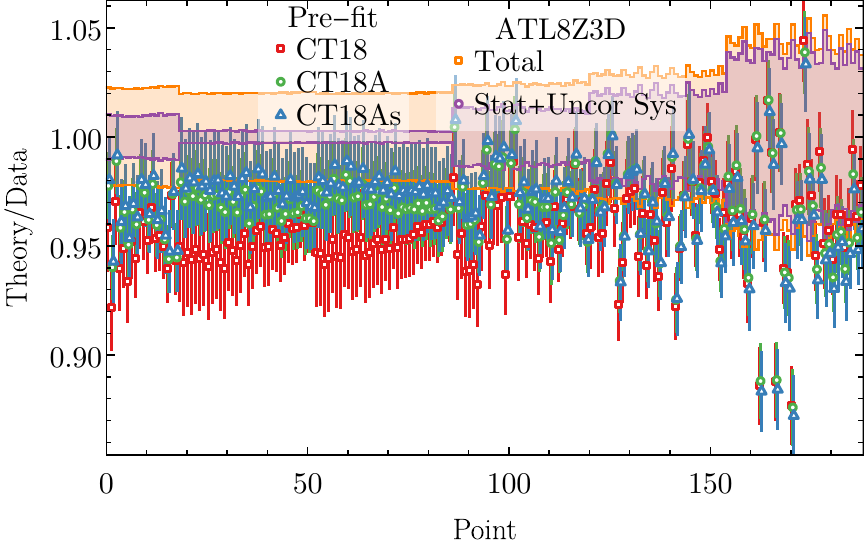}
    \includegraphics[width=0.49\textwidth]{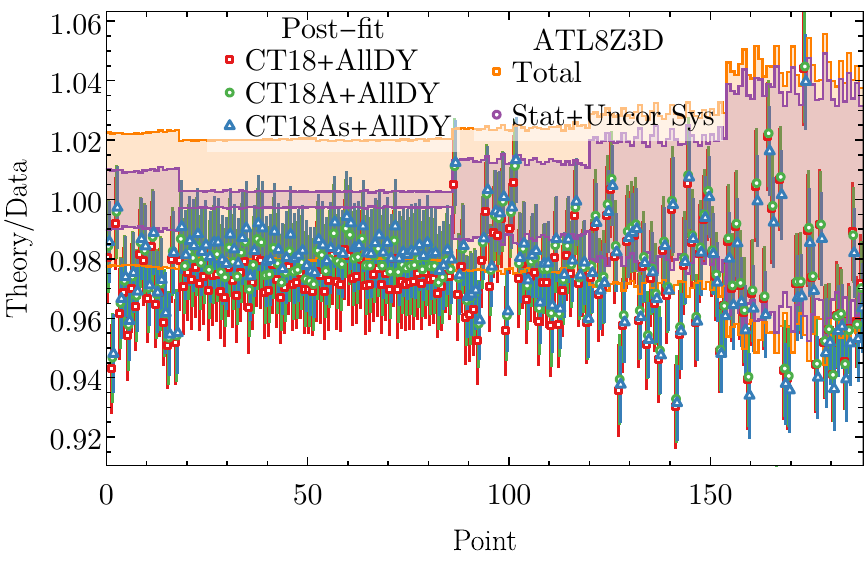}
    \caption{Similar to Fig.~\ref{fig:thATL}, but for the ATLAS 8 TeV triple differential distributions.}
    \label{fig:thATL8Z3D}
\end{figure}

\begin{figure}[!h]
    \centering
    \includegraphics[width=0.49\textwidth]{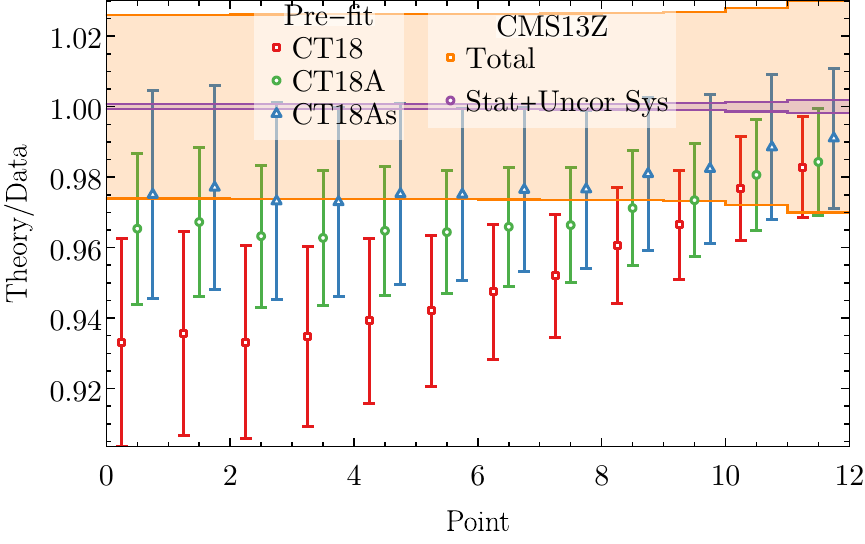}
    \includegraphics[width=0.49\textwidth]{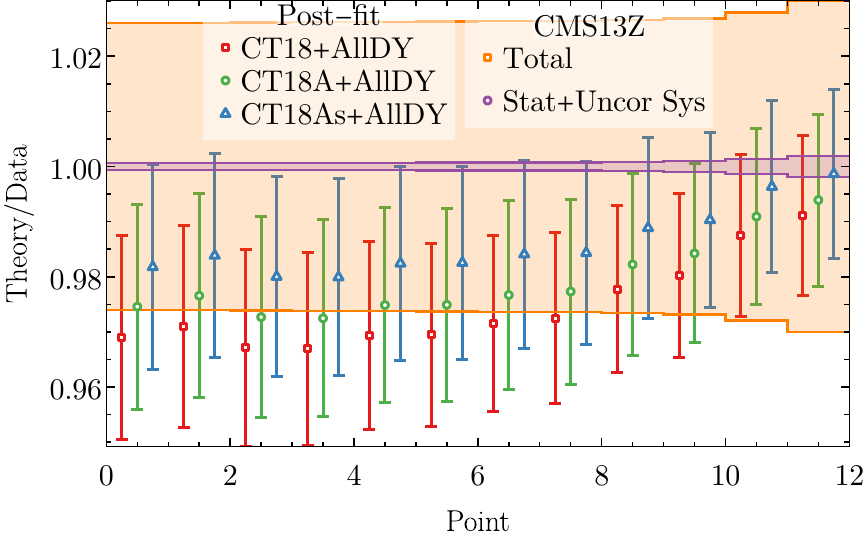}      
    \caption{Similar to Fig.~\ref{fig:thATL}, but for CMS 13 TeV $Z$ production.}
    \label{fig:thCMS}
\end{figure}

\begin{figure}[!h]
    \centering
    \includegraphics[width=0.49\textwidth]{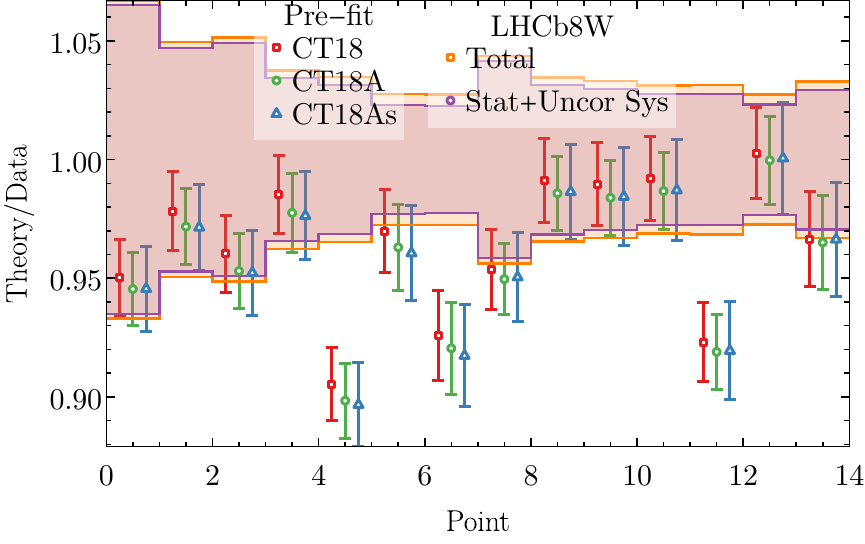}
    \includegraphics[width=0.49\textwidth]{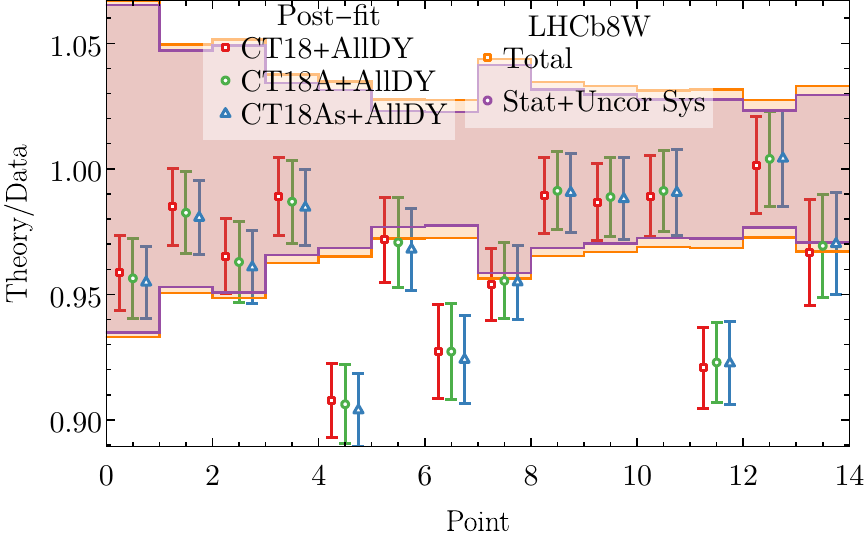}
    \includegraphics[width=0.49\textwidth]{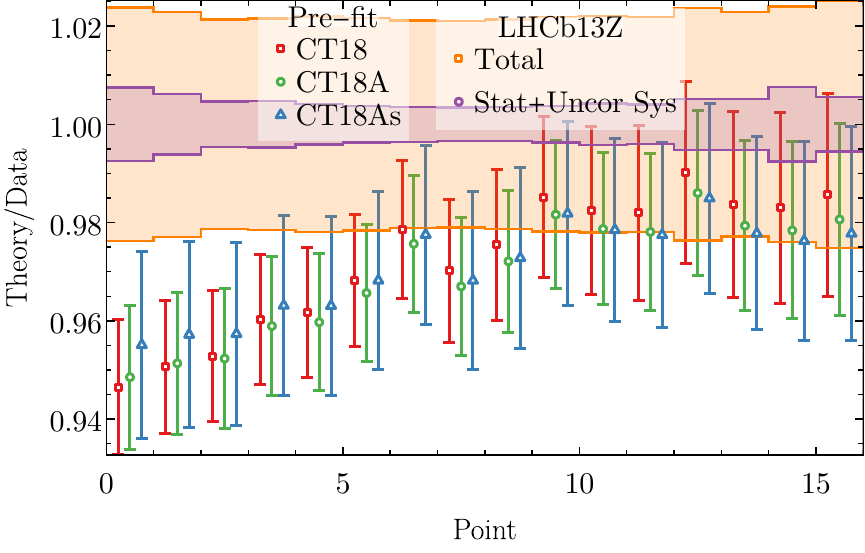}
    \includegraphics[width=0.49\textwidth]{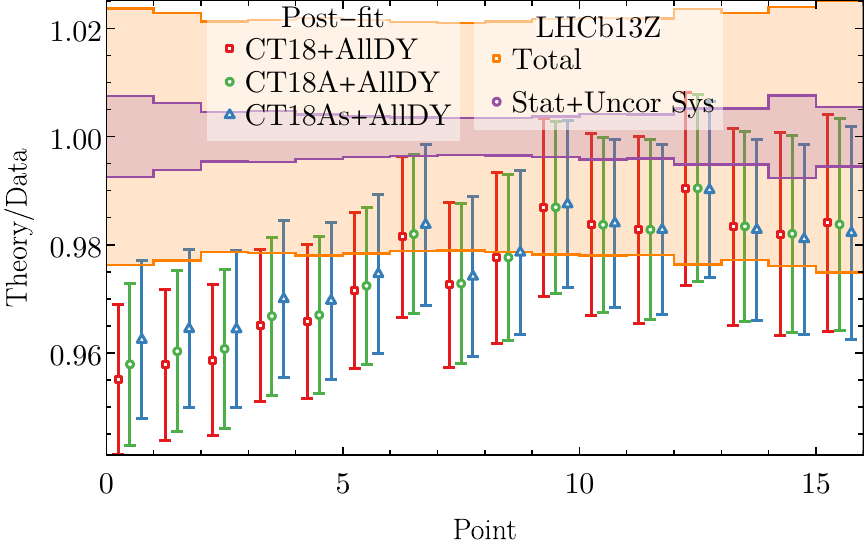}      
    \caption{Similar to Fig.~\ref{fig:thATL}, but for the LHCb 8 TeV $W$ and 13 TeV $Z$ data sets.}
    \label{fig:thLHCb}
\end{figure}

\begin{figure}[!h]
    \centering
    \includegraphics[width=0.49\textwidth]{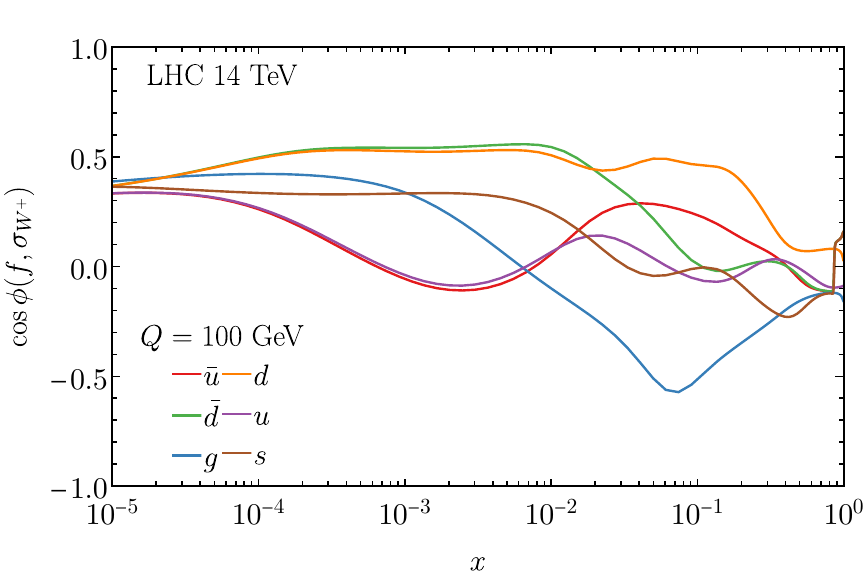}    
    \includegraphics[width=0.49\textwidth]{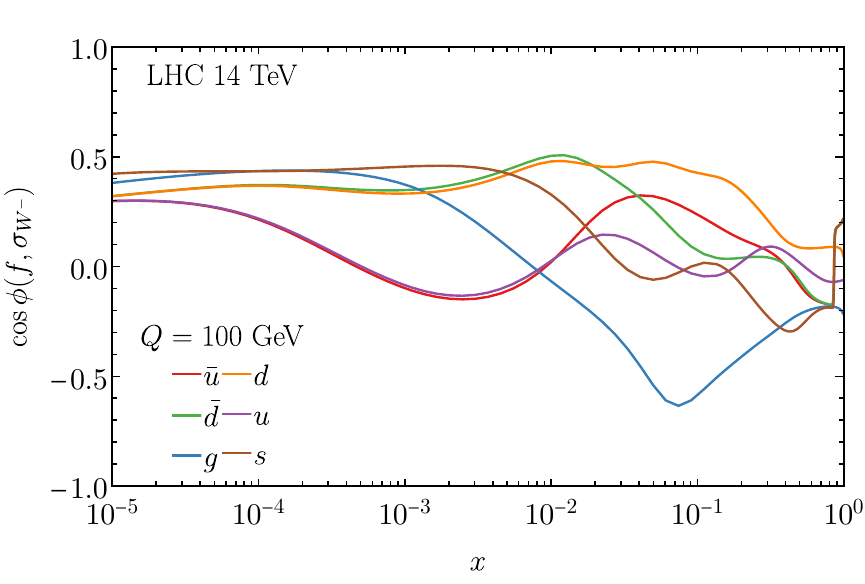} 
    \includegraphics[width=0.49\textwidth]{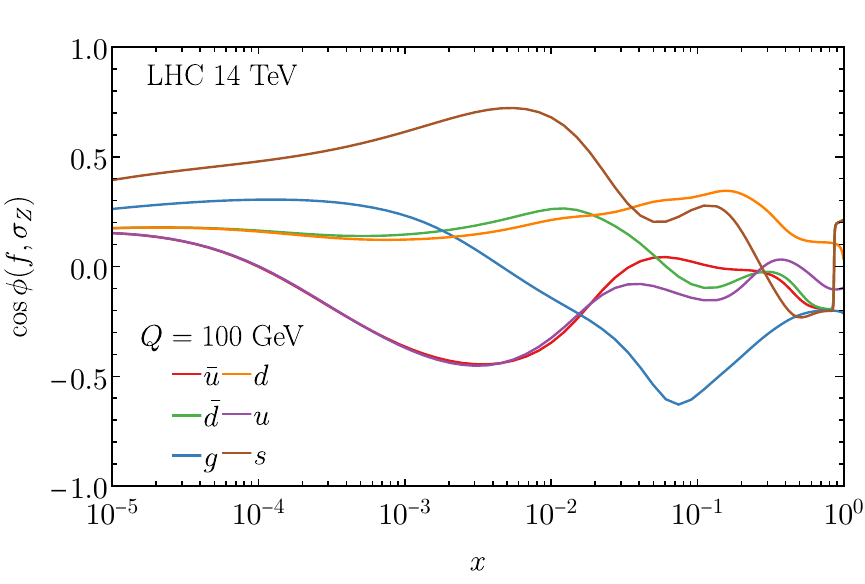}    
    \includegraphics[width=0.49\textwidth]{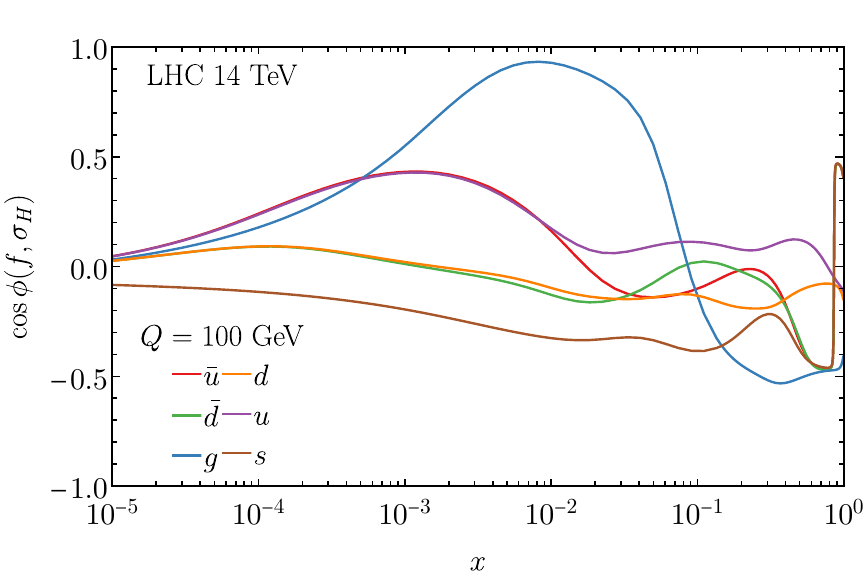}
    \includegraphics[width=0.49\textwidth]{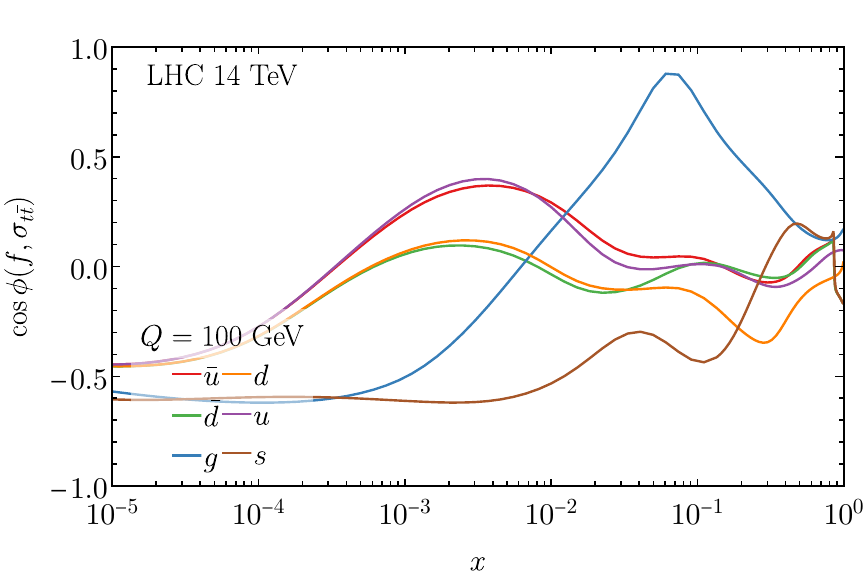}
    \includegraphics[width=0.49\textwidth]{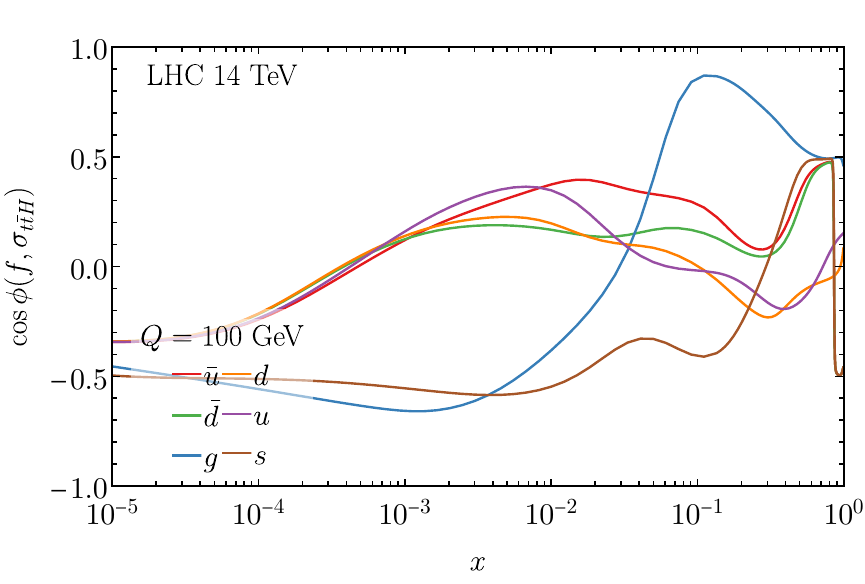}   
    \caption{The correlation cosine between the fiducial $W^\pm,Z$ and inclusive $t\bar{t},H,t\bar{t}H$ cross sections with the CT18 PDF flavors at $Q=100~\GeV$.}
    \label{fig:CosPhiInc}
\end{figure}

\begin{figure}[!h]
    \centering
    \includegraphics[width=0.49\textwidth]{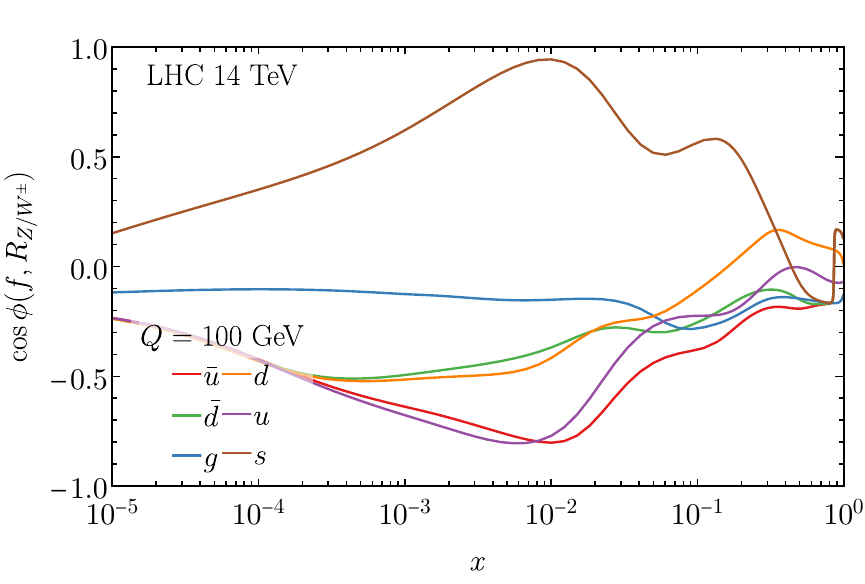}    
    \includegraphics[width=0.49\textwidth]{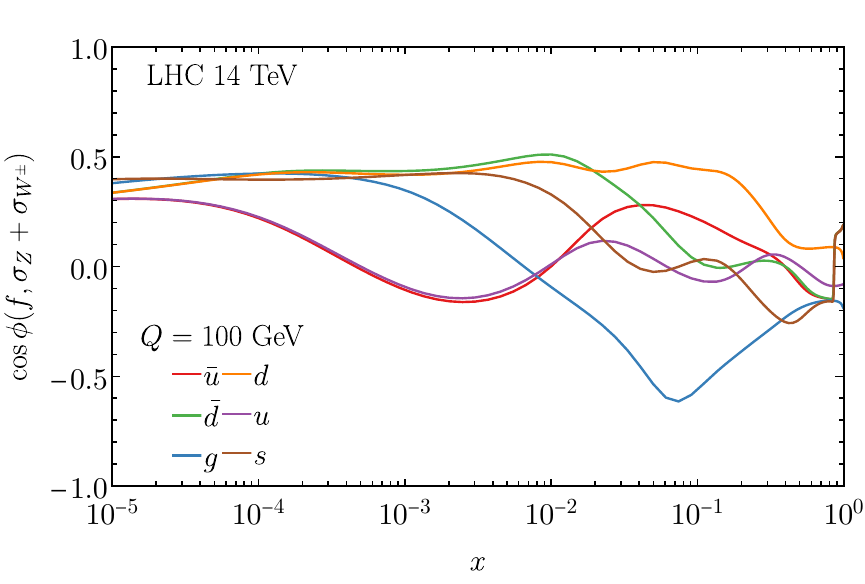}        
    \caption{The correlation cosine of the cross-section ratio $R_{Z/W^\pm}$ and sum $\sigma_{Z}+\sigma_{W^\pm}$ with the CT18 PDF flavors at $Q=100~\GeV$.}
    \label{fig:CosPhiRS}
\end{figure}

\begin{figure}
    \centering
    \includegraphics[width=0.49\textwidth]{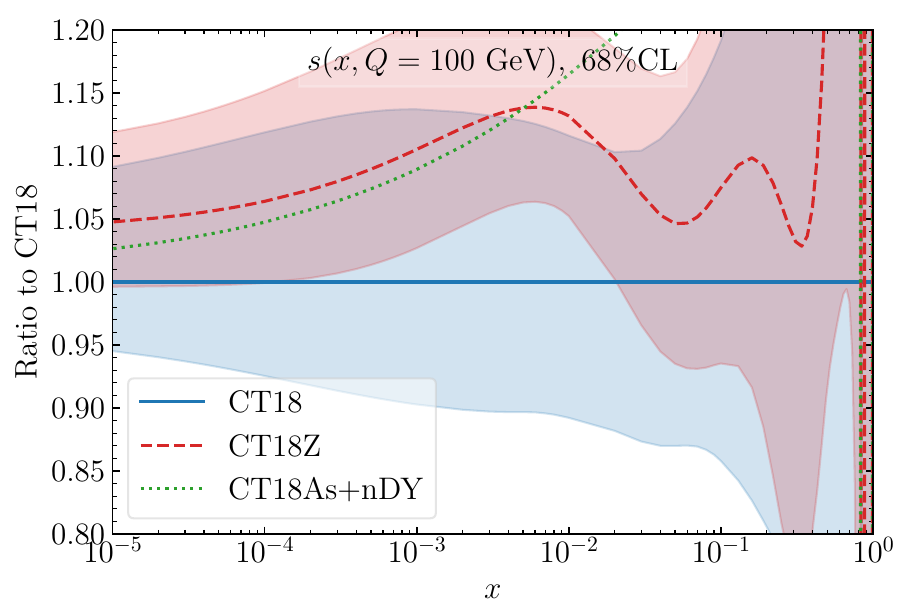}
    \includegraphics[width=0.49\textwidth]{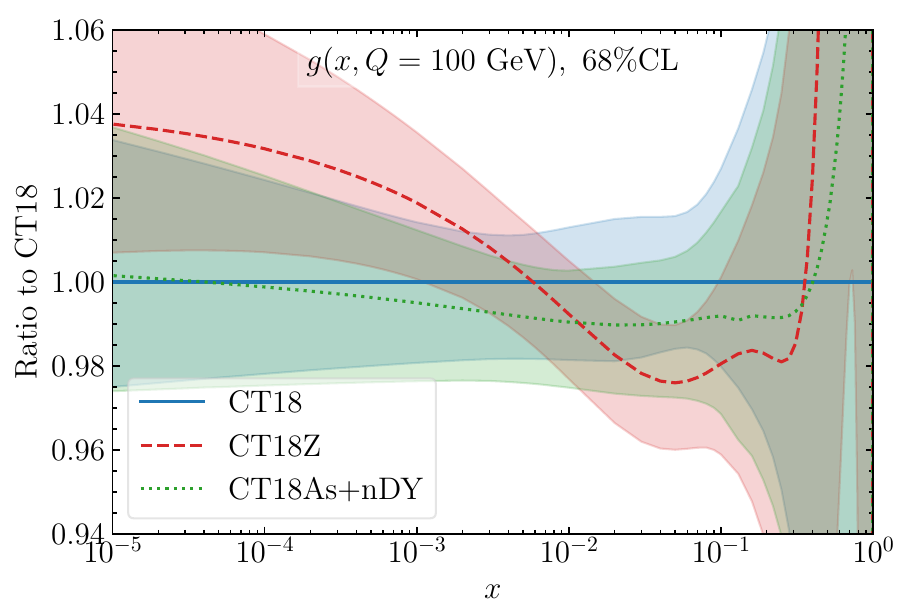}    
    \caption{The comparison of strangeness and gluon PDFs in the CT18, CT18Z, and CT18As+nDY fits at $Q=100~\GeV$.}
    \label{fig:CT18Z}
\end{figure}


\providecommand{\href}[2]{#2}\begingroup\raggedright\endgroup

\end{document}